\UseRawInputEncoding

\documentclass[twocolumn]{aastex62}
\usepackage{natbib}
\usepackage{graphicx}
\usepackage{array}
\usepackage[caption=false]{subfig}
\usepackage{makecell}

\graphicspath{{./}{}}

\received{7/26/2022}
\revised{11/2/2022}
\accepted{9/12/2022}
\submitjournal{AJ}

\shorttitle{The GAPlanetS Campaign}
\shortauthors{Follette et al.}

\begin{document}

\title{The Giant Accreting Protoplanet Survey (GAPlanetS) - Results from a Six Year Campaign to Image Accreting Protoplanets}

\correspondingauthor{Katherine B. Follette}
\email{kfollette@amherst.edu}

\author[0000-0002-7821-0695]{Katherine B. Follette}
\affiliation{Department of Physics \& Astronomy, Amherst College, 25 East Drive, Amherst, MA 01002, USA}

\author[0000-0002-2167-8246]{Laird M. Close}
\affiliation{Steward Observatory, University of Arizona, Tucson, 933 N Cherry Ave, Tucson, AZ 85721, USA}

\author{Jared R. Males}
\affiliation{Steward Observatory, University of Arizona, Tucson, 933 N Cherry Ave, Tucson, AZ 85721, USA}

\author[0000-0002-4479-8291]{Kimberly Ward-Duong}
\affiliation{Department of Astronomy, Smith College, Northampton MA 01063 USA}

\author[0000-0001-6396-8439]{William O. Balmer}
\affiliation{Department of Physics \& Astronomy, Johns Hopkins University, 3400 N. Charles Street, Baltimore, MD 21218, USA}
\affiliation{Space Telescope Science Institute, 3700 San Martin Drive, Baltimore MD 21218, USA}
\affiliation{Department of Physics \& Astronomy, Amherst College, 25 East Drive, Amherst, MA 01002, USA}

\author[0000-0002-4489-3168]{J\'ea Adams Redai}
\affiliation{Center for Astrophysics, Harvard \& Smithsonian, 60 Garden Street, Cambridge, MA 02138, USA}
\affiliation{Department of Physics \& Astronomy, Amherst College, 25 East Drive, Amherst, MA 01002, USA}

\author[0000-0001-7525-7423]{Julio Morales}
\affiliation{Department of Astronomy, University of Massachusetts, Amherst, MA, 01003, USA}

\author{Catherine Sarosi}
\affiliation{Department of Physics \& Astronomy, Amherst College, 25 East Drive, Amherst, MA 01002, USA}

\author{Beck Dacus}
\affiliation{Department of Physics \& Astronomy, Amherst College, 25 East Drive, Amherst, MA 01002, USA}

\author[0000-0002-4918-0247]{Robert J. De Rosa}
\affiliation{European Southern Observatory, Alonso de C\'{o}rdova 3107, Vitacura, Santiago, Chile}

\author[0000-0002-0618-1119]{Fernando Garcia Toro}
\affiliation{Department of Physics \& Astronomy, Amherst College, 25 East Drive, Amherst, MA 01002, USA}

\author{Clare Leonard}
\affiliation{Department of Physics \& Astronomy, Amherst College, 25 East Drive, Amherst, MA 01002, USA}

\author{Bruce Macintosh}
\affiliation{Kavli Institute for Particle Astrophysics and Cosmology, Department of Physics, Stanford University, Stanford, CA, 94305, USA}

\author[0000-0002-1384-0063]{Katie M. Morzinski}
\affiliation{Steward Observatory, University of Arizona, Tucson, 933 N Cherry Ave, Tucson, AZ 85721, USA}

\author{Wyatt Mullen}
\affiliation{Kavli Institute for Particle Astrophysics and Cosmology, Department of Physics, Stanford University, Stanford, CA, 94305, USA}

\author{Joseph Palmo}
\affiliation{Department of Physics \& Astronomy, Amherst College, 25 East Drive, Amherst, MA 01002, USA}

\author{Raymond Nzaba Saitoti}
\affiliation{Department of Physics \& Astronomy, Amherst College, 25 East Drive, Amherst, MA 01002, USA}


\author{Elijah Spiro}
\affiliation{Department of Physics \& Astronomy, Amherst College, 25 East Drive, Amherst, MA 01002, USA}

\author[0000-0003-0660-9776]{Helena Treiber}
\affiliation{Department of Physics \& Astronomy, Amherst College, 25 East Drive, Amherst, MA 01002, USA}

\author{Kevin Wagner}
\affiliation{Steward Observatory, University of Arizona, Tucson, 933 N Cherry Ave, Tucson, AZ 85721, USA}
\altaffiliation{NASA Hubble Fellowship Program $-$ Sagan Fellow}

\author[0000-0003-0774-6502]{Jason Wang (王劲飞)}
\altaffiliation{51 Pegasi b Fellow}
\affiliation{Department of Astronomy, California Institute of Technology, Pasadena, CA 91125, USA}

\author{David Wang}
\affiliation{Department of Physics \& Astronomy, Amherst College, 25 East Drive, Amherst, MA 01002, USA}

\author{Alex Watson}
\affiliation{Department of Physics \& Astronomy, Amherst College, 25 East Drive, Amherst, MA 01002, USA}

\author[0000-0001-6654-7859]{Alycia J. Weinberger}
\affiliation{Earth \& Planets Laboratory, Carnegie Institution for Science, 5241 Broad Branch Rd NW, Washington, DC 20015, USA}



\begin{abstract}

Accreting protoplanets are windows into planet formation processes, and high-contrast differential imaging is an effective way to identify them. We report results from the Giant Accreting Protoplanet Survey (GAPlanetS), which collected H$\alpha$ differential imagery of 14 transitional disk host stars with the Magellan Adaptive Optics (MagAO) system. To address the twin challenges of morphological complexity and PSF instability, GAPlanetS requires novel approaches for frame selection and optimization of the Karhounen-Lo\'eve Image Processing algorithm \texttt{pyKLIP}. We detect one new candidate, CS Cha ``c", at a separation of 68mas and a modest $\Delta$mag of 2.3. We recover the HD 142527 B and HD 100453 B accreting stellar companions in several epochs, and the protoplanet PDS 70 c in 2017 imagery, extending its astrometric record by nine months. Though we cannot rule out scattered light structure, we also recover LkCa 15 ``b" at H$\alpha$; its presence inside the disk cavity, absence in continuum imagery, and consistency with a forward-modeled point-source suggest that it remains a viable candidate. Through targeted optimization, we tentatively recover PDS 70 c at two additional epochs and PDS 70 b in one epoch. Despite numerous previously reported companion candidates around GAplanetS targets, we recover no additional point-sources. Our moderate H$\alpha$ contrasts do not preclude most protoplanets, and we report limiting H$\alpha$ contrasts at unrecovered candidate locations. We find an overall detection rate of $\sim$36$^{+26}_{-22}$\%, considerably higher than most direct imaging surveys, speaking to both GAPlanetS’ highly targeted nature and the promise of H$\alpha$ differential imaging for protoplanet identification.
\clearpage
\end{abstract}



\section{Introduction} \label{sec:intro}

To date, the field of exoplanet direct imaging has largely and necessarily been focused on detection of several tens of Myr old ``adolescent" planets that are warm and self-luminous. Most detected planets have been massive ($M>1M_J$), nearby ($d<50pc$), and members of young moving groups (ages generally $>10$Myr). In this section, we present the motivation for protoplanet direct imaging, outline progress in this emerging subfield, and describe synergies with other techniques. 

\subsection{Protoplanet Imaging Challenges}

Detection of planets in very young ($<$10Myr) systems has the potential to inform where, when, and how planet formation occurs, but is notoriously difficult for several reasons. First, the nearest star forming regions are more distant ($d\ge140pc$) than the bulk of the directly imaged planet population. This increased distance means that host stars have higher apparent magnitudes, making natural guide star adaptive optics imaging (which relies on photons from the star for wavefront sensing) difficult, while laser guide star systems cannot yet achieve the requisite contrasts at small angular separation. Distant targets also require higher angular resolution to image the environs of these young stars on the scale of planetary systems ($\sim$10 au, $<0\farcs1$). Visible light adaptive optics imaging mitigates this difficulty somewhat, providing a resolution ($\sim\lambda$/D) advantage of approximately a factor of two over diffraction-limited near-infrared (NIR) imaging.

In the field of protoplanet direct imaging, the most widely studied systems host so-called ``transitional" disks, with central cavities cleared of a majority of dusty disk material. Large (tens of astronomical units) heavily dust-depleted transitional disk cavities are likely maintained by the presence of multiple orbiting planets in mean motion resonance whose Hill spheres overlap \citep{DodsonRobinson2011, Close2020} or by more disruptive eccentric or noncoplanar orbits of single companions \citep[e.g.][]{Price2018}. the outer environs of transitional disks are, however, morphologically complex. This presents a second obstacle to the direct detection of very young exoplanets embedded within them, namely the difficulty of unambiguously separating disk and planet signal.

The field of High-Contrast Imaging (HCI) relies on sophisticated processing algorithms for Point Spread Function (PSF) subtraction \citep[e.g., Locally Optimized Combination of Images (LOCI) and Karhounen-Lo\'eve Image Processing (KLIP)][]{Lafreniere2007, Soummer2012a}. These algorithms are powerful tools for isolating faint disk and planet signals, but have been demonstrated to affect the apparent morphology of disk features to varying degrees and to yield apparent point-sources when structures are in fact continuous \citep[e.g.,][]{Follette2017, Ligi2018}. For this reason, most reported detections of protoplanet candidates in transitional disks have been heavily debated. These include the planet candidates LkCa~15~b, c, and d \citep{Kraus:2012, Sallum2015, Thalmann2015, Currie2019}, HD~169142~b \citep{Biller2014, Reggiani2014, Ligi2018, Gratton2019}, T~Cha~b \citep{Huelamo2011}, MWC~758~b \citep{Reggiani2018}, HD~100546~b and c \citep{Quanz2013, Quanz2015, Currie2015, Follette2017, Rameau2017}, and AB Aur b \citep{Currie2022,Zhou2022}.

The only protoplanets considered unambiguous at present are PDS~70~b and PDS~70~c. PDS~70~b's nature as a \textit{bona fide} accreting protoplanet is supported by multiwavelength characterization, namely its detection in thermal emission at a variety of infrared wavelengths \citep{Keppler2018, Muller2018,Wang2021}, at H$\alpha$ \citep{Wagner2018,Haffert2019}, and in ultraviolet accretion continuum emission \citep{Zhou2021}. Likewise, PDS~70~c has been detected in IR thermal emission \citep{Mesa2019,Wang2021}, at H$\alpha$ \citep{Haffert2019}, and in sub-mm continuum emission \citep[evidence of the presence of a circumplanetary disk;][]{Benisty2021, Isella2019}.

Prior to the discovery of PDS~70~b and c, the only object imaged \textit{inside} of a transitional disk gap was the stellar-mass companion HD~142527~B \citep{Biller2014,Close2014}. Though it is not a planetary-mass object \citep[$\mathrm{M_B}=0.11\pm0.06\mathrm{M_{\odot}}$][]{Claudi2019}, its tight separation \citep[a$\sim$15au][]{Balmer2022}, ongoing accretion \citep[$\sim10^{-9}M_{\odot}/yr$][]{Balmer2022} and extreme mas ratio of $\sim$0.05 relative to the primary \citep[$\mathrm{M_A}=2.0\pm0.3\mathrm{M_{\odot}}$][]{Mendegutia2015} make it an important benchmark system and test case for detections of accreting companions embedded in transitional disks. 

Evidence of ongoing accretion onto more widely separated imaged companions has also been found. These include the substellar companions DH Tau b, GSC 6214-210 b, GQ Lup b, Delorme 1 (AB) b, and SR 12 c \citep[e.g.][]{Zhou2014, Betti2022,SM2018}. Although their ages and separations do not make them direct analogs of protoplanets in transitional disk gaps, these more easily recovered and characterized substellar companions provide a fruitful testing ground for planetary accretion models.

In addition to the high required resolution and morphological complexity of transitional disk systems, a third obstacle to direct protoplanet detection is achieving the requisite contrasts at extremely tight separations. For example, in NIR thermal emission, contrasts of $\sim10^{-6}-10^{-4}$ are required at $\sim$0$\farcs$1-0$\farcs$25 in order to image protoplanets inside of disk gaps, and this region lies beneath the coronagraphic masks of most extreme adaptive optics imagers. 

\subsection{H$\alpha$ Differential Imaging}
One way to mitigate the obstacles outlined in the previous section is to leverage (a) the higher resolutions of visible light adaptive optics imaging, and (b) the lower contrasts required to detect protoplanets at wavelengths where accretion emission is present. 

The resolution of a 6.5m telescope is 21mas at H$\alpha$ (656.3nm, the brightest accretion emission line), small enough to resolve companions inside of the 0$\farcs$1-0$\farcs$25 cavities of most transitional disks. In young objects, emission at H$\alpha$ is often associated with ongoing accretion, and is predicted to elevate the flux of planetary-mass objects into a modest contrast range of $\sim$10$^{-3}$-10$^{-2}$ for super-Jovian planets \citep[][]{Mordasini2017}. Accretion is expected even for protoplanets within cleared  transitional disk cavities because the primary stars in these systems are themselves still actively accreting gas \citep[e.g.][]{Salyk2013, Cieza2012}. ``Off" H$\alpha$, planet contrasts should drop below the detectability threshold, making differential imaging a powerful tool to separate direct planet light from signals that are equivalently bright in the two channels or whose brightnesses mimic the stellar H$\alpha$-to-continuum ratio, such as disk scattered light and PSF subtraction artifacts. 

To date, five accreting object candidates have been identified inside of transitional disk gaps. The assertion of accretion in each case is based, at least in part, on the presence of H$\alpha$ excess emission. These are: the stellar-mass companion HD~142527~B \citep{Close2014}, the protoplanets PDS~70~b and ~c \citep{Wagner2018, Haffert2019, Zhou2021}, and the protoplanet candidates LkCa~15~b \citep{Sallum2015} and AB Aur~b \citep{Currie2022}. Three of these objects (HD~142527~B, PDS~70~b, and LkCa~15~b) were first detected at H$\alpha$ as part of the Giant Accreting Protoplanet Survey (GAPlanetS). 

\subsection{Multiwavelength Disk Imaging}
High-resolution submillimeter (sub-mm) imagery of transitional disks obtained with the Atacama Large Millimeter/sumbillimeter Array (ALMA) strongly complements these initial accreting object detections \citep[e.g.,][]{Perez2014,vanderMarel2016}, as does NIR scattered light imagery obtained by extreme adaptive optics imagers such as the Gemini Planet Imager \citep[GPI;][]{Macintosh:2014js}, Spectro-Polarimetric High-contrast Exoplanet REsearch instrument \citep[SPHERE;][]{Beuzit2008}, Subaru Coronagraphic Extreme Adaptive Optics System \citep[SCExAO;][]{Jovanovic2015}, and MagAO-X \citep{Males2018}. The proliferation of multiwavelength imagery of transitional disks over the past decade revealed the initially puzzling fact that transitional disk cavities often appear at substantially different radii in the NIR and sub-mm \citep[e.g., ][]{Villenave2019,Follette2013, Dong2012}. The prevailing theories for this now well-established discrepancy invoke so-called ``dust filtration" processes, whereby the location and degree of clearing of variously sized grains is controlled by the location and mass of planets embedded in the disk \citep[e.g.][]{Zhu2012a, Fung2014}. More specifically, because large grains settle toward the midplane, protoplanets may effectively clear a sub-mm cavity while allowing higher scale-height optical/NIR-scattering small grains to filter through the cavity at the disk surface layer. Although differential clearing of large vs. small grains is suggestive of the presence of planets inside disk gaps and cavities, direct inference of planet masses based on gap radii is difficult. Since scale height varies as a function of radius in the disk, so too does the ``gap clearing" mass threshold. 

Knowledge of disk morphology is important for protoplanet searches for several reasons. Practically speaking, cavities that are not substantially cleared of small optical/NIR-scattering grains will be poor candidates for protoplanet detections, as H$\alpha$ light emitted by planets at the disk midplane is likely to be extincted by that overlying material. Furthermore, although in principle features such as differential clearing and disk asymmetries (spiral arms, dust traps, annular rings, velocity kinks etc.) should predict the location of accreting protoplanets, detection of structure-inciting planets has not been widely successful. This indicates either that planet--disk interaction models still need to be refined or that we are not yet achieving the requisite contrasts, or both. 

\subsection{Outline of This Work}

The MagAO Giant Accreting Protoplanet Survey (GAPlanetS), whose results are reported in the remainder of this paper, surveyed 14 transitional disk host stars for evidence of accreting protoplanets. The search required development of a general framework for robust, uniform high-contrast image processing in complex, embedded planetary systems. This paper aims to present both the results of the survey and the details of the data processing framework.

The GAPlanetS survey sample is described in Section \ref{sec:obs}. Detailed summaries of known disk morphologies and previously reported evidence for the presence of protoplanets in each system appear in Appendix \ref{sec:targets}. Image preprocessing procedures, which are somewhat more complex than for NIR HCI surveys due to the instability of visible light AO PSFs, are detailed in Section \ref{sec:preproc}. Post-processing procedures that allow data-dfor riven optimization of PSF subtraction algorithmic parameters, necessary for analyzing these highly morphologically complex systems, are detailed in Section \ref{sec:postproc}. Strategies adopted to ensure uniform, data-driven comparison among objects for the full survey sample are described in Section \ref{sec:analysis}.

Leveraging the techniques outlined in Sections \ref{sec:preproc}--\ref{sec:analysis}, GAPlanetS searched for embedded planets responsible for inciting disk structures and clearing the observed central cavities of the 14 targeted transitional disks. Individual object-by-object results are detailed in Section \ref{sec:results} and in Appendix \ref{sec:allepochs}. In Section \ref{sec:survey}, we present constraints on the astrometry and accretion rates of the five accreting companions and companion candidates detected planet candidates, limits on the contrasts/accretion rates of protoplanet candidates undetected in our survey, and aggregate survey statistics.

\section{Observations} \label{sec:obs}

\subsection{The GAPlanetS Sample \label{sec:sample}}

In Table \ref{tab:sample}, we outline the physical properties of the GAPlanetS transitional disk sample. 
Because of the importance of varied gap radii to the interpretation of evidence for embedded planets in these systems, we report both the NIR and sub-mm gap radii in Table \ref{tab:sample}. Reviews of the literature on individual objects in the sample are provided in Appendix \ref{sec:targets}, including details about previous evidence for and characterization of protoplanets.

Of the known transitional disks, relatively few are around sufficiently bright stars for natural guide star adaptive optics imaging. The selection criteria for the initial GAPlanetS sample were: (a) an $r$'-band magnitude brighter than 11, (b) a declination less than +30$^\circ$ such that the object is visible from Magellan, and (c) a previously resolved NIR gap or cavity at a separation of at least 0$\farcs$1 from the central star. After a successful pilot observation of LkCa~15 ($R$=10.7), the guide star magnitude requirement was relaxed to $R<$11 and five additional objects were added to the sample (including PDS~70, which was not at the time known to host protoplanets). Only two of these additional objects were successfully observed due to weather and time constraints, for a total of 14 targets in the GAPlanetS sample. 

\subsection{Data Collection}
All data for the GAPlanetS campaign were taken with the Magellan Clay Telescope's Adaptive Optics System (MagAO, \citealp{Close:2013, Morzinski2014, Morzinski2016}) and the VisAO instrument (VisAO, \citealp{Males2014a,Males2013diss}) between April 2013 and May 2018. The number of observations per object ranges from one to seven, with a median of three. Some objects were attempted multiple times in the same semester, either because conditions were poor during a first attempt or because observers using the MagAO infrared camera Clio2 \citep{Morzinski2013} were observing GAPlanetS targets and contributed their visible light data to the campaign. Any imaging sequence with fewer than 10$^\circ$ of rotation on the sky ($N$=9), fewer than 10 minutes of total integration time ($N$=1), or an FWHM of greater than $\sim$0$\farcs$15 for the registered and median-combined PSF ($N$=7) was not analyzed.
Finally, seven datasets were of sufficiently low or variable quality that we were unable to extract false planets injected into the images at a contrast of $1\times10^{-1}$, and these were not analyzed further. In total, 23 datasets were discarded, representing 18 hr, or 20\% of the total campaign time. This is consistent with typical weather and seeing site statistics for Las Campanas \citep[e.g.][]{Duhalde1984}, where these observations were conducted classically. Table \ref{tab:data} summarizes the datasets that were retained and analyzed for this paper, which amount to a total of 60 hr of open shutter time. Raw, preprocessed, and post-processed data for all GAPlanetS datasets are available at \url{https://doi.org/10.7910/DVN/LW9WJJ}

\subsection{Image Preprocessing \label{sec:preproc}}
The VisAO camera cannot take zero second bias exposures, so dark frames were taken interspersed with the science frames to track both bias drifts and dark current contributions. Median levels in each CCD channel for all dark frames taken in a given sequence were inspected for bias drifts, and the level was found to be static to within 3 Analog to Digital Units (ADU, ``counts"). We find that flux in dark frames is constant to $<$1\%, so we utilize a median combination of all dark frames for a given dataset in our calibrations, subtracting the median dark from each raw image frame. 

\movetabledown=2in
\begin{rotatetable*}
\begin{deluxetable*}{cccccccccccccc}
\tabletypesize{\scriptsize}
\tablecaption{GAPlanetS Target Properties}
\tablehead{\colhead{Object} & \colhead{R.A.$^{1}$}   & \colhead{Dec.$^{1}$}  & \colhead{r' mag}$^{2,3}$ & \colhead{A$_{r'}^{2,4}$} & \colhead{SpT} & \colhead{\makecell[t]{d$^{1}$ \\ (pc)}} & \colhead{Assn$^{16}$} & \colhead{\makecell[t]{Member \\ Prob.$^{16}$}} & \colhead{\makecell[t]{Age\\(Myr})} & \colhead{\makecell[t]{$\log{\dot{M}}$ \\($\frac{M_{\odot}}{yr}$)}} & \colhead{mm $r_{gap(s)}$} & \colhead{NIR $r_{gap(s)}$} & \colhead{gap refs}}
\startdata
HD~142527                                                   & 239.174469 & -42.323241 & 8.2    & 0.8 & F6$^5$        & 159.3$\pm$0.7                                         & UCL                    & 92.10\%                                                      & 16$\pm$2$^{15}$                                     & -7.16$^{23}$ & 0$\farcs$26-1$\farcs$2 & 0$\farcs$1-1$\farcs$0 & 38,50 \\
PDS~70                                                   & 212.0421   & -41.39804  & 11.7 & 0 & K7$^{15}$    & 112.4$\pm$0.2                                         & UCL                    & 98.70\%                                                      & 5.4$\pm$1$^{36}$ & -10.22$^{37}$   & 0$\farcs$11-0$\farcs$43 & 0$\farcs$2-0$\farcs$4   & 65,66\\ 
LkCa~15                                                 & 69.82418   & 22.35086   & 11.6 & 0.5 & K5$^{14}$    & 157.2$\pm$0.7                                         & TAU                    & 88.20\%                                                      & 1-2$^{22}$                                          & -9.13$^{35}$ & \makecell[t]{$<$0$\farcs$27 \\ 0$\farcs$33-0$\farcs$41 \\ 0$\farcs$45-0$\farcs$58} & 0$\farcs$21-0$\farcs$42 & 49,64\\
HD~169142                                                   & 276.1241   & -29.78054  & 8.2 & 0  & F1$^6$       & 114.9$\pm$0.4                                         & Field                  & 98.90\%                                                      & 6$^{+6}_{-3}{}^{20}$            & -8.68$^{27}$ & \makecell[t]{0$\farcs$02-0$\farcs$23 \\ 0$\farcs$28-0$\farcs$48 \\ 0$\farcs$51-0$\farcs$56 \\ 0$\farcs$57-0$\farcs$65}	& \makecell[t]{ 0$\farcs$004-0$\farcs$17 \\ 0$\farcs$28-0$\farcs$48} & 38,42,43,55\\
HD~100546                                                  & 173.3555   & -70.19479  & 6.8 & 0.2  & A0$^6$       & 108.1$\pm$0.4                                         & LCC                    & 98.90\%                                                      & 15$\pm$3$^{15}$                                     & -6.44$^{24}$ & 
0$\farcs$02-0$\farcs$19 &
0$\farcs$007-0$\farcs$14 & 39, 51 \\
SAO~206462                                                 & 228.9518   & -37.15455  & 8.6 & 0.1 & F8$^7$       & 135.0$\pm$0.4                                         & UCL                    & 99.50\%                                                     & 16$\pm$2$^{15}$                                 & -7.7$^{28}$   & \makecell[t]{ $<$0$\farcs$38 \\0$\farcs$48-0$\farcs$55} & 0$\farcs$18 & 44,56\\
TW~Hya                                                      & 165.4659   & -34.70479  & 10.5 & 0.5 & K7$^{12}$    & 60.1$\pm$0.1                                          & TWA                    & 99.90\%                                                      & 10$\pm$3$^{21}$                                     & -8.58$^{33}$ & \makecell[t]{0$\farcs$0083-0$\farcs$033 \\ 0$\farcs$2-0$\farcs$22 \\ 0$\farcs$35-0$\farcs$38 \\ 0$\farcs$45-0$\farcs$48 \\ 0$\farcs$62-0$\farcs$65 \\ 0$\farcs$7-0$\farcs$73} & \makecell[t]{0$\farcs$083-0$\farcs$17 \\ 0$\farcs$27-0$\farcs$42 \\ 1$\farcs$16-1$\farcs$5} & 48,62\\
HD~100453                                                   & 173.273    & -54.32462  & 7.8 & 0.2 & A9$^5$       & 103.8$\pm$0.2                                         & LCC                    & 99.30\%                                                      & 15$\pm$3$^{15}$                                     & \textless -8.85$^{26}$ & \makecell[t]{ 0$\farcs$09-0$\farcs$22 \\ 0$\farcs$40-0$\farcs$48} & 0$\farcs$09-0$\farcs$14 & 41,54\\
CS~Cha                                                     & 165.6032   & -77.55989  & 11.1 & 1.0 & K2$^{13}$    & 168.8$\pm$1.9                                         & Cha I$^{18}$          & n/a                                                          & 2$\pm$2$^{18}$                                      & -8.3$^{34}$ & $<$0$\farcs$21 &  $<$0$\farcs$0925 & 38,63\\
HD~141569                                                   & 237.4905   & -3.921291  & 7.2  & 0.2 & A2$^6$       & 111.6$\pm$0.4  & Field & 99.90\% & 5$\pm$3$^{19}$ & -8.13$^{25}$ &  0$\farcs$9$\sim$1$\farcs$9 & \makecell[t]{ 0$\farcs$25-0$\farcs$4 \\ 0$\farcs$43-0$\farcs$52 \\ 0$\farcs$60-0$\farcs$69 \\ 1-2"} & 40,52,53\\
PDS~66                                                   & 200.5309   & -69.63682  & 10.0  & 0.7 & K1$^{10}$    & 97.9$\pm$0.1                                          & LCC                    & 97.50\%                                                      & 15$\pm$3$^{15}$                                     & -8.3$^{31}$  & \nodata & 0$\farcs$46-0$\farcs$81 & 60 \\
UX~Tau~A                                                  & 67.5167    & 18.23033   & 11.3 & 0.5 & K2$^{11}$ & 142.2$\pm$0.7                                         & TAU                    & 98.10\%                                                      & 1-2$^{22}$                                          & -8$^{32}$ & 0$\farcs$18	&  0$\farcs$16 & 47,61\\
V1247Ori                                                   & 84.52188   & -1.25603   & 9.9 & 0.3 & F0$^8$        & 401.3$\pm$3.2                                         & $\epsilon$ Ori$^{17}$ & n/a                                                          & 5-10$^{17}$                                         & -8$^{29}$ & 0$\farcs$04-0$\farcs$15 & $<$ 0$\farcs$07-0$\farcs$11 & 45,57,58\\
V4046~Sgr                                                  & 273.5437   & -32.79316  & 10.0 & 0 & K5-7$^9$     & 71.5$\pm$0.1                                          & BPMG                   & 98.40\%                                                      & 24$\pm$3$^{21}$                                     & -9.3$^{30}$ & 0$\farcs$08-0$\farcs$3 & 0$\farcs$10-0$\farcs$19 & 46,59\\
\enddata
\tablecomments{\scriptsize Relevant stellar properties for the GAPlanetS sample of young, transitional disk systems. The targets are ordered according to order of appearance in Section \ref{sec:results}. References: $^1$ \citet{Gaia2022},  $^2$ \citet{Gaia2022} DR3 photometry converted to the SDSS $r$' magnitude per $^3$\citep{Alam2015}, $^4$ \citep{Pecaut2013a},$^5$ \citet{Houk1994}, $^6$ \citet{Gray2017}, $^7$ \citet{Coulson1995}, $^8$ \citet{Vieira2003}, $^9$ \citet{Nefs2012}, $^{10}$ \citet{daSilva2006}, $^{11}$ \citet{Kraus2009}, $^{12}$ \citet{Wichmann1998}, $^{13}$ \citet{Appenzeller1977}, $^{14}$ \citet{Herbig1986}, $^{15}$ \citet{Pecaut2016}, $^{16}$ \citet{Gagne2018}, $^{17}$ \citet{Caballero2008}, $^{18}$ \citet{Luhman2008}, $^{19}$ \citet{Weinberger2000}, $^{20}$ \citet{Grady2007}, $^{21}$ \citet{Bell2015}, $^{22}$ \citet{Kenyon1995}, $^{23}$ \citet{GarciaLopez2006}, $^{24}$ \citet{Mendegutia2015}, $^{25}$ \citet{Salyk2013}, $^{26}$ \citet{Collins2009}, $^{27}$ \citet{Wagner2015}, $^{28}$ \citet{Sitko2012}, $^{29}$ \citet{Willson2019}, $^{30}$ \citet{Curran2011}, $^{31}$ \citet{Pascucci2007}, $^{32}$ \citet{Andrews2011}, $^{33}$ \citet{Robinson2019}, $^{34}$ \citet{Manara2014}, $^{35}$ \citet{Alencar2018}, $^{36}$ \citet{Muller2018}, $^{37}$ \citet{Thanathibodee2020} $^{38}$\citet{Francis2020}, $^{39}$\citet{Pineda2019}, $^{40}$\citet{Miley2018}, $^{41}$\citet{VanderPlas2019}, $^{42}$\citet{Fedele2017}, $^{43}$\citet{Perez2019}, $^{44}$\citet{Cazzoletti2018}, $^{45}$\citet{Kraus2017}, $^{46}$\citet{Kastner2018}, $^{47}$\citet{Pinilla2014}, $^{48}$\citet{Andrews2016}, $^{49}$\citet{Facchini2020}, $^{50}$\citet{Avenhaus2014}, $^{51}$\citet{Follette2017}, $^{52}$\citet{Konishi2016}, $^{53}$\citet{Perrot2016}, $^{54}$\citet{Benisty2017}, $^{55}$\citet{Monnier2017}, $^{56}$\citet{Stolker2016a}, $^{57}$\citet{Ohta2016}, $^{58}$\citet{Willson2019}, $^{59}$\citet{Rapson2015}, $^{60}$\citet{Wolff2016}, $^{61}$\citet{Tanii2012}, $^{62}$\citet{vanBoekel2017}, $^{63}$\citet{Ginski2018b}, $^{64}$\citet{Oh2016}, $^{65}$\citet{Long2018}, $^{66}$\citet{Keppler2018} \label{tab:sample}}
\end{deluxetable*}
\end{rotatetable*}

\begin{deluxetable*}{cccccccccc}
    
    \tablewidth{700pt}
    \tabletypesize{\scriptsize}
    \tablecaption{Descriptive Statistics for GAPlanetS Campaign Data}
    \tablehead{
       \colhead{Object Name} & \colhead{Date} & \colhead{n$_{\rm{ims}}$} & \colhead{t$_{\rm{exp}}$ (sec)} & \colhead{t$_{\rm{tot}}$ (min)} & \colhead{rot ($^\circ$)} & \colhead{FWHM (pix)} & \colhead{r$_{\rm{sat}}$ (pix)} & \colhead{Avg. Seeing (")} & \colhead{Scale Factor}}
         \startdata
HD~142527 & 4/11/13 & 1961 & 2.27 & 74.2 & 65.3 & 4.56$^{G}$ & 6 & 0.56 & 0.88$\pm$0.04 \\
HD~142527 & 4/8/14 & 68 & 45 & 51.0 & 100.7 & 4 & N/A & \nodata & 1.14$\pm$0.02 \\
HD~142527 & 4/8/14 & 1758 & 2.27 & 66.5 & 101.7 & 5.58$^{G}$ & 10 & \nodata & 1.13$\pm$0.03 \\
HD~169142 & 4/8/14 & 2796 & 2.27 & 105.8 & 180.1 & 5.5 & N/A & 0.72 & 0.99$\pm$0.03 \\
TW~Hya & 4/8/14 & 1958 & 2.27 & 74.1 & 82.5 & 6.36 & N/A & \nodata & 8.79$\pm$0.11 \\
HD~169142 & 4/9/14 & 178 & 15 & 44.5 & 171.6 & 5.01 & N/A & \nodata & 0.98$\pm$0.05 \\
HD~141569 & 4/9/14 & 2402 & 2.27 & 90.9 & 55.7 & 17.2$^{G}$ & 4 & 0.70 & 0.94$\pm$0.06 \\
HD~141569 & 4/10/14 & 1364 & 2.27 & 51.6 & 36.6 & 6.85$^{G}$ & 4 & \nodata & 0.96$\pm$0.10 \\
HD~141569 & 4/11/14 & 2340 & 2.27 & 88.5 & 58.9 & 8.57 & N/A & \nodata & 0.95$\pm$0.05 \\
HD~100546 & 4/12/14 & 4939 & 2.27 & 186.9 & 71.6 & 3.92$^{G}$ & 4 & \nodata & 1.43$\pm$0.06 \\
SAO~206462 & 4/12/14 & 3993 & 2.27 & 151.1 & 143.7 & 5.16$^{G}$ & 3 & \nodata & 1.22$\pm$0.10\\
V4046~Sgr & 4/12/14 & 1414 & 5 & 117.8 & 156.1 & 7.86 & N/A & \nodata & 1.79$\pm$0.40 \\
UX~Tau~A & 11/15/14 & 52 & 45 & 39.0 & 13.8 & 9.55 & N/A & \nodata & 1.42$\pm$0.04 \\
V1247~Ori & 11/15/14 & 893 & 7/10 & 113.0 & 46.5 & 5.39 & N/A & \nodata & 1.13$\pm$0.02 \\
LkCa~15 & 11/16/14 & 308 & 30 & 154.0 & 48.6 & 8.67 & N/A & \nodata & 1.81$\pm$0.03 \\
HD~141569 & 5/28/15 & 723 & 5/10 & 84.1 & 54.2 & 5.79$^{G}$ & 5 & \nodata & 0.94$\pm$0.02 \\
HD~142527 & 5/15/15 & 2387 & 2.27 & 90.3 & 117.4 & 5.5 & N/A & \nodata & 1.13$\pm$0.10 \\
CS~Cha & 5/15/15 & 143 & 30 & 71.5 & 31.4 & 9.17 & N/A & 0.59 & 2.26$\pm$0.05 \\
HD~142527 & 5/16/15 & 1143 & 2.27 & 43.2 & 34.8 & 5.01 & N/A & 0.55 & 1.14$\pm$0.12 \\
V4046~Sgr & 5/17/15 & 720 & 5 & 60.0 & 146.5 & 6.44 & N/A & 0.66 & 1.80$\pm$0.10 \\
HD~142527 & 5/18/15 & 159 & 30 & 79.5 & 76.8 & 5.24$^{G}$ & 2 & 0.80 & 1.12$\pm$0.06 \\
HD~169142 & 5/18/15 & 1731 & 2.27 & 65.5 & 180.6 & 8.17$^{G}$ & 6 & \nodata & 1.06$\pm$0.09 \\
SAO~206462 & 5/26/15 & 408 & 10 & 68.0 & 15.9 & 6.21 & N/A & 0.70 & 1.22$\pm$0.10 \\
HD~141569 & 5/29/15 & 404 & 10 & 67.3 & 56.1 & 12.1 & N/A & 0.80 & 0.92$\pm$0.03 \\
HD~100546 & 5/30/15 & 2459 & 2.27 & 93.0 & 43.7 & 5.31$^{G}$ & 6 & \nodata & 1.59$\pm$0.20 \\
V1247~Ori & 12/11/15 & 878 & 7 & 102.4 & 21.7 & 6.93 & N/A & 0.69 & 1.12$\pm$0.04 \\
LkCa~15 & 11/18/16 & 252 & 30 & 126.0 & 36.03 & 12.2 & N/A & 0.47 & 1.58$\pm$0.10 \\
PDS~66 & 2/7/17 & 243$^{+}$ & 30 & 121.5 & 42.7 & 6.74 & N/A & 0.61 & 1.91$\pm$0.02 \\
TW~Hya & 2/7/17 & 452$^{+}$ & 30 & 226.0 & 139.6 & 7.98 & N/A & 0.69 & 7.17$\pm$0.20 \\
PDS~70 & 2/8/17 & 188$^{+}$ & 45 & 141.0 & 92.9 & 8.36 & N/A & 0.47 & 1.32$\pm$0.02 \\
HD~142527 & 2/10/17 & 242 & 12 & 48.4 & 16.1 & 4.49$^{G}$ & 8 & 0.66 & 1.22$\pm$0.10 \\
HD~100453 & 2/17/17 & 2947 & 3/5 & 160.0 & 61.4 & 4.26$^{G}$ & 4 & 0.60 & 1.10$\pm$0.03 \\
HD~169142 & 8/30/17 & 1658$^{+}$ & 2.27 & 62.7 & 171.3 & 5.83$^{G}$ & 3 & 0.70 & 1.13$\pm$0.03 \\
HD~142527 & 4/27/18 & 580 & 5 & 48.3 & 49.2 & 4.37 & 3 & \nodata &  1.28$\pm$0.06\\
HD~100453 & 5/2/18 & 563$^{+}$ & 15 & 140.8 & 83.3 & 4.34 & N/A & 0.64 & 1.26$\pm$0.03 \\
HD~100453 & 5/2/18 & 356$^{+}$ & 2.27 & 13.5 & 86.3 & 4.90$^{G}$ & 3 & 0.64 & 1.05$\pm$0.02 \\
PDS~70 & 5/2/18 & 209$^{+}$ & 30 & 104.5 & 90.9 & 7.08 & N/A & 0.52 & 1.29$\pm$0.02 \\
PDS~70 & 5/3/18 & 284$^{+}$ & 30 & 142.0 & 111.7 & 6.82 & N/A & 0.50 & 1.36$\pm$0.02 \\
HD~100453 & 5/3/18 & 2831$^{+}$ & 2.27 & 107.1 & 66.18 & 6.17 & N/A & 0.45 & 1.04$\pm$0.10 \\
\enddata
\tablecomments{\scriptsize These statistics represent the datasets before the data quality cut step described in the text in Section \ref{sec:dqcuts}. Statistics for the final post-processings of each dataset are given in Table \ref{tab:optsumm}. The + superscript indicates a dataset observed in SDI$^+$ mode, which utilizes a spinning half-waveplate to mitigate polarization effects. A G superscript indicates datasets for which the stable instrumental ghost was used to estimate the FWHM of the saturated central star. Seeing statistics were measured by either the site's DIMM or by the neighboring Baade telescope, and were averaged where both measures were available. In some cases, no seeing data were available from the Magellan site monitors. Scale factors reported are the median value of the ratio of the flux of the primary (unsaturated data) or ghost (saturated data) at H$\alpha$ relative to the contemporaneous continuum in individual images, as determined by aperture photometry and described in the text in Section \ref{sec:SDI}. Uncertainties on scale factors represent the standard deviation of the scale measurements for individual images. Raw, preprocessed, and postprocessed data is available for all GAPlanetS datasets at \url{https://doi.org/10.7910/DVN/LW9WJJ}}.
\label{tab:data}
\end{deluxetable*}

$r$' band (which spans H$\alpha$) twilight sky flats were used to calibrate most of the datasets. Due to scheduling constraints at twilight, flat frames were not collected on every night of GAPlanetS observations. 
In cases where more than one $r$' band flat dataset was available in a given semester, we selected the flat that was closest in time to the observations, so long as it was of high quality. During the 2017A semester, no $r$' band flats were collected, so we utilized a $z$ band flat for calibration. 
During the 2013A semester (containing a single GAPlanetS dataset), no appropriate flat datasets were available, so we applied a 2014A flat to these data. The median number of days between flat and science exposures for all GAPlanetS datasets was five.

We find that the VisAO detector's sensitivity is flat to better than 1\% in both space (across the detector) and time (from semester to semester), so the primary purpose of flat field correction is removal of near-focus dust spots on the CCD window, which attenuate light by a few to a few tens of percents. The influence of these dust spots on final images is mitigated both by dividing by the flat image for a given semester and by dithering the star on the detector during observations. For one dataset (UX~Tau~A), the observations were aborted before dithers were completed. 

In developing the GAPlanetS pipeline, we experimented with masking dust spots and with interpolating over them using various methodologies; however, standard unmasked flat fields resulted in the highest-quality reductions of the HD~142527~B companion, both qualitatively (final image appearance) and quantitatively (signal-to-noise ratio, SNR). The majority of VisAO dust spots are static from semester to semester, and in all cases individual images were inspected for poorly corrected dust spots, and such images were discarded before final PSF subtraction.  

GAPlanetS data were taken in the VisAO camera's Simultaneous Differential Imaging (SDI) mode. The MagAO wavefront sensor, like the VisAO camera, operates in visible light. For all GAPlanetS observations, a 50/50 beamsplitter was used to send half of the incoming light to the wavefront sensor and half to the VisAO science camera. In SDI mode, a Wollaston prism is used to further split the VisAO science beam into two equal components. One Wollaston beam is passed through a narrowband filter centered on the H$\alpha$ emission line ($\lambda_{\rm{central}}$=656nm, $\Delta\lambda$=6nm) and the other beam is passed through a narrowband continuum filter centered nearby at a wavelength of 642nm ($\Delta\lambda$=6nm). The dark-subtracted, flat-fielded 1024 x 1024 pixel images are therefore split into two 1024 x 512 ``channels", representing the images of the star at H$\alpha$ and the continuum. The proximity of these filters in wavelength and the minimal non-common path makes the PSFs of the two channels extremely similar, with a few caveats, outlined in Section \ref{sec:SDI}.

Following the splitting of the wavelength channels, all GAPlanetS images are then registered using Fourier cross-correlation against a single representative science image selected from within the sequence. The reference image itself is centered via cross-correlation with a Gaussian of equivalent FWHM. This simple registration method was found to yield the highest average SNR for the known companion HD~142527~B across many datasets. 

In the case of saturated images, both registration and photometry are computed relative to an optical ghost that is present at the same location in all images. The ghost has been found to be astrometrically stable to within 1 pixel and to have a stable brightness ratio relative to the primary star of $179.68\pm4.59$ in the H$\alpha$ filter and $196.31\pm3.56$ in the continuum filter. The ghost also has an FWHM 7\% larger than the central PSF, indicating that it is slightly out of focus and is likely produced by a reflection off of the backside of the 6 millimeter thick MagAO 50/50 beamsplitter. A full description of MagAO astrometric and ghost calibration is provided in \citet{Balmer2022}.

All GAPlanetS images contain a bright ring of emission at the boundary of the AO system's ``control radius", or ``dark hole". The location of the control radius is defined by the boundary between spatial frequencies that are sensed versus unsensed by the wavefront sensor.  When imaging guide stars with $r$'-band magnitudes fainter than $\sim$8, the pyramid WFS camera is binned to 2x2 pixels in order to obtain sufficient signal for wavefront correction. This effectively halves the control radius in such images. The MagAO control radius is $\sim$0$\farcs$25 (30 pixels) in bin 1 and $\sim$0$\farcs$12 in bin 2. Only five of 14 GAPlanetS targets (HD~142527, HD~100546, HD~141569, HD~100453, and HD~169142) have $r$'-band magnitudes brighter than 8, therefore, an r=30~pixel control radius. The remaining nine GAPlanetS systems were imaged with a control radius of 15 pixels. This is relevant in that the dark hole is the region in which the adaptive optics system most effectively concentrates starlight into the central PSF, and we concentrate our search and optimization algorithms in this region of the images, which happens to also correspond approximately to the size of the cleared central cavities of most GAPlanetS systems. 
 
The 1024x512 pixel channel images are cropped after registration to reduce processing time. The size of the cropped region was chosen to be 451x451 pixel (3$\farcs$5) square, slightly smaller than the 521 pixel channel width so that images would be equivalently sampled across dithers.

The final step before PSF subtraction is a by-eye inspection of the registered image frames. Images where the adaptive optics control loop is fully open are rejected in the initial phase of the pipeline. An additional by-eye rejection step allows us to discard images where the loop is formally closed, but in the process of breaking or re-closing. At this stage, we also reject several types of artifacts, namely any images with: (a) cosmic rays within 50 pixels of the central star, (b) severe instrumental artifacts such as mid-image dithers, and (c) incompletely removed dust spot artifacts within the AO control radius ($r\sim$30pixels). A median of 97.8\% of H$\alpha$ images are retained following this rejection step, with a standard deviation of 4.6\%. For continuum images, a median of 95.3\% of images are retained with a standard deviation of 7.4\%; this larger proportion of rejected images is due to an increase in the number of dust spots on the bottom half of the detector. Dataset statistics reported in Table \ref{tab:data} record the total number of closed-loop images and total integration times prior to this rejection step. The statistics for the proportion of images used in final analyses are reported in Table \ref{tab:optsumm}. They reflect an additional frame selection step described in detail in Section \ref{sec:dqcuts}.

A coarse grid search of \texttt{PyKLIP} parameters defining (a) the geometry of separately modeled annular zones in the images, (b) the size of the PSF reference library, and (c) the complexity of the model is then conducted for this subset of images. The details of this methodology, including motivation for the choice of parameters to optimize, is provided in Section \ref{sec:optimize}, and the optimized values for pyKLIP parameters are provided in Table \ref{tab:optsumm}.

\subsection{SDI Processing and Mitigation of Possible Sources of Line vs. Continuum Mismatch \label{sec:SDI}}

Since a Wollaston prism operates by splitting light according to polarization state, and scattered light from circumstellar disks (which is present in all GAPlanetS targets) is polarized, there is a reasonable expectation that individual pairs of line and continuum images will contain some differences in scattered light contributions. This is mitigated in several ways. 

First, data are highpass filtered to remove low spatial frequency structures (including extended disk emission and the AO control radius) before PSF subtraction. Second, the majority of the cavities where the search for accreting protoplanets was concentrated are sufficiently cleared of small grain optical scatterers that this contribution is minimal. Nevertheless, we mark the location of known scattered light structures in our final images when there is reason to be concerned about the fidelity of observed signals. Finally, on-sky rotation tends to reduce residual polarization structure across full sequences. 

We also note that most GAPlanetS data taken after the 2017A semester are free of differential polarization effects because of the addition of a spinning half waveplate to the instrument. This half waveplate spins at 2Hz, modulating the polarization state of the two Wollaston channels at 8Hz throughout the image sequence and attenuating polarization noise by a factor of 40 for a 5 s image \citep{Close2018}. These ``GAPlanetS$^+$" datasets are indicated in Table \ref{tab:data} with a + symbol. 

A second concern regarding comparison of line and continuum channels stems from the fact that the central stars of many of the GAPlanetS targets are themselves actively accreting, and are thus measurably brighter in H$\alpha$. Indeed, this is the primary reason to expect that protoplanets within their disk gaps will also be accreting. In such cases, high spatial frequency scattered light disk features that survive the highpass filtering algorithm that is applied before PSF subtraction should be brighter in H$\alpha$ than in the continuum and may be mistaken for accreting protoplanets. We mitigate this effect by quantifying and compensating for the brightness differential at the two wavelengths directly. 

We measure the H$\alpha$/continuum brightness ratio of the star for each image in the sequence using aperture photometry of the star (or the ghost in the case of saturated images).  We report the median and standard deviation of these line-to-continuum scale factors for each dataset in the rightmost column of Table \ref{tab:data}. We then use these computed scale factors to complete ``conservative" SDI reductions for all datasets. This is done by multiplying the KLIP-ADI reduced continuum image by the scale factor before subtraction from the H$\alpha$ image to create an SDI image. This compensates for the difference in the brightness of the primary star at the two wavelengths and should remove both stellar residuals and scattered light structures, including any scattered light emission from circumprimary and circumsecondary disk structures. If any apparent point-sources disappear when the continuum image is scaled and subtracted, scattered starlight may be the source of the emission. We also compute a direct subtraction of the continuum image from the H$\alpha$ image. This is most appropriate in cases where nearby disk structures are not a concern. 

\section{Post-processing \label{sec:postproc}}

\subsection{Data Quality Cuts \label{sec:dqcuts}}

\begin{figure*}
    \centering
    \includegraphics[width=\textwidth]{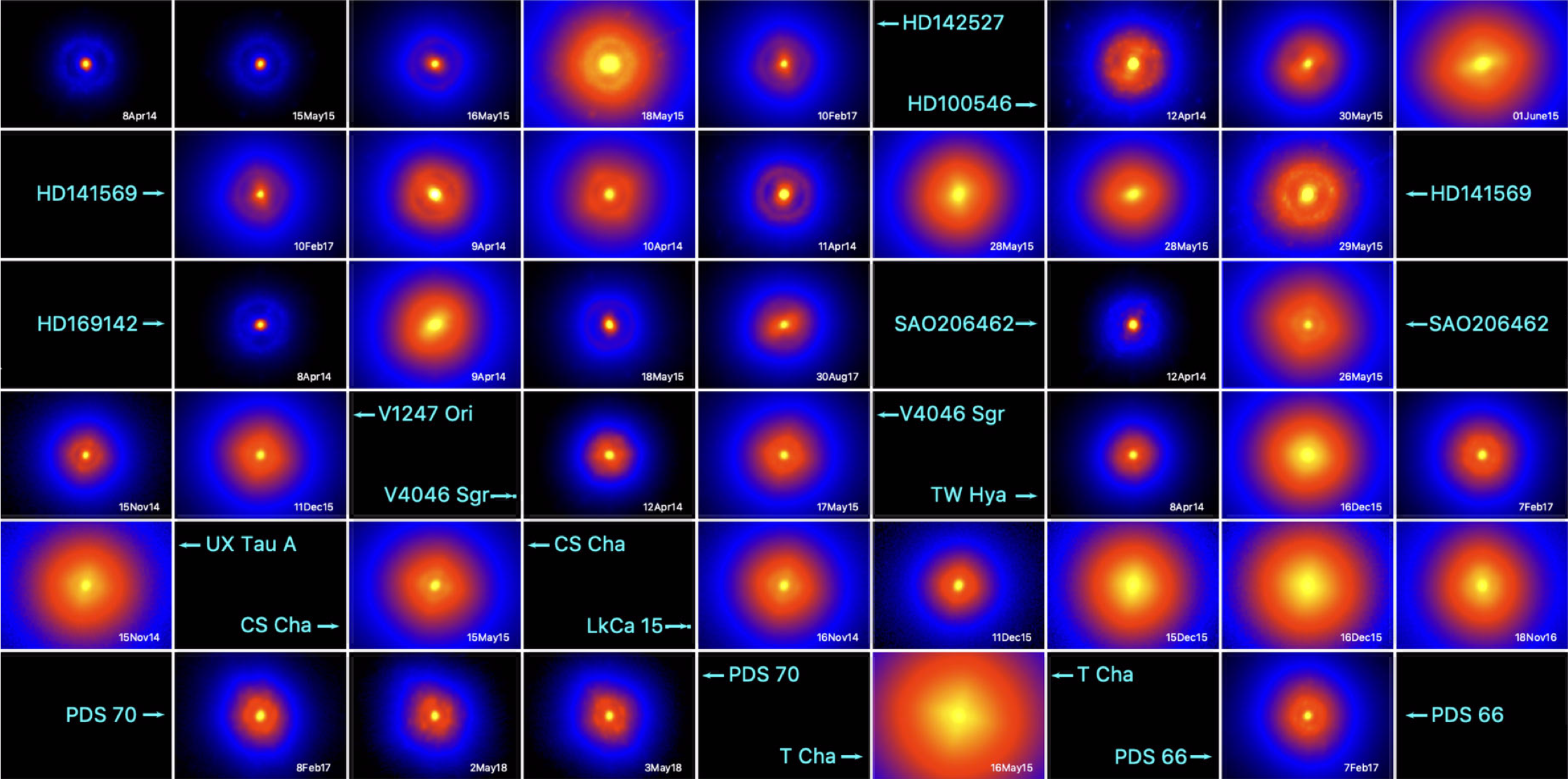}
    \caption{Median combinations of the images in each GAPlanetS image sequence, normalized to 1 by dividing by the peak pixel and arranged from brightest to faintest $r$' band magnitude. The extreme variability in the PSF between objects and among images of the same object acquired on different nights is apparent in the variable size of the PSFs. The VisAO PSF is highly dependent on weather conditions.} 
    \label{fig:PSFgallery}
\end{figure*}

 Operating at visible wavelengths, GAPlanetS PSFs are more unstable on short timescales than NIR high-contrast images. Figure \ref{fig:PSFgallery}, which shows the median-combined PSF of each GAPlanetS image sequence, demonstrates this. Although estimation of Strehl ratios is difficult in this regime ($<$20\%), we note that in cases where seeing data are available, the median value of $\textrm{FWHM}_{\rm{VisAO}}/\textrm{FWHM}_{\rm{BDAvg}}$ (where $\textrm{BDAvg}$ is the average of the reported seeing from the Magellan Baade telescope and the summit Differential Image Motion Monitor) is 11\%, indicating approximately a factor of 10 improvement over seeing-limited imaging. The image resolutions are, on the other hand, a median of 2.8 times the diffraction limit at H$\alpha$, indicating substantial room for improvement in visible light adaptive optics imaging technology (see Section \ref{sec:future}). 

\begin{figure*}
    \centering
    \includegraphics[width=\textwidth]{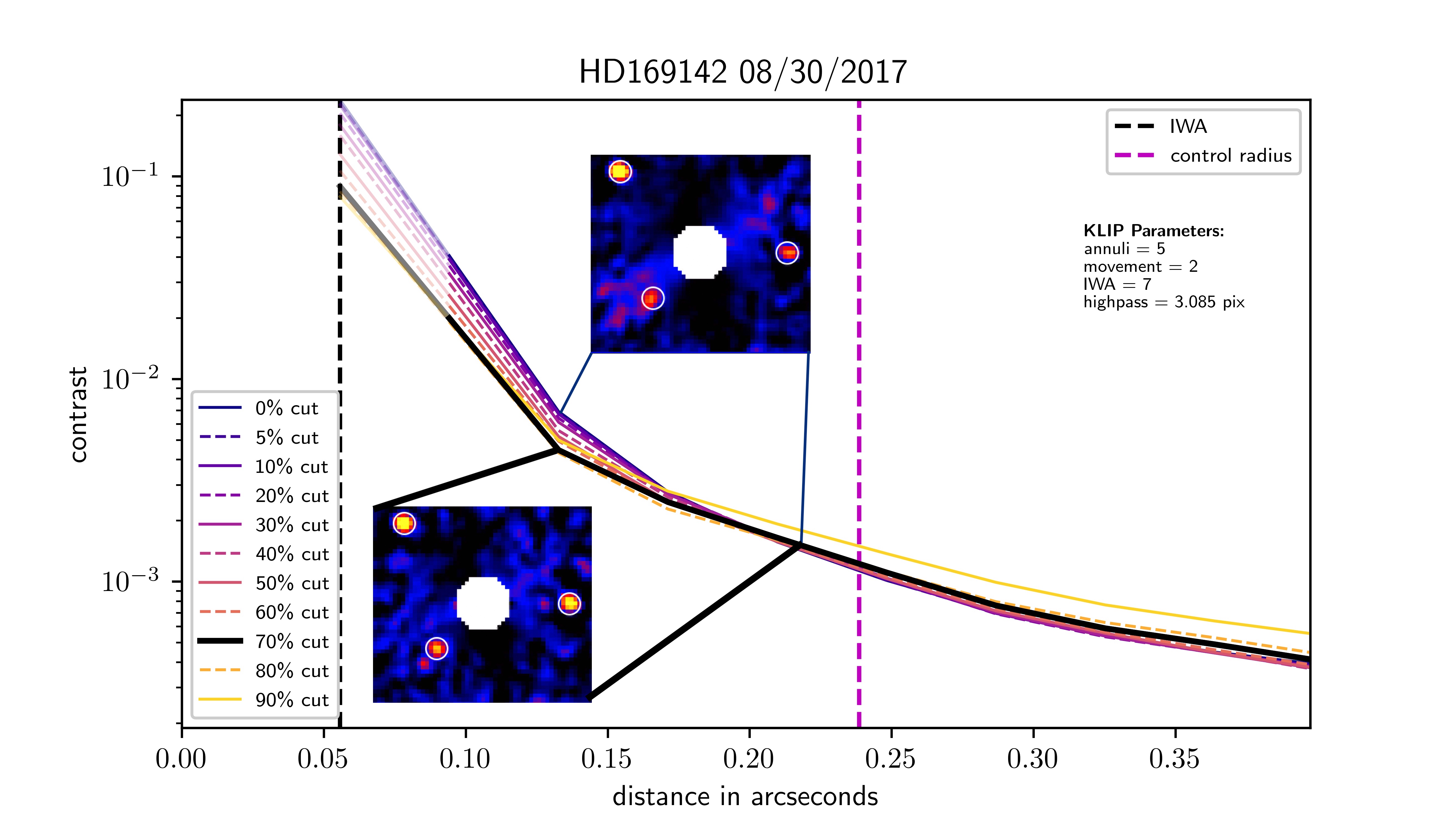}
    \caption{A representative set of contrast curves for a single dataset demonstrating the methodology used to choose a ``data quality cut". The colored curves reflect the contrast achieved after discarding varying proportions of the lowest-quality images. In this case, the curves converge after $\sim0\farcs2$, but the achieved contrast inside of this radius varies with the proportion of data discarded. Inset images are SNR maps showing three planets injected into raw images at separations of 0$\farcs$12, 0$\farcs$17, and 0$\farcs$22 with no data discarded (top, 0\% cut) and with the selected optimal amount of data discarded (bottom, 70\% cut). Recovery of the innermost two planets is markedly improved in the lower plot, increasing their recovered SNRs by $\sim$2. The outer planet is recovered at equivalent SNR in both reductions, reflective of the converging of the contrast curves at this separation.}
    \label{fig:qualitycuts}
\end{figure*}

 Large variations in the stellar PSF appear to decrease the quality of post-processed images in some cases. Lower-quality (higher FWHM) images also limit our ability to extract tightly separated point-sources. In order to mitigate the effects of this variation, we built on the concept of ``Lucky imaging" \citep{Fried1978} and developed a data-driven method to cull a proportion of the lowest-quality images for each dataset. Our contrast-curve based approach is outlined below, and its benefits are highlighted in Figure \ref{fig:qualitycuts}. 
\begin{enumerate}
    \item Eleven subsets of the full image sequence were created for each GAPlanetS dataset by culling 0\%, 5\%, 10\%, 20\%, 30\%, 40\%, 50\%, 60\%, 70\%, 80\% and 90\% of the lowest-quality images. The metric for image quality was the peak value of a 2D Moffat fit to the central star (unsaturated data) or the ghost (saturated data). This peak value should closely trace instantaneous wavefront error in the absence of significant variability on the timescale of individual exposures.
    \item False planets were injected into the raw images at a contrast of $10^{-2}$ or $5\times10^{-2}$ (this value for each dataset is indicated in Table \ref{tab:optsumm}) in a spiral pattern separated by 85$^{\circ}$ in azimuth and 0.5 FWHM radially.
    \item Raw images were highpass filtered with a 0.5xFWHM Gaussian highpass filter to remove low spatial frequency structure.
    \item These images with injected false planets were passed through the KLIP algorithm with a fixed set of KLIP parameters that experimentation indicated would yield high-quality reductions for all datasets (namely \texttt{annuli}=5, \texttt{movement}=2), and the ratio of their recovered-to-injected brightness was used to determine the throughput of the KLIP algorithm as a function of separation.
    \item Steps 2-3 were repeated twice, with the locations of the innermost injected false planet rotated by 75 degrees each time.
    \item Throughput values (which correct for KLIP self-subtraction, a strong function of angular separation) for the three sets of false planet injections were averaged in order to capture azimuthal variation in the PSF. Cases where the innermost planets were not recovered at a contrast of 10$^{-2}$ in all three injections resulted in a repetition of steps 2-4 at a brighter injected planet contrast ($5\times10^{-2}$). If the innermost planets were not robustly recovered at a contrast of 10$^{-1}$, the dataset was excluded from the sample.
    \item The unadulterated (no false planet) images were also passed through KLIP with the same parameters (\texttt{annuli}=5, \texttt{movement}=2), and the noise level was estimated as the standard deviation at each separation \citep[corrected to reflect a t-distribution following][]{Mawet2014}.
    \item The noise level was multiplied by 5 to represent a 5$\sigma$ detection and divided by the throughput to compute the detection limit at each separation, resulting in a contrast curve for each data quality cut.
    \item As the optimal cut varies radially for many datasets, the proportion of images to discard was determined by eye, prioritizing the inner $\sim$0$\farcs$25 where planets are most likely to be found.
\end{enumerate}

We note that several of the choices outlined above may have substantial influence on the ``answer" for the optimal cut, most notably the choice of KLIP parameters and the aggressiveness of the highpass filter applied before PSF subtraction. We also note that our by-eye choice of the ``optimal" cut is somewhat subjective, as there are several competing concerns. 

First, the cut with the lowest contrast varies radially. The most common pattern (11 datasets) appears to be a crossing of curves near the AO control radius, perhaps due to a shift in the dominant noise source at this boundary. We choose to minimize the contrast curve in the inner regions, where, notably: (a) accreting protoplanets are most likely to reside \citep{Close2020}, and (b) moderate improvements in contrast are likely to yield pronounced differences in detectability. In cases where the lowest curve was only marginally lower than others inside the control radius, but was substantially higher outside the control radius, we selected a curve that balanced these two regions. 

We computed contrast curves for each of the 11 cuts under 5, 10, 20, and 50 KL mode PSF subtractions. In most cases, the optimal contrasts agreed across KL modes; however, in some cases, we were forced to balance variations among them in selecting the optimum. 

From a  practical standpoint, the KLIP algorithm is computationally intensive, and removing some subset of images from the analysis can lead to substantial improvements in processing time. Since we apply a grid search algorithm to optimize \texttt{pyKLIP} parameters (see Section \ref{sec:optimize}), culling the datasets was important in making processing and optimization more tractable. For this reason, in cases where contrasts were equivalent among data quality cuts, we chose the most aggressive cut.

\begin{figure*}
    \centering
    \includegraphics[width=\textwidth]{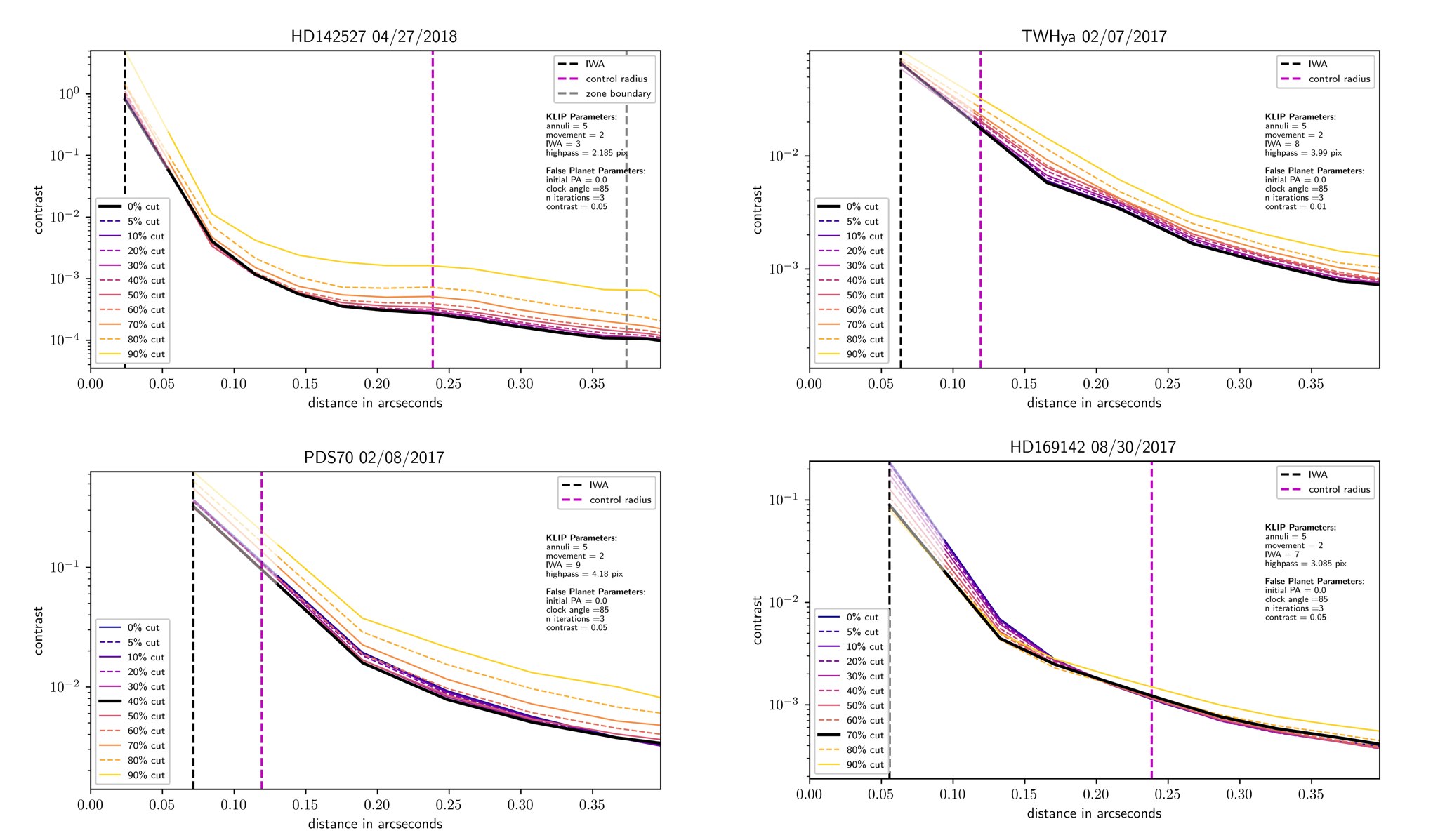}
    \caption{A representative sample of contrast curves used to cull GAPlanetS datasets before KLIP optimization. Each of our four classifications is represented here, namely (clockwise from upper left: clustered, sequential, turnover, and down-and-up). Each colored line represents the post-KLIP contrast achieved by culling different proportions (0\%, 5\%, 10\%, 20\%, 30\%, 40\%, 50\%, 60\%, 70\%, 80\%, and 90\%) of the lowest-quality images from the sequence, where the FWHM of a Moffat fit to the central star (for unsaturated data) or ghost (for saturated data) is used as a proxy for image quality. The curve is extrapolated inward to the inner working angle following the slope of the two closest points at which throughput was computed. The curves are given higher transparency in this extrapolated region. Curves were computed with a standard set of KLIP parameters (\texttt{annuli}=5, \texttt{movement}=2), and are shown here for 10 KL modes (though 5, 10, 20, and 50 mode reductions were also generated and compared before making a final choice). The optimal cut was determined by eye, prioritizing achieved contrast inside of the AO control radius, though several concerns were balanced, as described in the text. Plots for all GAPlanetS datasets are available as a figure set (39 images) in the online journal.}
    \label{fig:cutsfig}
\end{figure*}

The final adopted values for data quality cuts for each dataset are given in Table \ref{tab:optsumm}. Figure \ref{fig:cutsfig} shows a representative sample of common patterns seen in these data. 

By-eye examination of the contrast curves that we used to determine data quality cuts resulted in four basic classifications among the 39 datasets. We hypothesize that these classifications are driven by a combination of (a) overall atmospheric quality, (b) variability in atmospheric quality, (c) the dominant noise regime at each separation in the image, and (d) the preservation of rotational space for the PSF library in the image sequence. An in-depth exploration of these trends is beyond the scope of this work. A sample of each of the four families of curves can be seen in Figure \ref{fig:cutsfig}. 

\paragraph{Clustered} datasets ($N$=16) show equivalent, overlapping contrast curves at all separations for a range of data quality cuts from 0\% to $N$\% where $N$ is usually in the range of 30-50\%. After this cutoff, the curves generally evolve upward (toward poorer contrasts) as the cuts get more aggressive. We hypothesize that this coincides with the point at which the total amount of rotation in the dataset begins to decrease as more data are discarded.

\paragraph{Turnover} datasets ($N$=9) show a marked crossing of contrast curves at some turnover point that ranges in distance from 0$\farcs$15 to 0$\farcs$3. Inside of the turnover, the contrast generally improves as more data are discarded. The magnitude of this improvement varies. Outside of the crossover, the opposite is true, though the curves are generally more tightly clustered as distance from the star increases. We hypothesize that this turnover corresponds primarily to a switch from a speckle-limited regime at close separation to a photon-noise-limited regime at greater distance. 

\paragraph{Dipping} curves show an improvement in contrast as an increasing proportion of data is discarded up to a certain threshold, after which the contrast gets poorer again as more data are discarded. Sometimes this is true at all separations ($N$=2), but this pattern is seen more often in the inner ($r<$0$\farcs$2) regions only while the outer regions exhibit some other pattern ($N$=8). 

\paragraph{Sequential} curves show a global evolution in which the contrast is either improving ($N$=1) or worsening ($N$=4) steadily as the amount of data discarded increases. 

Further optimization of this approach is warranted, and includes incorporation into our broader post-processing grid search and more rigorous exploration of the patterns outlined above as they relate to dataset properties and noise regimes.

\begin{figure*}
    \centering
    \includegraphics[width=\textwidth]{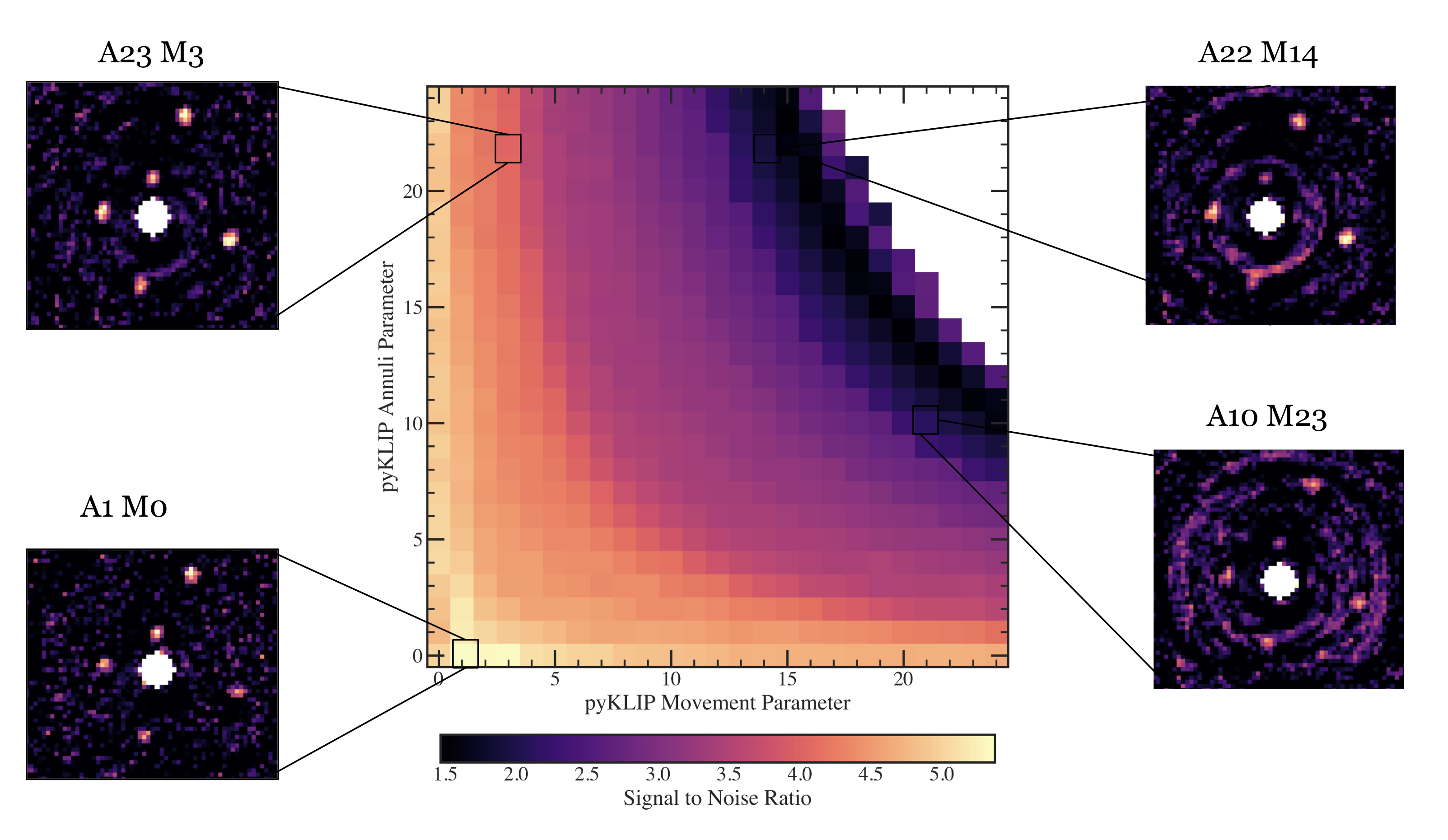}
    \caption{A representative \texttt{pyKLIP-PE} output heat map depicting the quality of recovery of the \textit{innermost} false planet (of five total) injected at $r\sim$0$\farcs$06 into the continuum imagery of HD~169142. The $x$ and $y$-axes of the central heatmap represent the values of the tunable KLIP \texttt{annuli} and \texttt{movement} parameters, respectively. The value in each cell of the heatmap is equal to the peak SNR of the innermost false planet when reduced under those \texttt{pyKLIP} parameters and with a 10 KL mode PSF model. Each pixel therefore represents one KLIP reduction; this is further shown in the inset images, which depict the full SNR maps for four of the pixels in the heatmap. There is a clear optimal region for robust extractions of this innermost planet signal, and other combinations of KLIP parameters that are suboptimal. The white region in the upper-right corner reflects KLIP annuli/movement combinations for which no reference images remain in the PSF library and KLIP cannot be completed. The heatmap is shown here for 10 KL modes, the innermost planet, and the four selected annuli/movement combinations only. However, a fully interactive version of this figure is available in the online journal, allowing the user to select an arbitrary movement/annuli combination, injected planet, KL mode, and image quality metric.}
    \label{fig:paramexplore}
\end{figure*}

\subsection{Optimization of KLIP Parameters with \texttt{pyKLIP-PE} \label{sec:optimize}}

Variation in the final appearance of PSF-subtracted images according to algorithmic parameters is now a well-established fact \citep[e.g.,][]{Milli2012, Meshkat2013, Follette2017}, and can be easily seen in Figure \ref{fig:paramexplore}. In order to make well-justified data-driven decisions about optimal parameter choices for each morphologically complex GAPlanetS system, we completed a coarse grid search of select KLIP parameters for each dataset individually. 

This ``pyKLIP Parameter Explorer" (\texttt{pyKLIP-PE}) algorithm calculates a number of post-processed image quality metrics for real and/or injected point-sources with a range of \texttt{pyKLIP} \texttt{annuli}, \texttt{movement}, \texttt{subsections}, and \texttt{numbasis} (KL modes) parameters. In this work, we have chosen to optimize recovery of false planets injected into continuum images. Injection of synthetic planets allow us to balance post-processed image quality across a broad image region, optimize datasets without known point-sources, and avoid cognitive biases in the selection of parameters for recovery of controversial planet candidates. A companion paper to this work \citep{Jea} details the \texttt{pyKLIP-PE} algorithm and validation of its use for optimization of point-source recovery in GAPlanetS data. Here, we summarize the results in broad strokes, and refer the reader to \citet{Jea} for details.

To make \texttt{pyKLIP-PE} computational time tractable for the entire GAPlanetS database, we applied fixed choices for the \texttt{pyKLIP} \texttt{subsections} ($n$=1), \texttt{IWA} (IWA=FWHM), and \texttt{highpass} (0.5$\times$FWHM) parameters. These were selected to optimize point-source recovery (by applying an aggressive highpass filter) in a region where we might reasonably expect to resolve point-sources (beyond 1 FWHM of the central star). 

False planets are constructed by scaling images of the central star (unsaturated datasets) or ghost (saturated datasets) to a particular contrast. Individual exposures are used for this purpose so that the PSF of the injected companion in each image mirrors that of the star, as would be expected for a true planet. For each dataset, planets are injected between $r$=1.5$\times$FWHM and the AO control radius with an angular separation of 85$^{\circ}$ and a radial separation of 1FWHM for AO bin 1 (control radius=30~pixels) datasets, and 0.5$\times$FWHM for bin 2 (control radius=15~pixels) datasets. 

Contrasts for injected planets are iterated upon until their recovered SNRs under a conservative choice of KLIP parameters (\texttt{annuli}=5, \texttt{movement}=2) has an average across 5, 10, 20 and 50 KL modes of 6.5-7.5. This SNR$\sim$7 threshold was selected to be somewhat higher than the canonical detection threshold of SNR=5 so that a range of KLIP parameter combinations would result in robust (SNR$>$5) detections. 

The result of the \texttt{pyKLIP-PE} algorithm is a multidimensional array of image quality metrics (planet SNR, false-positive pixel count, etc.) for various combinations of KLIP parameters. In principle, there could be one dimension to this grid for each of the more than 20 parameters of the \texttt{pyKLIP} algorithm. However, we have chosen to focus on optimizing only a few key parameters and have made data-driven decisions about reasonable fixed choices for others, as described in detail in \citet{Jea}. 

\subsubsection{Optimized Parameters}
The principal \texttt{pyKLIP} parameters optimized for the final GAPlanetS reductions were the \texttt{annuli}, \texttt{movement}, and \texttt{numbasis} (KL mode) parameters, described briefly below. 
\begin{enumerate}
    \item The \texttt{pyKLIP} \texttt{annuli} parameter sets the number of concentric, equal width, annular zones that are analyzed separately by KLIP. The exact width of the annuli in pixels varies very slightly among datasets due to variation in IWA, but ranges from roughly 225 pixels wide for 1 annulus to $\sim$9 pixels wide for 25 annuli. 
    \item The \texttt{pyKLIP} \texttt{movement} parameter controls rotational masking. All images where a planet would have rotated by fewer than a given number of pixels between the target image and the reference image(s) are excluded when constructing a PSF for the target image, thereby limiting self-subtraction. Low values of the \texttt{movement} parameter are thus ``aggressive'' values, with very few images excluded from the reference set and more prominent self-subtraction. A single \texttt{movement} value applied in multiple zones across an image is also more aggressive for annuli near the center of the image, where a given number of pixels of rotation about the center translates to a larger angular exclusion criterion. When the \texttt{movement} parameter becomes high enough, there are no remaining reference images that meet the exclusion criterion, and KLIP reductions are impossible. 
    \item The \texttt{pyKLIP} \texttt{numbasis} parameter determines the number of principal components, or ``KL modes", used to construct the final PSF. KL modes are a set of orthogonal basis vectors constructed from the PSF reference library, where the first mode is the vector that describes the most variance, and each subsequent mode describes some additional (smaller) amount of variance in the dataset. Thus, increasing the number of KL modes increases the complexity of the PSF model. 
\end{enumerate}

\begin{figure*}
    \centering
    \includegraphics[width=\textwidth]{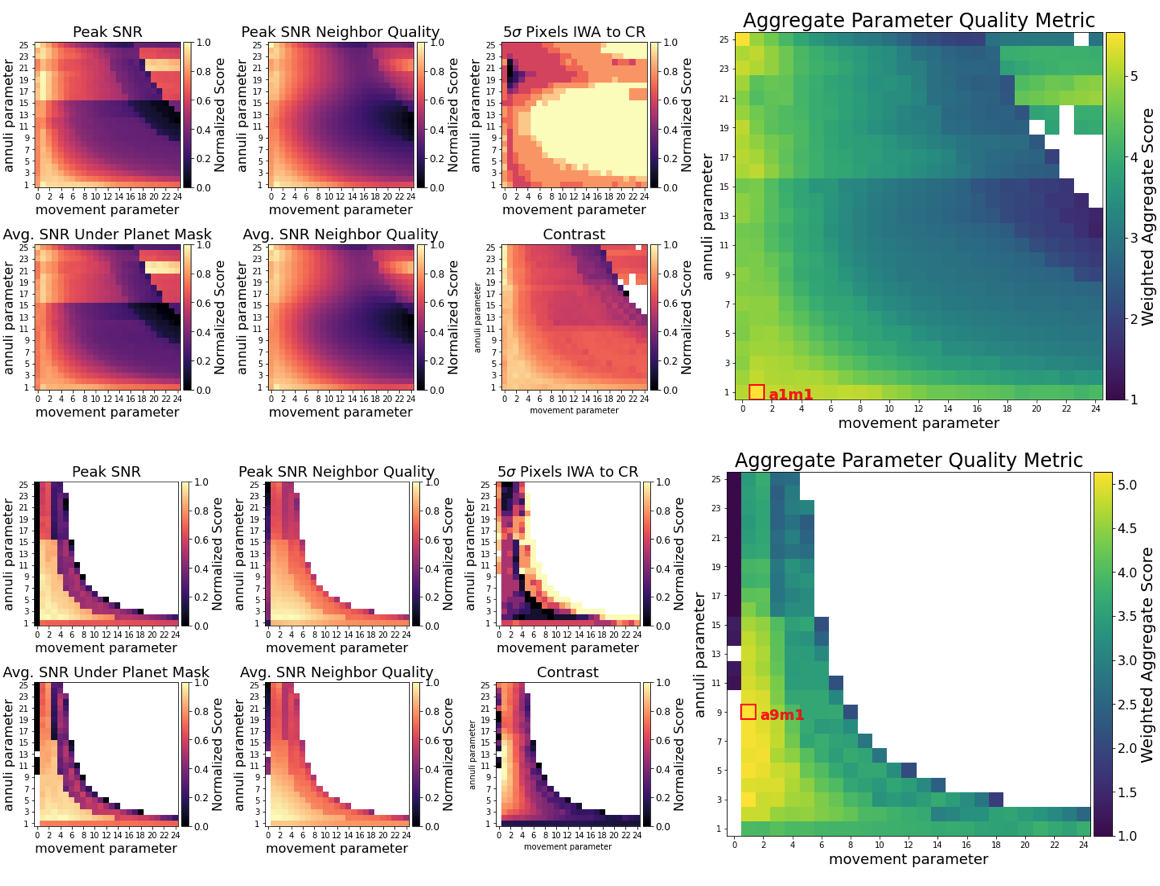}
    \caption{Sample \texttt{pyKLIP} parameter quality heatmaps for two representative datasets (top: HD~169142 04/08/2014, bottom: CS~Cha 05/15/2015) and for all six post-processed image quality metrics. Each small subpanel maps the normalized value of a different image quality metric, and each pixel represents the value of that metric averaged across all of the false planets in the \texttt{pyKLIP} post-processed image with the \texttt{movement} and \texttt{annuli} values indicated on the $x$ and $y$-axes, respectively. In this case, the heatmaps for 5 and 20 KL modes have also been averaged to ensure stability among low and moderate KL modes. The individual metrics in the six subpanels are weighted (equally in this case) and combined to form an aggregate parameter quality metric (rightmost panel), from which an ``optimum" parameter combination is selected (indicated in red). The nature of each individual metric is described in detail in the text. White pixels represent either parameter combinations for which metrics were not able to be extracted (often the case for \texttt{movement}=0) or where the aggressiveness of the rotational mask leaves no reference images in the PSF library (upper right of each plot). Plots for all GAPlanetS datasets are available as a figure set (39 images) in the online journal.}
    \label{fig:metric}
\end{figure*}

\subsubsection{Image Quality Metrics}
The \texttt{pyKLIP-PE} algorithm extracts four image quality metrics for injected (or real) companions. These are: (1) the peak pixel values of each planet in the SNR map (``peak SNR metric"), (2) the average SNR of all positive pixels within 0.5xFWHM radially and 5$^{\circ}$ azimuthally of each planet in the SNR map (``average SNR metric"), (3) the achieved contrast at the location of each planet, and (4) the number of false-positive ($>$5$\sigma$) pixels between the IWA and the AO control radius. An additional two metrics, which we call ``neighbor quality" metrics, are computed by smoothing the peak and average signal-to-noise metrics across \texttt{pyKLIP}'s \texttt{movement} and \texttt{annuli} parameters. These metrics are based on the anecdotal understanding prevalent in the community that the most robust KLIP detections are those where small variations in KLIP parameters do not substantially affect the recovered planetary SNR. 

In this work, we choose a simple equally weighted sum of all six normalized metrics. We then average the sum of these normalized metrics across all false planets, thereby balancing recovery of signals throughout the region of interest. As a further measure to ensure robustness, we average the 5 and 20 KL mode aggregate parameter quality maps before selection of a final \texttt{movement}, \texttt{annuli} combination. As described in detail in \citet{Jea}, these particular KL modes were chosen based on the statistical distribution of optimal parameter choices for HD~142527~B recovery under a range of optimization scenarios. 

Effectively, this methodology means that the algorithm attempts to select an \texttt{annuli} and \texttt{movement} combination that results in robust (according to all six metrics) extraction of injected planets throughout the region of interest for both low ($n$=5) and moderate ($n$=20) PSF model complexity. Once this combination of \texttt{annuli} and \texttt{movement} parameters is chosen, the final choice of ``optimal" KL mode is made by maximizing the sum of all six metrics for that dataset averaged over all injected planets. Optimal \texttt{annuli}, \texttt{movement}, and \texttt{numbasis} (KL mode) parameters selected using this methodology are reported for each dataset in Table \ref{tab:optsumm}. Unless otherwise indicated, all post-processed images shown in this work have \texttt{pyKLIP} parameters selected via this methodology.

Figure \ref{fig:metric} shows two examples of normalized parameter quality maps for all six image quality metrics averaged across all injected continuum planets and among 5 and 20 KL modes. The aggregate parameter quality map is shown in the large panel at right, and the optimal \texttt{annuli} and \texttt{movement} parameters are indicated. A complete description of how we arrived at this method, as well as detailed discussion of the features of these maps, can be found in \citet{Jea}.  

\begin{table*}[]
\tablecaption{Summary of Optimization Results for Each GAPlanetS Dataset}
\footnotesize
\begin{tabular}{cccccc|cccc|cccc}%
\multicolumn{6}{c}{\textbf{General Dataset Parameters}}&
\multicolumn{4}{c}{\textbf{Data Quality Cut Parameters}}&
\multicolumn{4}{c}{\textbf{KLIP Optimization Parameters}}\\
Object&
Date&
N$_{\rm{total}}$&
Bin&
r$_{\rm{sat}}$&
IWA&
C$_{\rm{fakes}}$&
Cut&
FWHM&
Cut Classification&
$N_{\rm{false}}$&
Annuli&
Movement&
KL Modes\\
\hline \hline 
CS~Cha    & 5/15/15  & 143  & 2 & \nodata & 7 & 0.05 & 30 & 7  & down and up inside 0$\farcs$2  & 2 & 6  & 2 & 3\\
HD~100453  & 2/17/17  & 2947 & 1 & 3 & 4 & 0.01 & 10 & 4  & clustered, increasing & 5 & 5 & 3 & 10\\
HD~100453  & 5/2/18   & 356  & 1 & 2 & 3 & 0.01 & 40 & 3  & clustered, increasing & 8 & 8  & 1 & 20\\
HD~100453  & 5/2/18   & 563  & 1 & 4 & 4 & 0.01 & 30 & 3  & clustered, increasing & 7 & 13 & 1 & 20\\
HD~100453  & 5/3/18   & 2831 & 1 & 4 & 4 & 0.01 & 70 & 3  & turnover at 0$\farcs$07 & 7 & 19 & 1 & 20\\
HD~100546  & 4/12/14  & 4939 & 1 & 8 & 8 & 0.01 & 10 & 5  & clustered, increasing & 3 & 13 & 1 & 10\\
HD~100546  & 5/30/15  & 2459 & 1 & 2 & 4 & 0.05 & 50 & 4  & turnover at 0$\farcs$12 & 5 & 25 & 2 & 5\\
HD~141569  & 4/9/14   & 2402 & 1 & 5  & 5 & 0.01 & 70 & 5  & turnover at 0$\farcs$15 & 4 & 1  & 1 & 5\\
HD~141569  & 4/10/14  & 1364 & 1 & 2 & 6 & 0.01 & 50 & 6  & turnover at 0$\farcs$15 & 3 & 2  & 4 & 1\\
HD~141569  & 4/11/14  & 2340 & 1 & 4 & 4 & 0.01 & 10 & 4  & clustered, increasing & 5 & 4  & 5 & 20\\
HD~141569  & 5/28/15  & 723  & 1 & \nodata & 7 & 0.05 & 70 & 7  & decreasing & 2 & 2  & 4 & 2\\
HD~141569  & 5/29/15  & 404  & 1 & 9 & 9 & 0.01 & 0  & 4  & increasing & 4 & 1  & 5 & 5\\
HD~142527  & 4/11/13  & 1961 & 1 & \nodata & 3 & 0.01 & 10 & 3  & clustered, increasing & 8 & 14 & 1 & 20\\
HD~142527  & 4/8/14   & 1758 & 1 & \nodata & 4 & 0.01 & 0  & 4  & clustered, increasing & 5 & 25 & 1 & 100\\
HD~142527  & 4/8/14   & 68 & 1 & 11 & 11 & 0.01 & 0  & 5  & increasing & 2 & 21 & 1 & 10\\
HD~142527  & 5/15/15  & 2387 & 1 & 2 & 4 & 0.01 & 50 & 4  & clustered, increasing & 5 & 2  & 2 & 20\\
HD~142527  & 5/16/15  & 1143 & 1 & 2  & 6 & 0.05 & 80 & 4  & turnover at 0$\farcs$3 & 5 & 20 & 1 & 20\\
HD~142527  & 5/18/15  & 159  & 1 & 6 & 6 & 0.01 & 0  & 4  & clustered, increasing & 4 & 2 & 7 & 5\\
HD~142527  & 2/10/17  & 242  & 1 & 2 & 3 & 0.05 & 0  & 4  & increasing & 5 & 25 & 1 & 5\\
HD~142527  & 4/27/18  & 580  & 1 & 3 & 3 & 0.05 & 0  & 3  & clustered, increasing & 8 & 22 & 1 & 10\\
HD~169142  & 4/8/14   & 2796 & 1 & \nodata & 4 & 0.01 & 80 & 4  & turnover at 0$\farcs$  & 5 & 1  & 1 & 20\\
HD~169142  & 4/9/14   & 178  & 1 & 6 & 6 & 0.01 & 50 & 5  & clustered, increasing & 3 & 20 & 1 & 4\\
HD~169142  & 5/18/15  & 1731 & 1 & 3 & 5 & 0.01 & 20 & 5  & clustered, increasing & 4 & 6  & 2 & 20\\
HD~169142  & 8/30/17  & 1658 & 1 & 2 & 4 & 0.05 & 70 & 4  & turnover at CR & 5 & 6  & 2 & 20\\
LkCa~15   & 11/16/14 & 308  & 2 & \nodata & 6 & 0.05 & 5  & 7  & clustered, increasing & 2 & 4  & 6 & 20\\
LkCa~15   & 11/18/16 & 252  & 2 & \nodata & 7 & 0.05 & 10 & 11 & mostly increasing & 1 & 10 & 1 & 20\\
PDS~66    & 2/7/17   & 243  & 2 & \nodata & 5 & 0.01 & 40 & 5 & clustered, increasing & 2 & 2 & 1 & 1 \\
PDS~70     & 2/8/17   & 188  & 2 & \nodata & 7 & 0.05 & 40 & 7  & down and up in inner 0$\farcs$15 & 2 & 1  & 1 & 1\\
PDS~70     & 5/2/18   & 209  & 2 & \nodata & 6 & 0.05 & 5  & 6  & clustered, increasing & 2 & 4  & 1 & 4 \\
PDS~70     & 5/3/18   & 284  & 2 & \nodata & 6 & 0.05 & 30 & 6  & down and up in inner 0$\farcs$1  & 2 & 3  & 7 & 50 \\
SAO~206462 & 4/12/14  & 3993 & 2 & 3 & 4 & 0.01 & 30 & 4  & down and up & 5 & 1  & 1 & 3\\
SAO~206462 & 5/26/15  & 408  & 2 & \nodata & 4 & 0.05 & 30 & 4  & clustered, increasing & 3 & 1  & 2 & 2\\
TW~Hya    & 4/8/14   & 1958 & 2 & \nodata & 5 & 0.01 & 50 & 5  & turnover at 0$\farcs$15  & 2 & 5  & 2 & 1\\
TW~Hya    & 2/7/17   & 452  & 2 & 2 & 6 & 0.01 & 0  & 6  & increasing   & 2 & 6  & 6 & 50\\
UX~Tau~A  & 11/15/14 & 52 & 2 & \nodata & 8 & 0.05 & 8 & 30 & clustered, increasing & 2 & 6 & 2 & 50\\
V1247~Ori & 11/15/14 & 893  & 2 & \nodata & 4 & 0.01 & 50 & 4  & turnover at CR & 3 & 2  & 9 & 10\\
V1247~Ori & 12/11/15 & 878  & 2 & \nodata & 5 & 0.01 & 60 & 5  & turnover at CR & 2 & 2  & 6 & 50\\
V4046~Sgr & 4/12/14  & 1414 & 2 & 2 & 5 & 0.01 & 50 & 5  & turnover at 0$\farcs$15  & 2 & 21 & 2 & 20\\
V4046~Sgr & 5/17/15  & 720  & 2 & \nodata & 4 & 0.01 & 50 & 4  & down and up & 3 & 13 & 1 & 50\\
\end{tabular}%
\tablecomments{\scriptsize The leftmost block of columns give general dataset parameters, where N$_{\rm{total}}$ is the total number of images in the dataset prior to implementing the data quality cut and "Bin" is the binning of the wavefront sensor, which determines the location of the control radius (30 pixels for bin 1, 15 for bin 2). The central block of columns give the parameters for the data quality cuts, where C$_{\rm{fakes}}$ is the contrast of injected fake planets used to compute contrast curves. The rightmost block of columns are the derived optimal values from pyKLIP-PE, where  N$_{\rm{false}}$ is the number of injected false planets between the Inner Working Angle (IWA) and control radius used to compute the optimal parameters. \label{tab:optsumm}}
\end{table*}

\section{Survey Analysis Methodologies \label{sec:analysis}}

In this section, we provide explanations of various methods used to generate the final, optimized images and sensitivity limits for the GAPlanetS companion candidate search, results of which appear in Section \ref{sec:results} and \ref{sec:survey}. We also detail the tools used for candidate characterization, including the procedure for obtaining a final estimation of mass accretion rate.

\paragraph{KLIP-ADI images} are shown in Section \ref{sec:results} and Appendix \ref{sec:allepochs} for all GAPlanetS targets. For those targets with known companions or companion candidates detected in GAPlanetS data, all epochs (including nondetection epochs) are shown in the main body of the text for completeness. As there are less robust 2$\sigma$-4$\sigma$ excess signals at a number of locations in most datasets, additional epochs for objects without recovered candidate companions are shown in Appendix \ref{sec:allepochs}. Some of these signals may prove in the future to be protoplanets upon higher-contrast follow-up and/or additional epochs of observation. 

\paragraph{Optimization strategies} One important consideration in extracting point-source signals from these data is whether to conduct analyses on post-processed images that have undergone conservative KLIP parameter optimization (on false planets injected into the continuum data, as described and shown in Section \ref{sec:optimize}), or to optimize on the known location of the companion(s) in H$\alpha$ images. The relative merits of each of these strategies is discussed in detail in \citet{Jea}. In short, optimization on false planets injected into the continuum data is a robust method that is substantially less likely to yield false-positive detections. We apply it to all datasets to achieve a uniform analysis. However, there is necessarily a penalty in the final signal-to-noise ratio of recovered H$\alpha$ companions by virtue of the optimizations being done on a different (albeit contemporaneous and close-in wavelength) dataset and averaged over planets located throughout the region of interest rather than at a particular PA and separation. In the case of very robust high-SNR recoveries, this penalty is of minimal concern as the companion is recovered using both strategies. However, in the case of detections at or near the detection threshold, this SNR penalty may result in nonrecovery of the companion under the continuum optimization method. In cases where the point-source nature of the companion is robustly established in the literature (HD~142527~B and PDS~70~b and c, all of which have been detected at continuum wavelengths in addition to H$\alpha$) and the companion is unrecovered under the standard survey optimization methodology, we report optimizations done directly on the H$\alpha$ imagery at the companion location as well. We note that such direct optimization on the H$\alpha$ images, though it is more likely to result in a recovery of planetary signal, risks overfitting and should be interpreted with caution. To mitigate this somewhat, we adopt a relatively conservative version of this direct optimization approach by averaging across several KL modes and image quality metrics (for difficult HD~142527~B recoveries) or across several known companions (for PDS~70). 

\paragraph {ASDI images} All KLIP image panels (figures \ref{fig:hd142527ims}--\ref{fig:hd169142} and all figures in Appendix \ref{sec:allepochs}) show H$\alpha$ (left) and continuum (middle left) reductions, as well as two ASDI images. The first (middle right) image is a conservative reduction computed by scaling the \texttt{pyKLIP}ed continuum image up by the median stellar H$\alpha$/continuum brightness ratio (as described in Section \ref{sec:SDI}). The second (right) ASDI reduction is computed by subtracting the two at 1:1 scale. The relative fidelity of the two types of SDI reduction is nuanced and is discussed in detail in Section \ref{sec:SDI}. To recap briefly here, scaling and subtracting by the H$\alpha$/continuum brightness ratio should effectively remove both scattered light and stellar residuals, where present. However, in the absence of such signals/residuals, the 1:1 scaled reduction is the more accurate indicator of H$\alpha$ excess. We show both, as well as the \texttt{pyKLIP}-reduced H$\alpha$ and continuum images in order to provide full context with which to judge the fidelity of any apparent signals. 

\paragraph{Robustness of signals among H$\alpha$ and SDI images} is an important concern in extracting accurate photometry and astrometry for GAPlanetS candidates. The highest-fidelity signals are present in \textit{both} H$\alpha$ and SDI images. While the SDI process helps to remove disk signal and stellar residuals, as well as the continuum contribution of objects like HD~142527~B, it can also induce false-positive signals into SDI imagery. This occurs when a negative speckle in the KLIP-processed continuum images is reversed during SDI subtraction, becoming positive. Negative continuum speckles with sufficient amplitude to mimic planetary signals are relatively uncommon, but are visible and appear point-source-like in several GAPlanetS datasets ($n$=2). These spurious point-sources are marked with yellow ``x" symbols in all SDI images where they are apparent. 

\paragraph{Multiepoch combinations} (post-KLIP mean combinations) are utilized in cases where independent datasets were acquired for a given object within several days of one another. Although this technique of combining post-processed KLIPed images from multiple near-in-time epochs has been shown to yield detections in some cases where single epoch images do not reveal a high SNR source \citep[e.g.;][]{Wagner2018}, we caution that accreting companions are also likely variable, so higher SNR detections will only result in cases where the object is accreting at a detectable rate in both epochs. Given the detectable level of H$\alpha$ variability seen on night-to-night timescales in GAPlanetS objects (see \citet{Balmer2022}), accretion rates derived from combined data should be interpreted with caution. 

\begin{figure}
    \centering
    \includegraphics[width=\columnwidth]{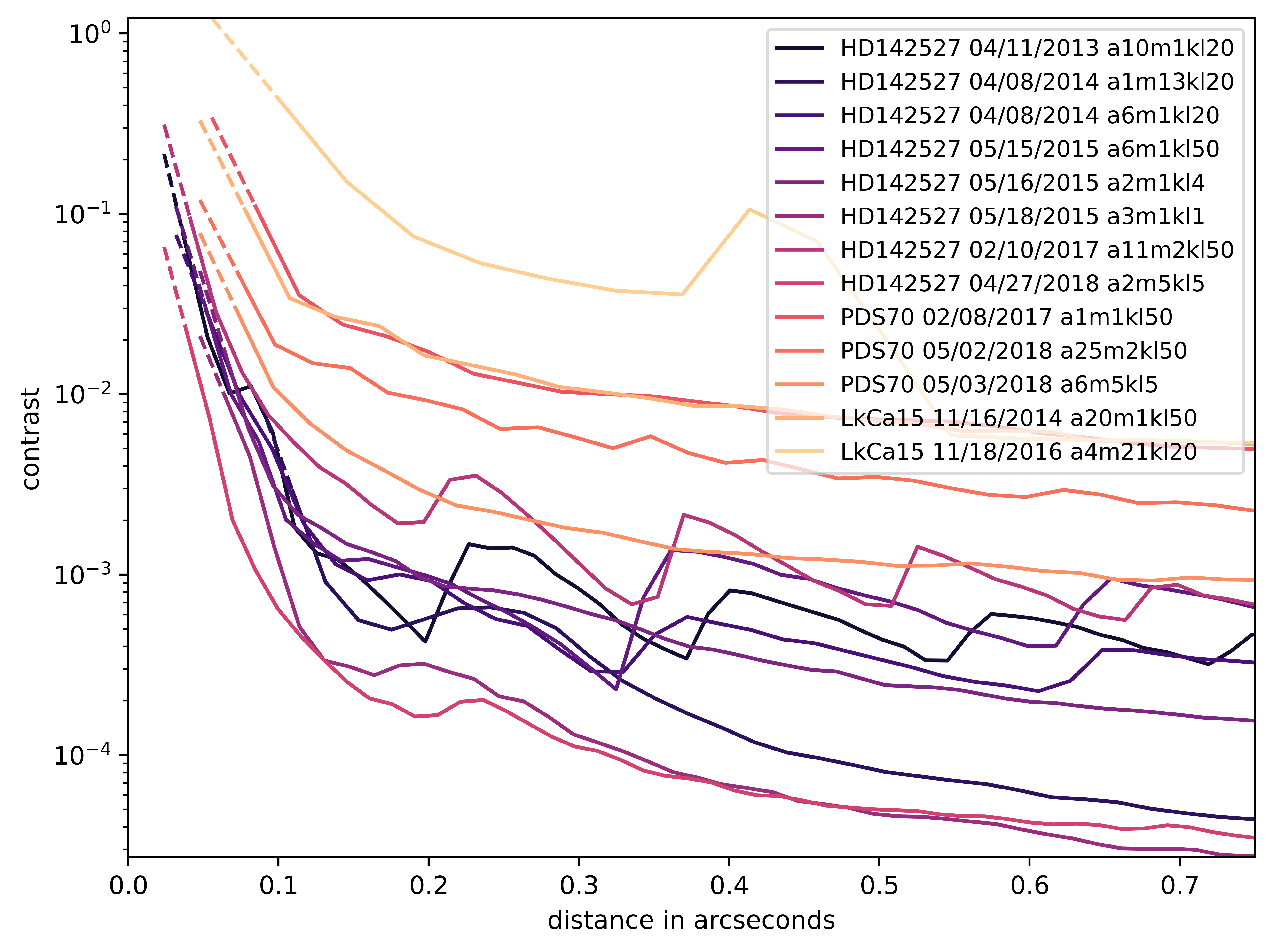}
    \caption{Throughput-corrected 5$\sigma$ contrast curves of all epochs for continuum false planet optimized \texttt{pyKLIP} reductions of the three previously reported GAPlanetS companion and companion candidate-hosts: HD~142527 (seven epochs), PDS~70 (three epochs), and LkCa~15 (two epochs). Throughput was computed as described in the text, with a correction for the small number of independent noise samples near the star following \citet{Mawet2014}. Solid curves indicate regions where throughput-corrected contrast was computed directly. The curves are also projected inward from the innermost throughput measurement to the inner working angle, and this extrapolated region is indicated with a dashed line.}
    \label{fig:detectctrst}
\end{figure}

\paragraph{Contrast curves} are shown at H$\alpha$ for all datasets in Figures \ref{fig:detectctrst}, \ref{fig:closecandctrst}, \ref{fig:widectrsts},  and \ref{fig:nondetectctrst}.  These contrast units are then translated into generalized mass accretion rate limits for the overall survey in Section~\ref{sec:survey}.

\paragraph{Astrometry of detected companions} is computed via the Bayesian KLIP Astrometry (BKA) technique described in detail in \citet{Wang2016} and implemented via \texttt{pyKLIP}. In short, the technique creates a forward model by projecting one or more PSFs onto the KL basis set.  This results in a post-processed PSF that replicates the complex shape of the planetary core and self-subtraction lobes unique to a given dataset, choice of KLIP parameters, and point-source location. This forward-modeled PSF is adjusted astrometrically and photometrically to produce quality-of-fit posterior distributions. BKA input can be either a single fixed PSF or a time-variable PSF cube. Given the demonstrated high degree of PSF variation in GAPlanetS data, we have opted for the latter. Our input PSF model is therefore a time series of normalized image stamps, one per image in the sequence. We report astrometric (and photometric) values for companions as the median of the BKA posterior distribution.

\paragraph{Contrasts of companion candidates} are also fit with BKA. In our case, we ensure that the PSF model (star or ghost image) has a fixed contrast by normalizing it and then multiplying by a fixed contrast that we know is reasonably close to that of the companion. Preprocessed images are also normalized (by dividing each image by the peak of a Moffat fit to the central star in the case of unsaturated data or the ghost multiplied by the established ghost-to-star scale factor in the case of saturated data) in preprocessing before injecting false planets and running KLIP. By normalizing the input images to a peak value of one and the input PSF to the ``best guess" contrast, this ensures that the simulated point-source is modeled at a fixed contrast relative to the time-varying central PSF. After computation of the forward model, the post-processed PSF brightness is iterated upon during BKA analyses with a scale factor parameter that we multiply by the initial contrast guess to get the final best-fit contrast value for a companion.

\paragraph{Uncertainties on companion astrometry and photometry} are reported as the 67\% credibility interval of the BKA posterior distributions for separation, position angle, and contrast. Uncertainty on absolute astrometric calibration of the instrument and the location of the central star in the images is incorporated into the error estimation for those quantities. However, the reported error bars for contrast encompass only the 67\% credibility interval for the scale factor posterior and not any uncertainties in photometric calibration, which we propagate separately into accretion rate estimates, as described in detail in Section \ref{sec:survey}.

\paragraph{Detection strengths} are characterized further by utilizing the \texttt{PlanetEvidence} class \citep{Golomb2019} within \texttt{pyklip} to conduct a Bayesian model comparison and make a more conservative SNR estimate for each companion detection. These values are reported in Table \ref{tab:detections}. \texttt{PlanetEvidence} uses the nested sampling implementation \texttt{pyMultiNest} \citep{Buchner2014, Feroz2009} to compare two models: $\mathrm{H_0}$, where the image contains only speckle noise, and $\mathrm{H_1}$, where the image contains a source at the position of the companion. \texttt{PlanetEvidence} returns marginal distributions of the parameters for the source and null cases, and calculates the SNR of the detection within the fitting region and the evidence values for $\mathrm{H_0}$ and $\mathrm{H_1}$ ($\mathrm{Z_0}$ and $\mathrm{Z_1}$). The log-ratio of these evidence values, $\log{B_{10}}=\log{Z_1/Z_0}$, enables us to quantify the confidence with which one model can be favored over the other. This framework provides a more robust estimate of the quality of the detection because it better captures asymmetric speckle noise, which can dominate at very close separations. Values for $\log{B_{10}}>5$ are considered ``strong" evidence against the null hypothesis. We note that \texttt{PlanetEvidence}-extracted SNR and $\log{Z_1/Z_0}$ values are sensitive to the size of the BKA fitting region, and can vary by $\sim$10\% based on this choice. We adopt a 15 pixel square fitting region for all forward model fits except in cases where nearby residual structure results in a clear under- or over-subtraction of the point-source candidate. 

\section{Individual Object Results \label{sec:results}}

This section outlines GAPlanetS results for each transitional disk in the sample and describes in basic terms the recovery (or lack thereof) of confirmed or candidate protoplanets. In each case, the data have been preprocessed as described in Section \ref{sec:obs}, culled as described in Section \ref{sec:dqcuts}, post-processed with \texttt{pyKLIP} with parameters optimized for false planets injected into continuum images as described in Section \ref{sec:optimize}, and images and contrast curves generated following the methods described in Section \ref{sec:analysis}. 

The object-by-object results are aggregated in this section into previously reported GAPlanetS detections (Section \ref{sec:detect}), new planet candidates (Section \ref{sec:newcand}), other objects for which known or candidate companions exist in the literature (Section \ref{sec:candidates}), and objects with no known or candidate companions (Section \ref{sec:nondetect}). Survey-level analyses follow in Section \ref{sec:survey}. Literature predictions for the locations of companions and companion candidates are shown in all images, and these predictions are compiled in Table \ref{tab:objects}. In most cases, literature candidate companions are not detected in GAPlanetS imagery, and detection limits at their predicted locations are summarized in Section \ref{sec:acclims}.

In cases where companions or companion candidates are successfully detected in H$\alpha$ and/or continuum GAPlanetS imagery, best-fit astrometry and photometry, computed as described in section \ref{sec:analysis}, is summarized in the text. Astrometric and photometric fit statistics and forward models follow in section \ref{sec:fms}.

\begin{table*}[]
\centering
\begin{tabular}{cccccccc}
\tablecaption{Reported and Predicted Positions for Companions and Companion Candidates}
\textbf{Object} & \textbf{Label} & \textbf{Epoch(s)}  & \textbf{Sep} & \textbf{Sep Error} & \textbf{PA} & \textbf{PA Error} & \textbf{Source} \\
 &   &   & (pix)   & (pix) & (deg)   & (deg)  &   \\
\hline
\multicolumn{8}{c}{\textit{Previous GAPlanetS Detected Objects with Epoch-Specific Location Measurements/Predictions}}\\
\hline
HD~142527 & B & 04/11/2013 & 10.2 & 0.1 & 126.0 & 0.5 & \citet{Balmer2022}\\
HD~142527 & B & 04/08/2014 & 9.7  & 0.2  & 115.9 & 1.0 & \citet{Balmer2022}\\
HD~142527 & B & 05/15/2015 & 8.4  & 0.2 & 109.6 & 0.8 & \citet{Balmer2022}\\
HD~142527 & B & 05/16/2015 & 8.6  & 0.3 & 108.6 & 1.0 & \citet{Balmer2022}\\
HD~142527 & B & 05/18/2015 & 8.6  & 0.2 & 110.2 & 0.7 & \citet{Balmer2022}\\
HD~142527 & B & 02/10/2017 & 6.4  & 0.2 & 77.8  & 2.0 & \citet{Balmer2022}\\
HD~142527 & B & 04/27/2018 & 4.8  & 0.3 & 58.0  & 1.8 & \citet{Balmer2022}\\
PDS~70    & WiPb & 02/08/2017 & 23.3 & 0.2 & 149.7 & 0.3 & \citet{Wang2021}\\
PDS~70    & WiPc & 02/08/2017 & 28.1 & 0.1 & 283.8 & 0.2 & \citet{Wang2021}\\
PDS~70    & WiPb & 05/02/2018 & 22.6 & 0.1 & 146.8 & 0.3  & \citet{Wang2021}\\
PDS~70    & WiPc & 05/02/2018 & 27.7 & 0.1 & 281.2 & 0.1 & \citet{Wang2021}\\
PDS~70    & WiPb & 05/03/2018 & 22.6 & 0.1 & 146.7 & 0.3 & \citet{Wang2021}\\
PDS~70    & WiPc & 05/03/2018 & 27.7 & 0.1 & 281.2 & 0.1 & \citet{Wang2021}\\
LkCa~15   & S14b & 11/16/2014 & 11.7 & 1.0 & 256   & 3   & \citet{Sallum2015}\\
LkCa~15   & S14c & 11/16/2014 & 10.1 & 1.5 & 318   & 11   & \citet{Sallum2015}\\
LkCa~15   & S14d & 11/16/2014 & 10.9 & 8.8 & 14    & 30  & \citet{Sallum2015}\\
LkCa~15   & S16b & 11/18/2016 & 12.2 & 0.8 & 248   & 2   & \citet{Sallum2016}\\
LkCa~15   & S16c & 11/18/2016 & 10.9 & 0.5 & 301   & 2   & \citet{Sallum2016}\\
LkCa~15   & S16d & 11/18/2016 & 10.3 & 2.0 & 348   & 5   & \citet{Sallum2016}\\
\hline
\multicolumn{8}{c}{\textit{Other Candidate Detections and Predictions from the Literature}}\\
\hline
PDS~70  & Z20d  & Feb-July 2020    & 13.8  & NR & 310  & NR & \citet{Zhou2021}\\
HD~169142  & O07 & June 2007  & 14.6  & 2.5 & 250.0 & 5.0      & \citet{Okamoto2017}\\
HD~169142  & O12-13 & 2012-2013  & 42.8  & NR  & 175.0 & NR      & \citet{Osorio2014}\\
HD~169142  & R13  & June 2013 & 19.6  & 4.0 & 7.4   & 11.3     & \citet{Reggiani2014}\\
HD~169142  & B13  & July 2013    & 13.8  & 3.8 & 0.0   & 14.0 & 
\citet{Biller2014}\\
HD~169142  & B14  & April 2014 & 22.6  & NR  & 33.0  & NR & \citet{Biller2014}\\
HD~169142  & G15-17A & 2015-2018 & 14.5  & 1.9 & 239.0 & 11.5    & \citet{Gratton2019}\\
HD~169142  & G15-18B & 2015-2018 & 23.8  & 1.0 & 17.0  & 8.0     & \citet{Gratton2019}\\
HD~169142  & G15-18C & 2015-2018 & 24.8  & 1.1 & 308.0 & 9.0     & \citet{Gratton2019}\\
HD~169142  & G15-18D  & 2015-2018 & 39.9  & 0.9 & 39.0  & 5.0    & \citet{Gratton2019}\\
HD~169142  & B18 & July 15, 2018  & 13.3  & 4.4  & 55.5 & 4.0      & \citet{Bertrang2020}\\
HD~100546  & Q11 & May 2011  & 59.0  & 1.5 & 7.0   & 1.4      & \citet{Quanz2013}\\
HD~100546  & C15  & Jan 2015  & 16.5  & 1.1 & 150.9 & 2.0     & \citet{Currie2015}\\
HD~100546  & S15-16  & 2015-2016  & 57.2  & 0.9 & 11.5 & 1.1    & \citet{Sissa2018}\\
HD~100546  & F15-16  & 2015-2016 & 121.3  & NR & 10 & NR    & \citet{Fedele2021}\\
SAO~206462 & C16 & March 2016 & 8.9   & 0.6 & 19.0  & 3.0     & \citet{Cugno2019}\\
SAO~206462 & C19 & July 13, 2019 & 53.6   & 0.2 & 212.4  & 0.7     & \citet{Casassus2021}\\
TW~Hya     & I16  & Dec 2016 & 105.7 & 8.5 & 242.5 & 2.1       & \citet{Ilee2022}\\
TW~Hya     & T17  & May 2017 & 108.9 & 0.1 & 237 & 1       & \citet{Tsukagoshi2019}\\
TW~Hya     & H19  & March 15, 2019 & 20.1 & 1.3 & 190 & 1  & \citet{Huelamo2022}\\
CS~Cha     & G17  & Feb-Jun 2017 & 165.6 & 0.6 & 261.4 & 0.2  & \citet{Ginski2018b}\\
HD~100453  & W17  & Feb 17, 2017  & 132.83 & 0.40 & 132.32 & 0.18    & \citet{Wagner2018b}\\
HD~100453  & G19  & April 7, 2019  & 135.1 & 4.0 & 132.7 & 0.8    & \citet{Gonzalez2020}\\
V1247Ori  & W12-13  & 2012-2013  & 5.2 & 0.8 & 305.5 & 5.5    & \citet{Willson2019}\\
\end{tabular}%
\tablecomments{Compilation of reported and predicted positions for objects with GAPlanetS detections (top) and nondetections (bottom). The ``Label" in column 2 corresponds to the text label at this candidates' location in Figures \ref{fig:hd142527diff}, \ref{fig:pds70cont}, \ref{fig:pds70ha}, \ref{fig:lkca15_inner}, \ref{fig:hd169142}, \ref{fig:widecomp_zoom}, and Appendis \ref{sec:allepochs} Figures \ref{fig:hd100546}, \ref{fig:sao206462}, \ref{fig:hd100453}, and \ref{fig:v1247ori}. Text labels correspond to the first letter of the last name of the study author and the epoch of \textit{observation} (not publication). In cases where the candidate was detected more than once, the full range of dates at which it was recovered is indicated. In these cases, the ``Sep Error" and ``PA Error" columns indicate the full range of possible positions due to both apparent orbital motion and astrometric uncertainty reported in the original reference, and the ``sep" and ``PA" columns are the central value for each of these ranges. An ``NR" designation indicates that the uncertainty value was not reported in the publication.}
\label{tab:objects}
\end{table*}

\subsection{Objects with Previously Reported GAPlanetS Detections\label{sec:detect}}

GAPlanetS data have already revealed three low-mass accreting companions and companion candidates: HD~142527~B, LkCa~15~b, and PDS~70~b, reported in \citet{Close:2013,Sallum2015,Wagner2018}, respectively. In this section, we present uniform reprocessings of the data for each of these targets under the GAPlanetS campaign framework, as well as contrast curves for all epochs (See Figure \ref{fig:detectctrst}). 
\begin{figure*}
    \centering
    \includegraphics[width=0.75\textwidth]{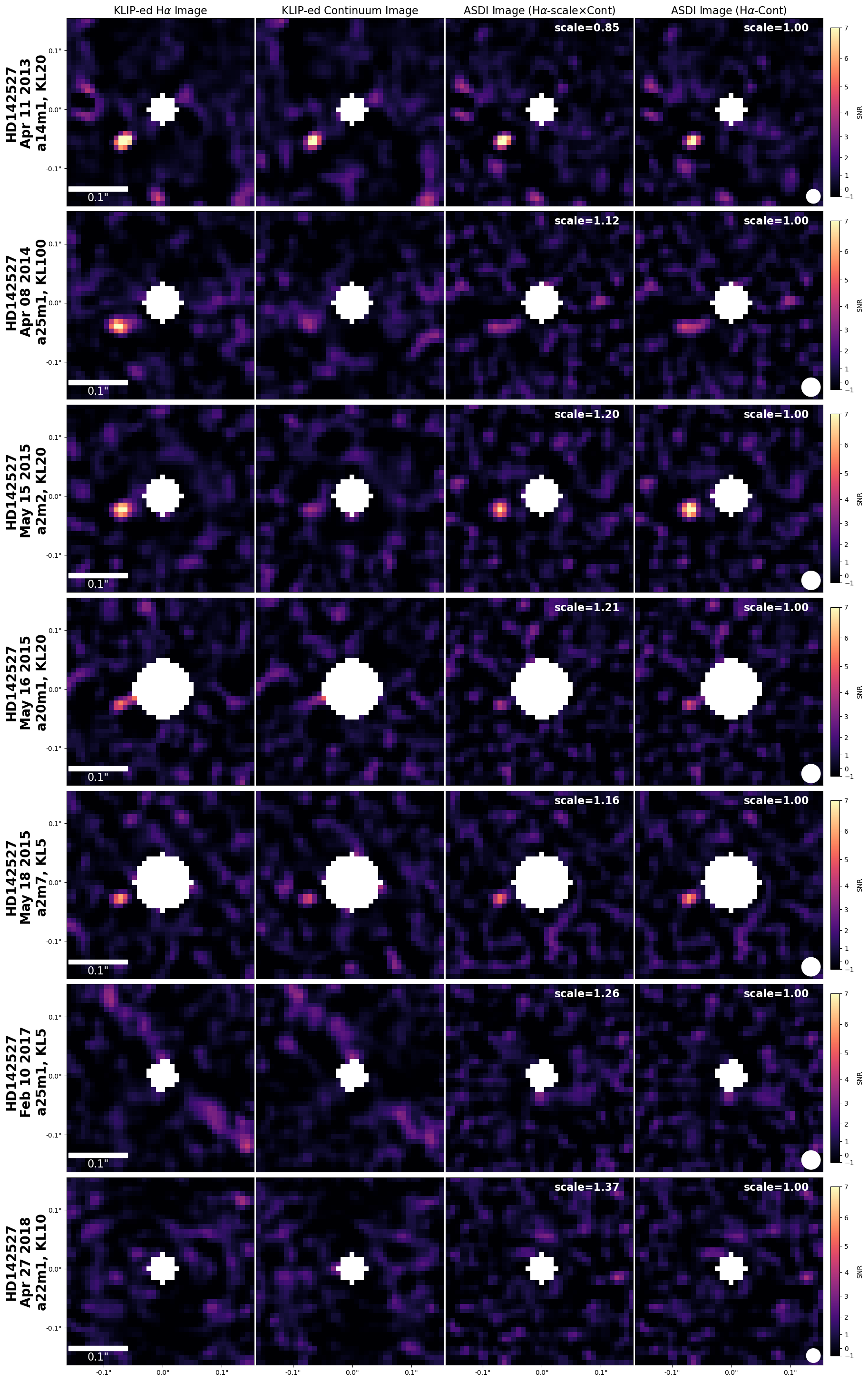}
    \caption{KLIP reductions of the H-alpha (left) and continuum (middle left) images for all HD~142527 epochs. KLIP parameters were optimized with pyKLIP-PE based on the recovered signal of false planets injected into the continuum images. The middle right panel shows a conservative SDI reduction created by multiplying the continuum KLIP image by the median H$\alpha$/continuum scale factor for the primary star (reported in table \ref{tab:data}) and subtracting it from H$\alpha$ imagery, which should effectively remove scattered light emission and continuum artifacts. The rightmost panel shows the unscaled H$\alpha$~--~continuum reduction, which is most appropriate in regions where no scattered light artifacts are present. The lack of a resolved inner disk component in HD~142527 makes scattered light from circumstellar material of minimal concern; however, there may be some contribution to these signals from a circumsecondary disk.}
    \label{fig:hd142527ims}
\end{figure*}
\begin{figure*}[htp]
\begin{centering}
{\includegraphics[clip,width=\textwidth]{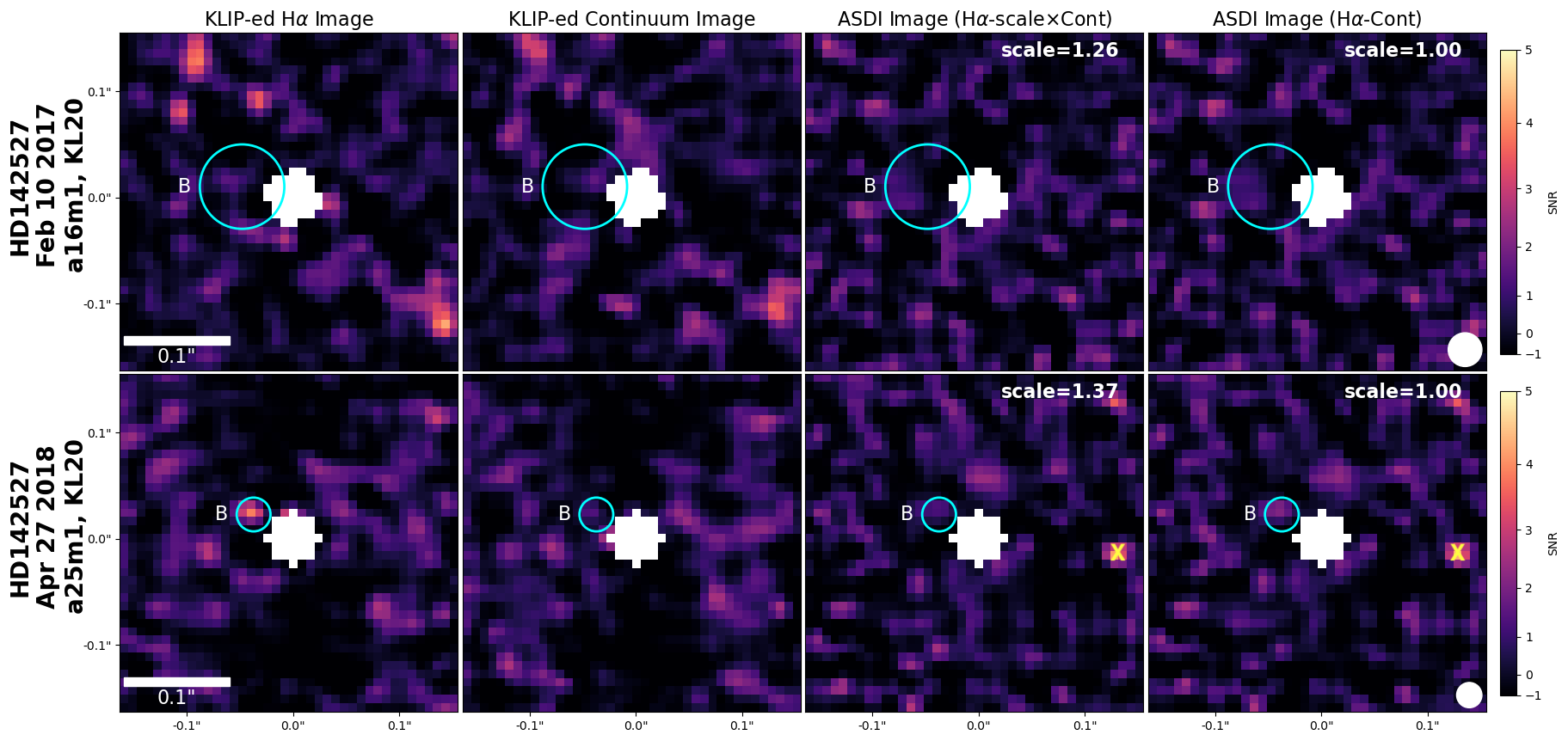}}
\caption{The same as Figure \ref{fig:hd142527ims}, except for the most difficult HD~142527~B detection epochs. Here, KLIP parameters have been optimized on the known location of the companion in the H$\alpha$ images \citep[cyan circle labeled "B";][]{Balmer2022}, resulting in a recovery of the companion in the 2018 data, but a nonrecovery in 2017. The yellow ``x" marks a strong negative continuum speckle.}
\label{fig:hd142527diff}
\end{centering}
\end{figure*}

\subsubsection{HD~142527}

The HD~142527 GAPlanetS datasets served as an excellent resource for optimizing and testing registration, centering, flat-fielding, dataset selection, and parameter optimization techniques for the GAPlanetS pipeline. The presence of the robust, high-SNR companion HD~142527~B allowed us to gauge the relative efficacy of various algorithmic choices on the recovery of the real companion at tight separation.

GAPlanetS data were collected for HD~142527 in 2013, 2014, 2015, 2017, and 2018.  We include continuum-optimized imaging and contrast results for all epochs in Figure \ref{fig:hd142527ims}, and H$\alpha$ optimized reductions for the most difficult detections in Figure \ref{fig:hd142527diff}. 

Photometric and astrometric monitoring and orbit fitting of HD~142527~B are the subject of a companion paper \citep{Balmer2022} and are not discussed in detail here. Importantly, \citet{Balmer2022} refined the orbit of the companion and demonstrated that it is significantly misaligned ($\theta>30^\circ$) with respect to both the inner and outer disk components. We touch briefly on consistency between our astrometric and photometric fits and those of \citet{Balmer2022} in Section \ref{sec:survey}. 

Figure \ref{fig:hd142527ims} demonstrates that HD~142527~B is easily recovered in H$\alpha$, continuum, and SDI imagery in datasets from 2013 (SNR=10.5/7.5/5.6 for H$\alpha$, continuum, and 1:1 scaled SDI imagery, respectively), 2014 (SNR=6.8/3.6/4.6), May 15 2015 (SNR=9.0/3.7/8.0), May 16, 2015 (SNR=5.7/3.5/4.1) and May 18 2015 (SNR=7.0/5.2/4.5) following the bulk survey strategy of optimizing on false planets injected into continuum images. 

The 2017 and 2018 HD~142527 datasets are substantially more difficult detections than the 2013--2015 epochs, as the predicted separation of HD~142527~B is much tighter (0$\farcs$05, $\sim$6pix). We attempted both continuum-optimized and H$\alpha$ companion-optimized reductions of these datasets. The companion is recovered in 2018 H$\alpha$ data at an SNR of 3.9 in the direct H$\alpha$ optimized reduction, but is not recovered in 2017 under either optimization method. The reason for this is readily visible in the contrast curves of Figure \ref{fig:detectctrst}, which reveal that even the most highly optimized \texttt{pyKLIP} reduction of the 2017 dataset boasts nearly an order-of-magnitude poorer contrast at the location of the companion than the next-lowest-quality dataset. This is likely a result of both the extremely tight separation of the companion at this epoch and the small amount of on-sky rotation obtained (16.1$^{\circ}$). More detail on the nature of the 2018 recovery is provided in \citet{Balmer2022}.

The source of the $\sim$ 140au \citep{Avenhaus2014} gap in HD~142527 has been debated, in particular whether the binary companion HD~142527~B can be solely responsible for carving the wide central cavity \citep{Fukagawa2006, Biller:2012, Casassus2015, Price2018}. In 2014 and 2015, we conducted deep imaging of the system to search for outer companions in the wide disk gap, allowing the detector to saturate out to near or beyond the companion's location. No additional candidates were found in either epoch. 

\begin{figure*}[h!]
\begin{centering}
\includegraphics[clip,width=\textwidth]{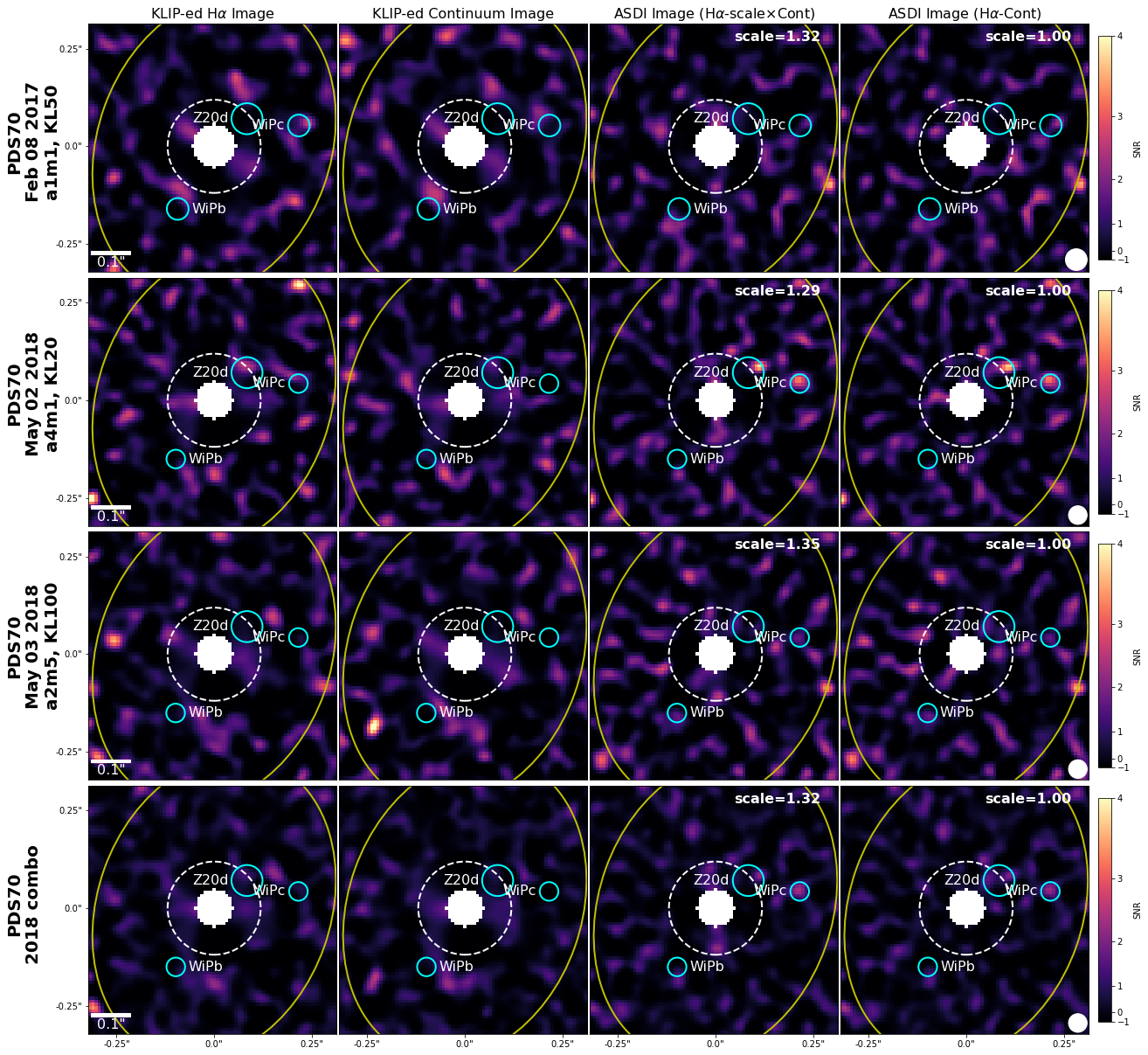}%
\caption{Post-processed GAPlanetS images for all PDS~70 epochs. KLIP parameters were optimized for false planets injected into the continuum channel, resulting in marginal (SNR$\sim$3) recoveries of PDS~70~c in 2017 and on 2018 May 2, but a nonrecovery of PDS~70~b in all epochs. The cyan ``WiPb" and ``WiPc" circles mark the predicted location of the planet at the precise epoch of our observations, computed using \citet{WiP2021}. The cyan circle labeled ``Z20d" marks the location of a third point-source candidate from \citet{Zurlo2020}. The yellow ellipse marks the inner edge of the known scattered light cavity in this system.}
\label{fig:pds70cont}
\end{centering}
\end{figure*}
\begin{figure*}
\begin{centering}
\includegraphics[clip,width=\textwidth]{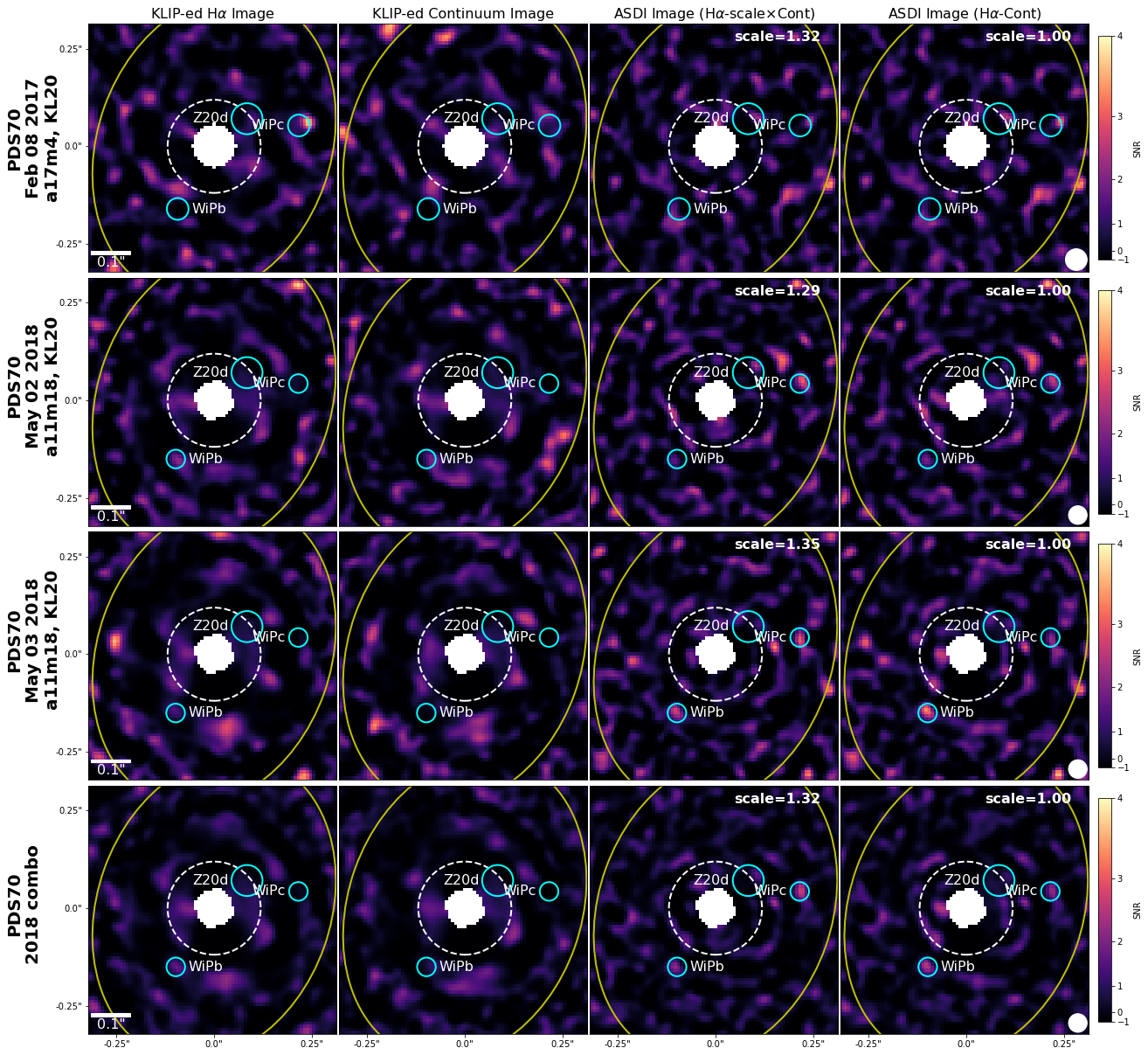}%
\caption{Post-processed GAPlanetS images for all PDS~70 epochs. KLIP parameters were optimized to maximize the average SNR metric at the locations of the PDS~70~b and c planets in H$\alpha$ (left-hand panel). Although SNRs are low at H$\alpha$, the parameters selected result in recovery of PDS~70~c in SDI imagery in all three epochs and recovery of PDS~70~b in the 2018 May 3 epoch. The cyan circle labeled ``Z20d" marks the location of a third point-source candidate in this system from \citet{Zurlo2020}. The cyan ``WiPb" and ``WiPc" circles mark the predicted location of the planet at the precise epoch of our observations, computed from \citet{WiP2021}. The yellow ellipse marks the inner edge of the known scattered light cavity in this system.}
\label{fig:pds70ha}
\end{centering}
\end{figure*}

\subsubsection{PDS~70}

We observed PDS~70 as part of the GAPlanetS campaign in 2017 and 2018. H$\alpha$ emission from PDS~70~b from data taken in 2018 on two consecutive nights was reported in \citet{Wagner2018}, establishing PDS~70~b as an accreting protoplanet. Frame selection, highpass filter, and KLIP parameters were tuned aggressively to allow for robust recovery of the companion at $\sim4\sigma$ in the combination of post-processed SDI imagery from the two 2018 nights. H$\alpha$ line emission was subsequently resolved in both PDS~70~b and c by \citet{Haffert2019} with the Very Large Telescope (VLT) Multi Unit Spectroscopic Explorer (MUSE) instrument, and in ultraviolet accretion continuum emission with the Hubble Space Telescope (HST) by \citet{Zhou2021}, lending additional credence to the original detection. 

Figure \ref{fig:pds70cont} shows the continuum false planet optimized reductions for all three PDS~70 epochs, as well as the combination of the two 2018 nights. Overplotted on these images are the predicted locations of the two known companions at the epoch of observation, derived from the orbital fits of \citet{Wang2021}. Contrast curves for all PDS~70 epochs are shown in Figure \ref{fig:detectctrst}. By optimizing on false continuum planets injected at a range of separations between the IWA and control radius of each dataset, the GAPlanetS reduction framework is intentionally conservative. Both PDS~70 planets lie considerably outside the control radius; thus, the lack of recovery under the standard pipeline is unsurprising. While neither companions is robustly recovered, a $\sim$3$\sigma$ excess signal appears within 1 FWHM of the predicted location of PDS~70~c in both H$\alpha$ and ASDI images for the 2017 epoch and in ASDI images for the 2018 May 2 epoch. 

Both known planets are recovered with more targeted optimization (see Figure \ref{fig:pds70ha}). As the PDS~70~b and c companions exist at known locations and their \textit{bona fide} planetary nature is well established, we follow the same procedure as for the more difficult HD~142527~B recoveries in order to recover their signals where possible. Namely, we optimize directly on the average SNR metric at the known planet location(s) in the H$\alpha$ images. We choose to average this SNR metric across both known planet locations in order to achieve a relatively conservative approach to this direct H$\alpha$ optimization. 

We note that this optimization is done using the H$\alpha$ images only and not SDI images. Because of the prevalence of false point-sources induced during subtraction of KLIP-ed continuum imagery, we believe that optimizing on the H$\alpha$ images is a more robust approach. 

Reductions tuned to maximize the average SNR metric across both planets in 20 KL mode H$\alpha$ post-processed imagery are shown in Figure \ref{fig:pds70ha}, with their optimized KLIP parameters given in the labels on the left-hand side of each subpanel. The companions are only readily visible in SDI imagery using this methodology, but both are marginally recovered. A signal consistent with PDS~70~c is present in all three SDI epochs, with classically computed SNRs of 3.8/2.6/2.9 and 3.1/3.0/3.4 in 1:1 scaled and conservatively scaled SDI images, respectively. It is also marginally recovered in H$\alpha$ at an SNR of 3.7 in the 2017 epoch. PDS~70~b is recovered only in the 2018 May 3 epoch at an SNR of 3.3 in 1:1 scaled SDI imagery and 2.8 in conservatively scaled SDI imagery.

We extract detailed astrometry and photometry only from the 2017 H$\alpha$ detection of PDS~70~c, as forward model fitting of SDI imagery is more complex and beyond the scope of this work. We note, however, that the positions of the marginal detections of PDS~70~b and PDS~70~c in 2018 SDI imagery are entirely consistent with their \texttt{whereistheplanet} predicted orbital positions (see Figure \ref{fig:pds70ha}, which incorporates the astrometry of \citet{Wagner2018} and other works. 

Our 2017 detection of PDS~70~c extends the time baseline of its astrometry by 9 months. The companion is detected at a separation of 246.9$\pm$4.4mas and a PA of 284.2$\pm$0.6$^{\circ}$. This is inconsistent at the $\sim$5$\sigma$ level with the  \texttt{whereistheplanet} \citep{WiP2021} prediction for the separation of the planet at this epoch (223.2$\pm$0.9mas), though the PA prediction (283.8$\pm$0.2$^{\circ}$) is consistent. Our best-fit photometry suggests a $\Delta$mag of 5.5$\pm$0.2 relative to the stellar continuum, brighter by nearly 2 magnitudes than the 7.7$\pm$0.2 reported from VLT MUSE observations taken a year later by \citet{Haffert2019}. Possible sources of these discrepancies are discussed in Section \ref{sec:fms}.
\begin{figure*}
    \centering
    \includegraphics[width=\textwidth]{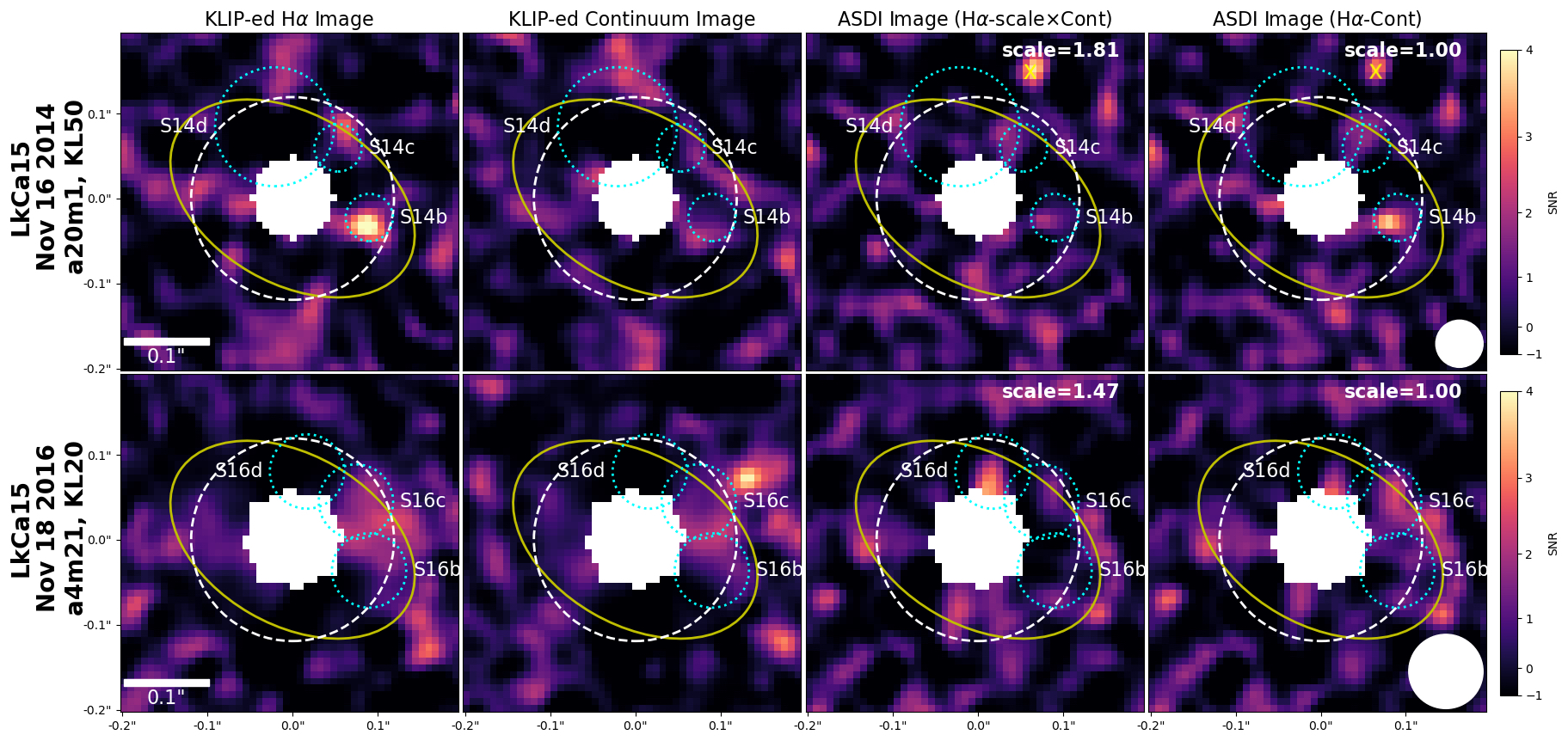}
    \caption{KLIP reductions of both LkCa~15 epochs, with parameters optimized by equally weighting all six post-processed image quality metrics across 5 and 20 KL modes for the single false planet injected into the continuum images inside of the control radius (at separation=0$\farcs$1, PA=0). The locations of candidate companions from the literature are indicated with cyan circles (``S14b", ``S14c", ``S14d":\citet{Sallum2015}; ``S16b", ``S16c", ``S16d":\citet{Sallum2016}. The yellow ``x" marks a strong negative speckle just outside the control radius in the continuum images, which appears bright in SDI imagery.}
    \label{fig:lkca15_inner}
\end{figure*}
\subsubsection{LkCa~15~b}

LkCa~15 was observed three times as part of the GAPlanetS campaign, in 2014, 2015, and 2016. The 2014 epoch is the original H$\alpha$ discovery epoch for LkCa~15~b, as well as the highest-contrast epoch (see Figure \ref{fig:lkca15_inner}). The 2015 data were of very poor quality and were discarded before KLIP optimization due to a failure to recover injected planets at a contrast of 10$^{-1}$. The 2016 dataset is of intermediate quality, though substantially poorer in contrast than the 2014 epoch, and the companion candidate is not recovered (see Figure \ref{fig:detectctrst}).

The existence of multiple protoplanet candidates in close proximity to the inner disk rim of LkCa~15 has been the source of some controversy. Recent work by \citet{Currie2019} resolved the inner disk component previously imaged by \citet{Thalmann2016} in polarized intensity, this time in NIR total intensity light with SCExAO on Subaru and in the thermal infrared with Keck/NIRC2. The authors interpret the smooth resolved inner disk rim as inconsistent with the existence of multiple protoplanets at similar separation; however, their results cannot explain observed orbital motion in the sparse aperture masking detections \citep{Sallum2016}, nor is an inner disk artifact consistent with the presence of H$\alpha$ excess emission in the companion. 

\citet{Mendigutia2018} attempted spectroastrometric detection of H$\alpha$ emission from the LkCa~15~b planet candidate and reported symmetric extended H$\alpha$ emission. However, their observation relied on long-slit spectroscopy that does not appear to have been well aligned with the predicted position of LkCa~15~b at that epoch 

Even if the companion was in the slit, the reported detection threshold was a contrast of 5.5 mag at H$\alpha$, close enough to the $\Delta$mag of 5.2$\pm$0.3 reported in \citet{Sallum2015} that even a small decrease in luminosity relative to the original 2014 November H$\alpha$ detection epoch (as might be expected if accretion onto the companion is stochastic) would render the planet undetectable. 

As the nature of the point-source candidates in the LkCa~15 disk has been debated and their proximity to the resolved inner disk rim firmly established, we adopt the most conservative approach in this work -- optimizing on false continuum planets only. We also overplot an ellipse in Figure \ref{fig:lkca15_inner} at the edge of the imaged inner disk rim. While the LkCa~15~``c" and ``d" candidates are coincident with the inner disk rim, the ``b" candidate lies significantly inside of it and is less likely to be a scattered light artifact, though we note that disk signal at similar separation to planet candidates can influence extracted KL modes \citep{Lawson2022} and further vetting of this candidate is warranted. 

Because of the tight separations of the companion candidates, the comparatively large PSF (FWHM$\sim$6pixels, 50mas), and the limited space available inside of the AO control radius for bin 2 data ($r$=15 pixels), we injected only one false planet inside the control radius (at a separation of 12 pixels/0$\farcs$10 and a PA of 0$^{\circ}$) to optimize these data.

Using our conservative optimization methodology, we recover the 2014 H$\alpha$ excess signal first reported in \citet{Sallum2015} with a classically computed SNR of 5.0 in the H$\alpha$ images, 4.9 in the 1:1 scaled H$\alpha$~--~continuum SDI reduction, and 2.9 in the conservatively scaled SDI reduction. 

Best-fit BKA astrometry from the 2014 H$\alpha$ epoch produces strong evidence ($Z_1/Z_0$=18) for a point-source with a separation of 69.7$\pm$6.1 mas, a PA of 242.6$\pm$2.7$^{\circ}$ and a $\Delta$mag of 3.1$\pm$0.3. This is inconsistent at the $\sim$1--2$\sigma$ level with the astrometry and photometry derived from the same data in  \citet{Sallum2015}, which placed the planet at a slightly wider separation (93$\pm$8mas), higher PA (256$\pm$3$^{\circ}$), and $\sim$1 magnitude fainter ($\Delta$Mag=5.2$\pm$0.3). This discrepancy is discussed in greater detail in Section \ref{sec:fms}.

While we recover a clear H$\alpha$ signal in the 2014 epoch at the location of the previously reported LkCa~15~``b" planet candidate that is well fit by a point-source forward model, its nature remains ambiguous. The ``b" candidate lies inside of the known scattered light inner disk rim, suggesting that it is not part of that structure. Although there is no apparent, comparable, point-source in the continuum images, scaling them up by the stellar H$\alpha$-to-continuum ratio and subtracting it from H$\alpha$ suppresses the signal heavily, meaning that it cannot be ruled out as a scattered light source. At the same time, a compact scattered light structure located inside the disk cavity would itself be a notable result. 

The optimized 2016 SDI reductions appear to show a low-SNR ($\sim$2.5) arc of excess emission to the west of the star consistent with the known inner disk rim. The predicted location of LkCa~15~``b" lies inside of this rim, but it is not recovered. The contrast of these data are a factor of 5 or more worse than the 2014 epoch at all separations, making the lack of recovery of the candidate unsurprising. The high \texttt{pyKLIP movement} value converged upon by the continuum false planet optimization equates to a small reference library for PSF subtraction with a large degree of rotation between reference and target images. High rotational mask reductions are often used to resolve disk structures in ADI imagery, and this is likely the reason why the apparent disk rim is seen in 2016 but not 2014. Its appearance in SDI imagery is, however, surprising and could suggest an additional H$\alpha$ emission source (i.e. the LkCa~15~b protoplanet) located to the west of the star, adding to the light being scattered by dust grains at this location.

There is also a $\sim$3$\sigma$ excess signal in the 2016 images at the predicted location of the ``d" protoplanet candidate; however, there is a comparable signal to the south opposite this feature, which is suggestive of a wavefront error (phase) induced speckle. 

\begin{figure}
    \centering
    \includegraphics[width=\columnwidth]{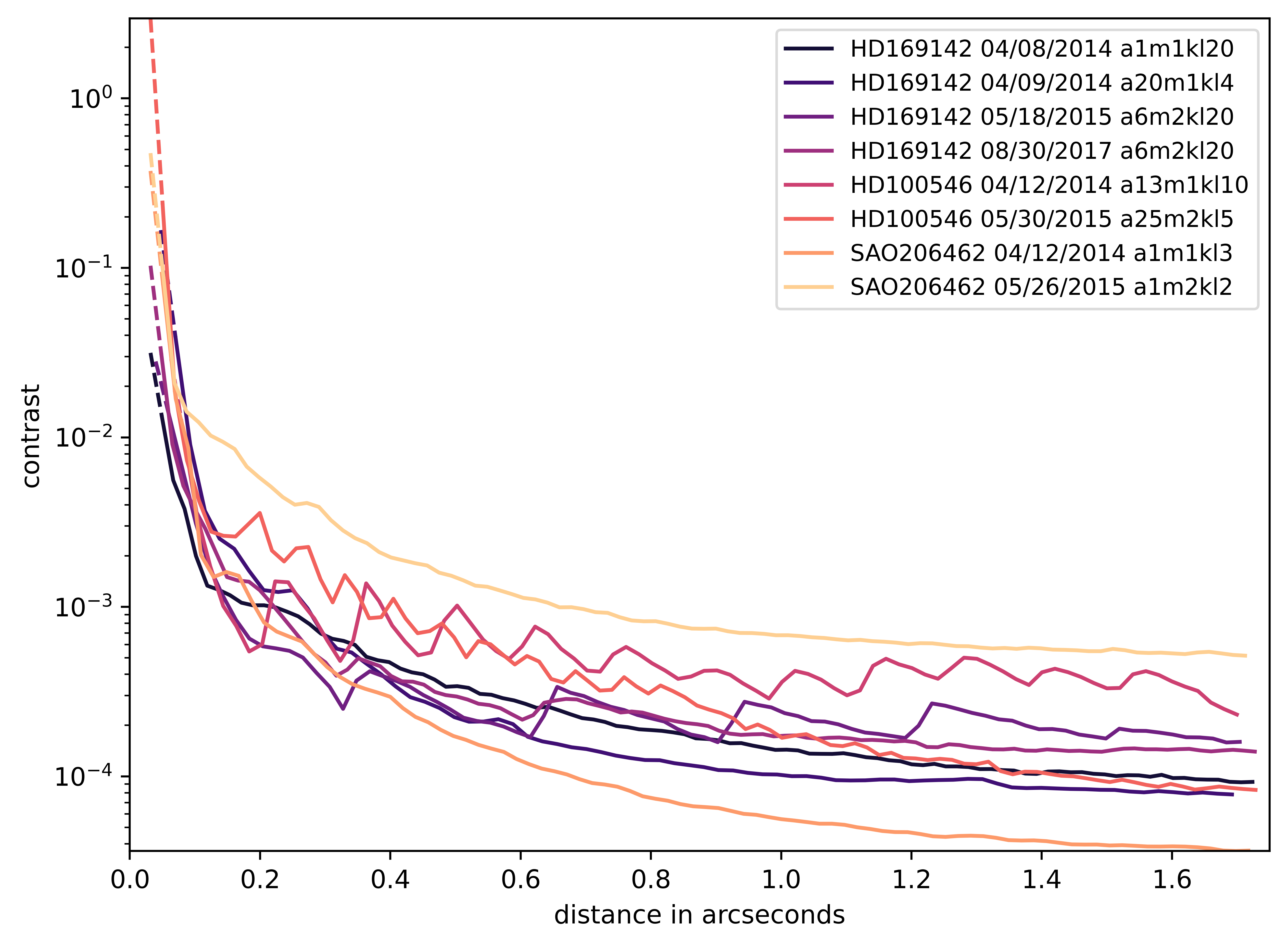}
    \caption{Throughput-corrected 5$\sigma$ contrast curves for continuum false planet optimized \texttt{pyKLIP} reductions of the three GAPlanetS targets with reported close ($<0\farcs5$) companion candidates in the literature:  HD~100546(two epochs), HD~169142 (four epochs) and SAO~206462 (two epochs). Throughput was computed as described in the text, with a correction for the small number of independent noise samples near the star following \citet{Mawet2014}. Solid curves indicate regions where throughput-corrected contrast was computed directly. The curves are also projected inward from the innermost throughput measurement to the inner working angle, and this extrapolated region is indicated with a dashed line.}
    \label{fig:closecandctrst}
\end{figure}

In summary, we reproduce here, with a systematic and robust pipeline designed to minimize false positives, the original LkCa~15 ``b" H$\alpha$ detection reported in \citet{Sallum2015}. We do not, however, recover the LkCa~15 ``c" or ``d" protoplanetary candidates, though this does not rule them out as protoplanets, as our achieved contrasts are modest at best. 

\subsection{Objects with New Protoplanet Candidates \label{sec:newcand}}

\subsubsection{CS~Cha}
\begin{figure*}
    \centering
    \includegraphics[width=\textwidth]{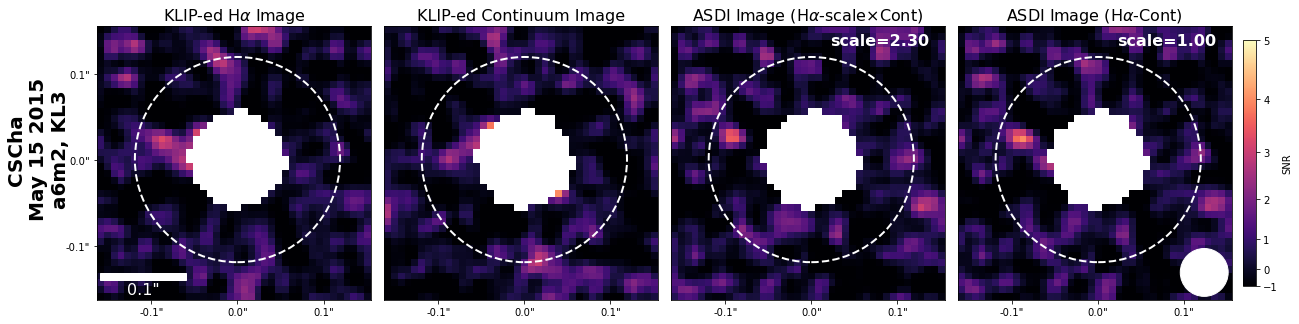}
    \caption{Final KLIPed H$\alpha$ (left), Continuum (middle left), stellar H$\alpha$/continuum-scaled ASDI (middle right),and 1:1 scaled ASDI (right) imagery for the single GAPlanetS CS~cha epoch. The new candidate companion, CS~Cha ``c" is apparent to the east of the star in H$\alpha$ and SDI imagery.}
    \label{fig:cscha}
\end{figure*}

One epoch of GAPlanetS data was obtained for the CS~Cha~AaAb spectroscopic binary \citep[unresolved, sep$<$44mas;][]{Kurtovic2022} in 2015. Although we do not see evidence of the wide polarized companion CS~Cha~B in post-processed images of this epoch (see Section \ref{sec:candidates}), we do find tentative evidence of a much more tightly separated point-source candidate with a classically computed SNR of 6.6 at H$\alpha$ (\texttt{PlanetEvidence} SNR=4.3) located at a separation of 68mas and a PA of 76$^{\circ}$, as seen in Figure \ref{fig:cscha}. 

The observed separation of this candidate is roughly twice the value of the spectroscopic binary CS~Cha~A's maximum predicted projected separation, suggesting that the imaged companion is not the other member of the binary. This detection places the candidate companion firmly within the disk's sub-mm continuum  \citep[$<$210mas;][]{Francis2020} and scattered light \citep[$<$92.5 - 337 mas;][]{Ginski2018b} cavities and interior to the probable inner edge of the gas cavity \citep[CO temperature peak at 128mas;][]{Kurtovic2022}. Simulations in \citet{Kurtovic2022} predict a Saturn-mass planet near the inner edge of the gas cavity, roughly consistent with the separation of the detected candidate. We note that their mass estimate is dependent on a viscosity assumption, and the presence of H$\alpha$ emission may be suggestive of a more massive companion.

Its moderate contrast of 0.05 (3.2mag) relative to the (spectroscopic binary) primary suggests that the CS Cha ``c" candidate may be massive or accreting at a very high rate. When corrected to a contrast relative to the stellar \textit{continuum} (rather than the actively accreting primary), the contrast is even more moderate -- 2.3mag. However, lack of detection in continuum imagery is more suggestive of a planetary nature. Due to these ambiguities and the lack of a second epoch, we refer to this detection as a protoplanet ``candidate" throughout the remainder of this work.

\subsection{Objects with Known Companions or Companion Candidates in the Literature \label{sec:candidates}}

Six of the remaining transitional disks in the GAPlanetS sample have previous reports of planetary, brown dwarf, or stellar-mass companions from NIR high-contrast imagery, a sub-mm point-source interpreted as evidence of a circum\textit{secondary} disk, or a so-called ALMA ``velocity-kink" indicative of possible inflow onto an embedded planet \citep[e.g.,][]{Rabago2021}. In each case, the reported location of the planet candidate(s) at the literature detection epoch is given in Table \ref{tab:objects} and shown in all KLIPed images of the system. True companions will in most cases have undergone some small amount of orbital motion since this original detection epoch.  

\begin{figure*}
    \centering
    \includegraphics[width=\textwidth]{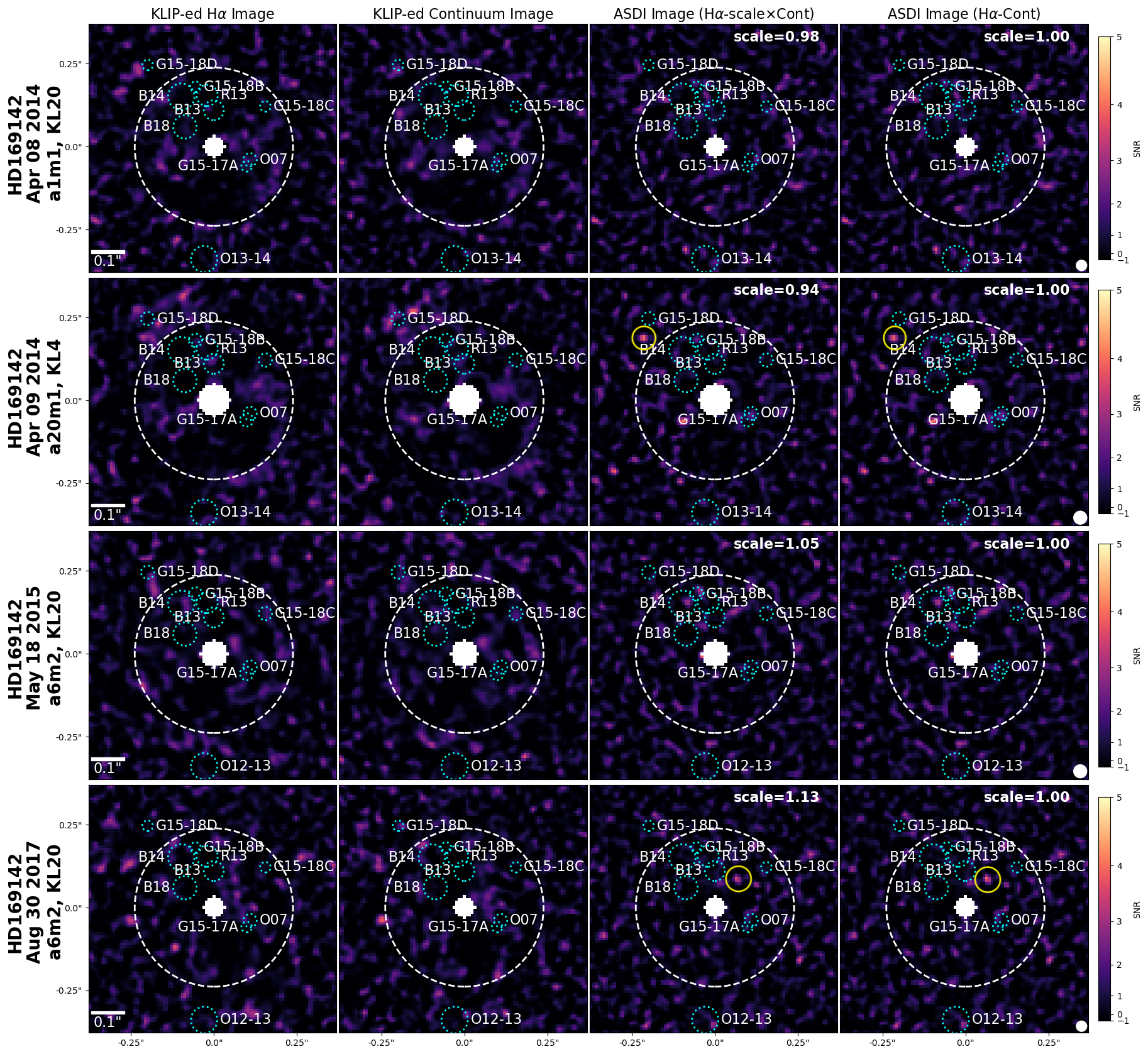}
    \caption{KLIPed H$\alpha$ (left), continuum (middle left), stellar H$\alpha$/continuum-scaled ASDI (middle right),and 1:1 scaled ASDI (right) imagery for all HD~169142 epochs. \texttt{pyKLIP} reduction parameters (indicated in the text labels to the left of each image panel) have been optimized for recovery of false continuum planets injected between the IWA and control radius, as described in detail in the text. The AO control radius of the images is indicated with a white dashed circle. The locations of candidate companions from previous work are indicated with dashed cyan circles at the literature detection epoch listed in Table \ref{tab:objects} (``O07":\citet{Okamoto2017}; ``O12-13":\citet{Osorio2014}; ``R13":\citet{Reggiani2014}; ``B13", ``B14":\citet{Biller2014}; ``G15-17A", ``G15-18B", ``G15-18C", ``G15-18D":\citet{Gratton2019}; ``B18":\citet{Bertrang2020}). Yellow ``x" symbols indicate locations where apparent companions are introduced into the SDI imagery through subtraction of a negative continuum speckle. The most compelling 3$\sigma$-4$\sigma$ point-source candidates are indicated with yellow circles, though there are reasons to be skeptical of each, as detailed in the text.}
    \label{fig:hd169142}
\end{figure*}

Three of the companions - HD~100546, HD~169142, and SAO~206462 - have previously reported planet candidates at low to moderate separation ($<0\farcs5$) from the central star. H$\alpha$, continuum, and ASDI images for all epochs of HD~169142 are shown in Figure \ref{fig:hd169142} in order to demonstrate the difficulty of assessing protoplanet candidates in highly morphologically complex systems, while the other two objects' reductions are shown in Appendix \ref{sec:allepochs}. The contrast curves for all HD~100546, HD~169142, and SAO~206462 epochs are shown in Figure \ref{fig:closecandctrst}. 

\begin{figure}[h!]
    \centering
    \includegraphics[width=\columnwidth]{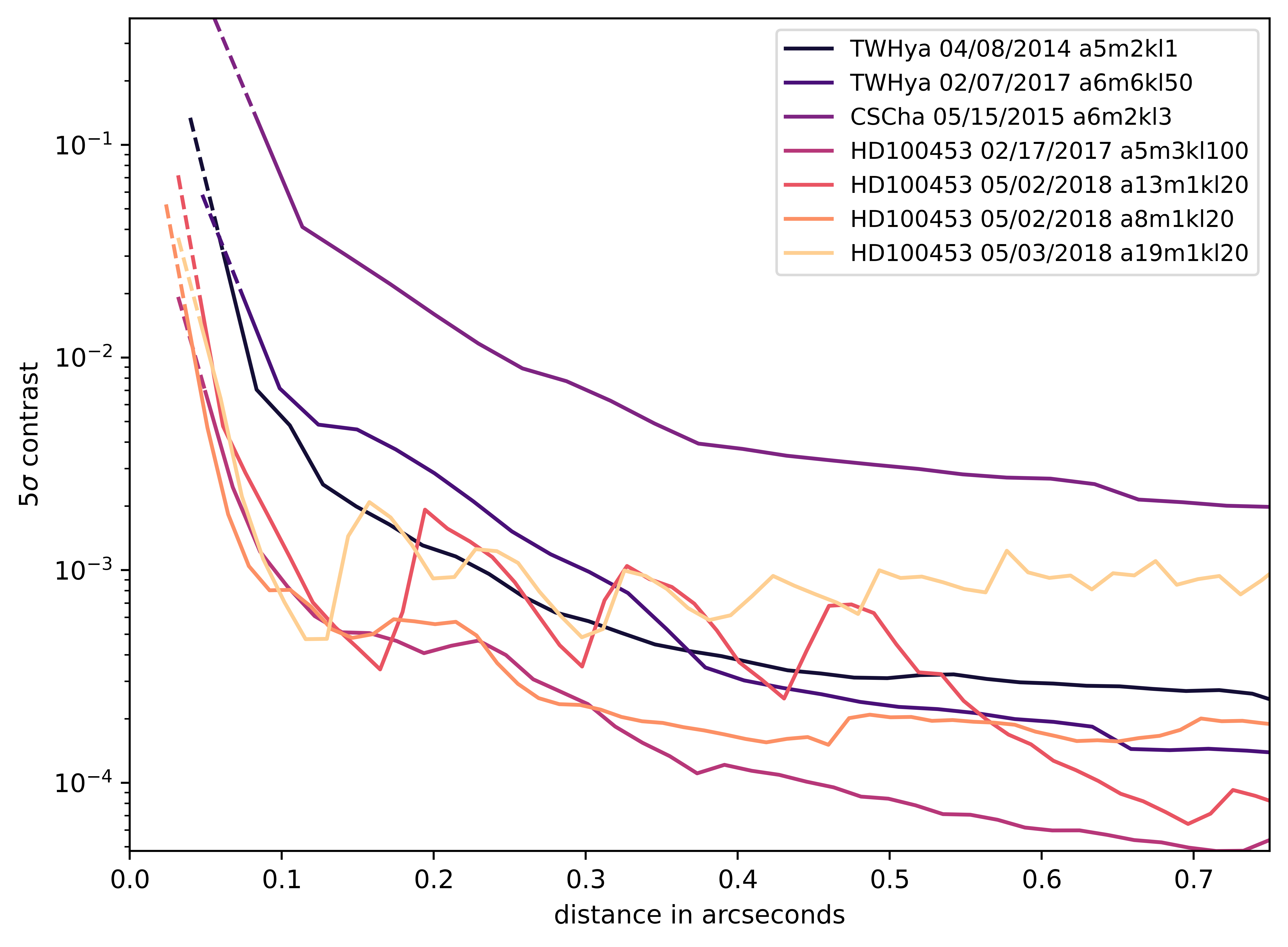}
    \caption{Throughput-corrected 5$\sigma$ contrast curves for continuum false planet optimized \texttt{pyKLIP} reductions of all epochs for the three GAPlanetS targets with reported wide ($>0\farcs5$) companions or companion candidates in the literature:  TW~Hya (two epochs), CS~Cha (one epoch) and HD~100453 (three epochs). Throughput was computed as described in the text, with a correction for the small number of independent noise samples present near the star implemented following \citet{Mawet2014}. Solid curves indicate regions where throughput-corrected contrast was computed directly. The curves are also projected inward from the innermost throughput measurement to the inner working angle, and this extrapolation region is indicated with a dashed line.}
    \label{fig:widectrsts}
\end{figure}

The remaining three objects - TW~Hya, HD~100453, and CS~Cha - have more distant ($>$0$\farcs$5) known or candidate companions, of which only HD~100453~B is recovered in GAPlanetS imagery. \texttt{PyKLIP} ASDI reductions for the highest-contrast epochs of these objects are shown in Figure \ref{fig:widecomp_zoom}. Optimized imagery of the inner regions of these three systems is shown in Appendix \ref{sec:allepochs}. Contrast curves for all TW~Hya, HD~100453, and CS~Cha epochs are given in Figure \ref{fig:widectrsts}.

\subsubsection{HD~169142}

GAPlanetS data of HD~169142 were collected on four nights - two consecutive nights in 2014, one night in 2015, and one night in 2017. All four epochs are shown in Figure \ref{fig:hd169142}. The data do not show a consistent excess at or near the location of any of the planet candidates across epochs; however, there are marginal signals consistent with H$\alpha$ excess near the location of several candidates in single epochs. These signals do not rise to the level of candidates in our analysis because they are neither consistent across epochs nor have sufficiently high SNR. However, some may later prove to be true planetary signals in light of future observations at higher contrast. 

We include all epochs of HD~169142 in the main body of the text as a demonstration of the difficulty of candidate identification in morphologically complex systems under variable conditions (and, potentially, intrinsic variability in protoplanet candidates' H$\alpha$ emission). For example, one of the more apparently compelling candidates in the images is a $\sim$3.5$\sigma$ point-source just outside the control radius near the \citet{Gratton2019} ``D" candidate in the 2014 April 9 epoch (marked with a yellow circle in Figure \ref{fig:hd169142}). The lack of similar signal in the higher-contrast 2014 April 8 epoch just 1 day earlier, as well as its proximity to the AO control radius (immediately outside of which a bright ring of variable signal induced by the wavefront control loop appears in raw images), means that it does not rise to the level of a candidate in our analysis.  Another $\sim$4$\sigma$ excess source appears to the northwest of the star in the 2017 August 30 epoch at similar separation as the \citet{Reggiani2014} and \citet{Biller2014} candidates (also indicated with a yellow circle in Figure \ref{fig:hd169142}). The 3 year time baseline between the original candidate identification epochs and this observation may allow for this degree of orbital motion; however, there is not a compelling excess at the same location in the H$\alpha$ imagery, and similarly strong excess is not seen near this location in the other epochs, so this candidate is also marginal. We conclude that it and the other candidates at similar separation are most likely scattered light features from a clumpy inner disk ring at this separation, consistent with their lack of recovery in SDI imagery. 

\subsubsection{HD~100546}

The GAPlanetS data for HD~100546 were analyzed in detail in \citet{Follette2017} and \citet{Rameau2017}; however, we have taken advantage of improvements to the GAPlanetS pipeline since initial publication and reprocessed the data, allowing us to place more stringent limits on the H$\alpha$ luminosity of the HD~100546 ``b" and ``c" planet candidates, as given in Table \ref{tab:nondetections}. No H$\alpha$ excess signals were seen in the vicinity of either candidate in either GAPlanetS epoch (see Appendix \ref{sec:allepochs}, Figure \ref{fig:hd100546}).
 
\subsubsection{SAO~206462}

GAPlanetS data were collected for SAO~206462 (HD~135344~B) in 2014 and 2015. No H$\alpha$ excess signals were detected at or near the location of the \citet{Cugno2019} or \citet{Casassus2021} candidates in either epoch, and the best limit on the contrast at the location of these candidates is provided in Table \ref{tab:nondetections}. Images of both SAO~206462 epochs are shown in Appendix \ref{sec:allepochs}, Figure \ref{fig:sao206462}.

\subsubsection{TW~Hya}

\begin{figure*}
    \centering
 \includegraphics[clip,width=0.85\textwidth]{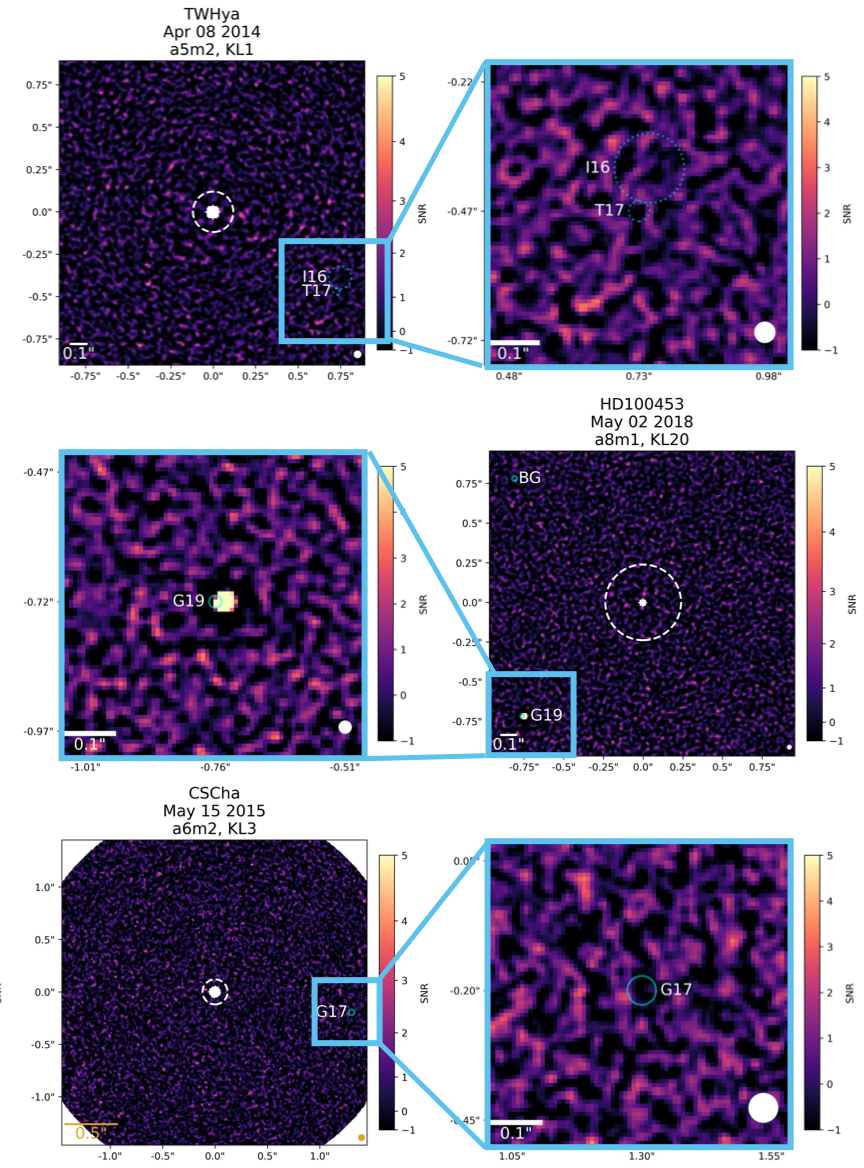}
    \caption{Wide-field and narrow-field views of the 1:1 scaled ASDI reduction from the highest-contrast epoch for all three GAPlanetS targets with known wide companions (CS~Cha~B, HD~100453~B) and companion candidates (TW~Hya). No clear H$\alpha$ excess signal is present at the location of the wide TW~Hya candidate(s) reported in \citet{Ilee2022} (``I16") or \citet{Tsukagoshi2019} (``T17"), nor is it apparent from the known, highly embedded CS~Cha~B companion \citep[``G17", ][]{Ginski2018b}. H$\alpha$ excess is, however, apparent from the HD~100453~B companion at a location consistent with its reported position in 2019, as reported in \citet{Gonzalez2020} (``G19").}
    \label{fig:widecomp_zoom}
\end{figure*}

GAPlanetS data of TW~Hya were collected in 2014 and 2017. The highest-contrast (2014) ASDI epoch is shown in a wide-angle view in Figure \ref{fig:widecomp_zoom}, together with a zoomed-in view of the region surrounding the location of the candidate previously reported by \citet{Tsukagoshi2019} and \citet{Ilee2022}. We note that TW~Hya is accreting at an enormously high rate compared to the other objects in the sample, and the primary star is 6-8 times brighter at H$\alpha$ than at the continuum in both epochs. This makes the difference between the 1:1 scaled SDI and the conservatively scaled SDI images striking and is an extreme example of the possible impact of incompletely removed stellar residuals on 1:1 scaled SDI images. At the same time, it is a clear demonstration of the power of measuring the stellar H$\alpha$/continuum ratio and scaling continuum imagery by it prior to SDI subtraction, as the residuals are very effectively removed in the conservatively scaled reduction shown in Figure \ref{fig:twhya}. We do not find any evidence of H$\alpha$ excess signals at the locations of the \citet{Tsukagoshi2019} or \citet{Huelamo2022} point-source candidates, nor elsewhere in the disk; however, both epochs are shown in Appendix \ref{sec:allepochs}, Figure \ref{fig:twhya}.

\subsubsection{HD~100453}

GAPlanetS data were collected for HD~100453 on one night in 2017 and two nights in 2018. The outer M dwarf companion HD~100453~B and a previously known background star at similar separation to the northeast are easily resolved, as seen in Figure \ref{fig:widecomp_zoom}. No additional point-source candidates are apparent in the imagery, including in the combination of the two 2018 datasets. A gallery of all HD~100453 epochs is shown in Appendix \ref{sec:allepochs}, Figure \ref{fig:hd100453}.

HD~100453~B is a wide M star companion, robustly recovered in both H$\alpha$ and continuum imagery in all GAPlanetS epochs, though it is saturated in all but the 2018 May 2 short exposure dataset. For this reason, we extract astrometry and photometry from the companion only at this epoch, and these values are given in Table \ref{tab:detections}.
As the M dwarf companion is well characterized \citep{Collins2009, Wagner2018b}, we do not examine it in detail in this work.

HD~100453~B exhibits H$\alpha$ excess, a clear indication of ongoing accretion, and an estimate for its accretion rate is given in Table \ref{tab:detections}. In contrast, the known background star to the northeast of the primary star is fully removed via ASDI except in the case of the deepest dataset, when some excess remains because the background star is saturated (see Appendix \ref{sec:allepochs}, Figure \ref{fig:hd100453}). 

\subsection{Objects without Companion Candidates in the Literature \label{sec:nondetect}}
\begin{figure}
    \centering
    \includegraphics[width=\columnwidth]{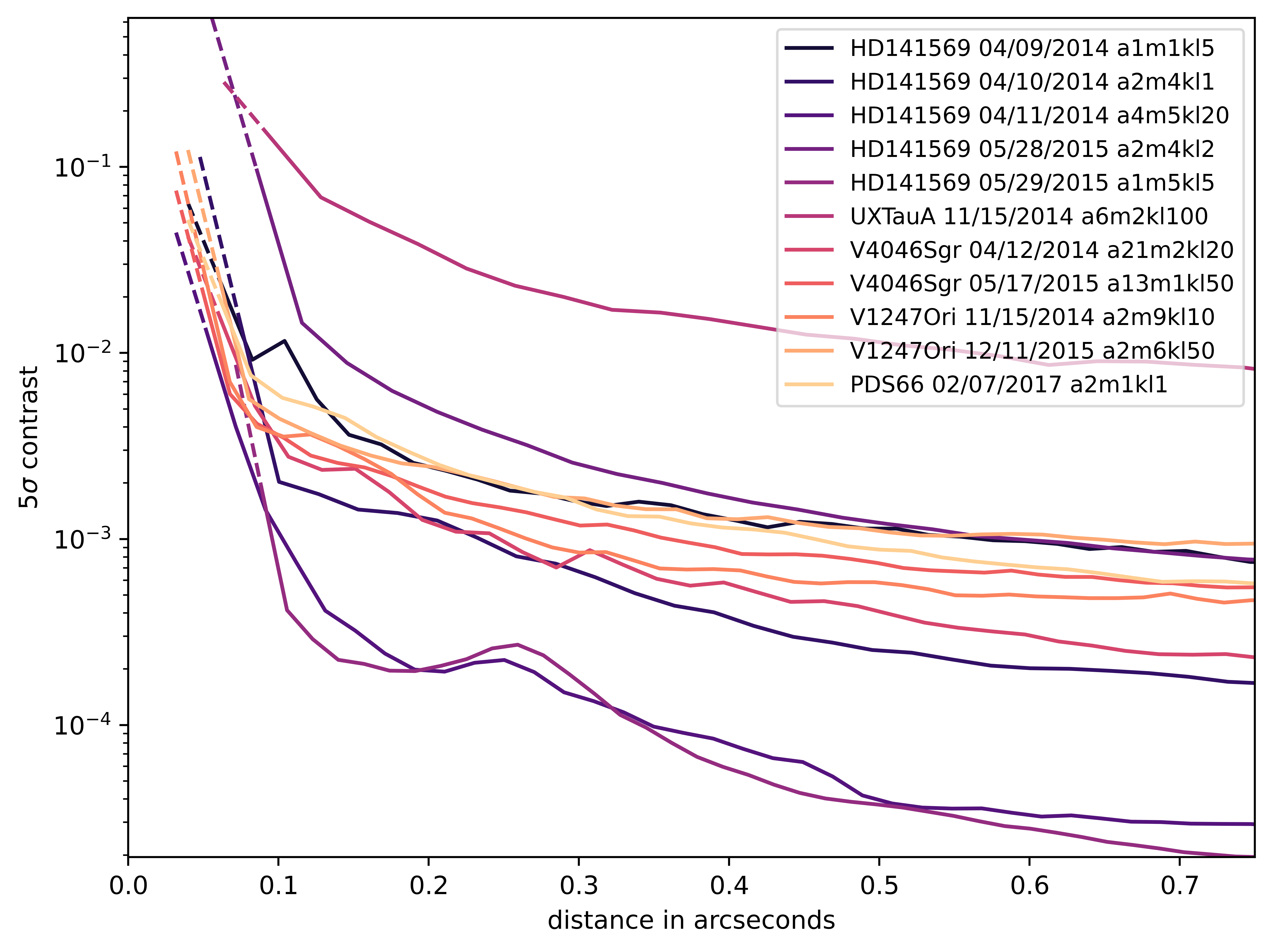}
    \caption{Throughput-corrected 5$\sigma$ contrast curves for continuum false planet optimized \texttt{pyKLIP} reductions of all epochs for the four GAPlanetS targets without reported companion candidates in the literature: HD~141569 (five epochs), UX~Tau~A (one epoch), V4046~Sgr (two epochs), V1247~Ori (two epochs), and PDS 66 (one epoch). Throughput was computed as described in the text, with a correction for the small number of independent noise samples present near the star implemented following \citet{Mawet2014}. Solid curves indicate regions where throughput-corrected contrast was computed directly. The curves are also projected inward from the innermost throughput measurement to the inner working angle, and this extrapolation region is indicated with a dashed line.}
    \label{fig:nondetectctrst}
\end{figure}

There are a number of GAPlanetS targets for which there are no previous reports of planet candidates at detectable separations (though there is a report of a very close companion to V1247~Ori, indicated in the images but under the inner mask), and there are no specific morphology-based predictions for protoplanet locations. These systems are HD~141569, PDS~66, UX~Tau~A, V1247~Ori, and V4046~Sgr. No compelling point-sources are detected in any of the epochs for these objects, though KLIP-ADI and ASDI images are shown for all epochs in Appendix \ref{sec:allepochs}. Contrast curves for all epochs are given in Figure \ref{fig:nondetectctrst}.

\section{Survey Results \label{sec:survey}}

\begin{figure*}
    \centering
    \includegraphics[width=0.85\textwidth]{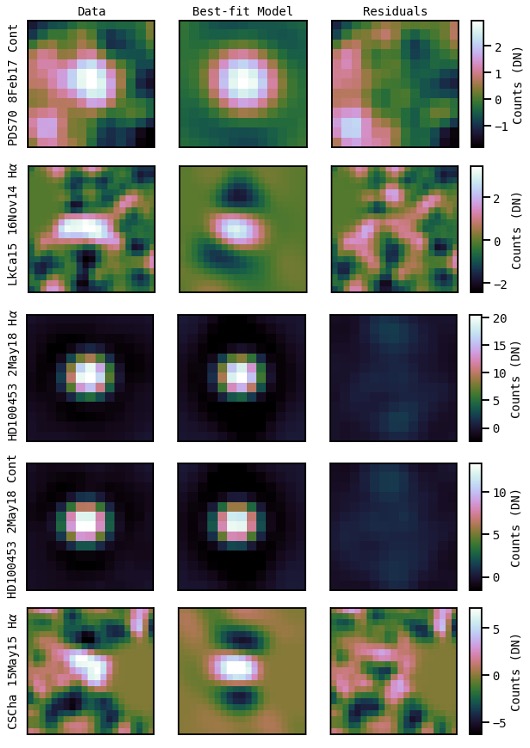}
    \caption{KLIPed data (left) compared to best-fit BKA forward models (middle), and the residuals of their subtraction (right). Fits are shown for all companions whose photometry and astrometry could be extracted from H$\alpha$ and/or continuum GAPlanetS imagery, except for HD~142527~B, as qualitatively and quantitatively similar fits to the same data are shown in \citet{Balmer2022}. Fit statistics are reported in Table \ref{tab:detections}.}
    \label{fig:allfms}
\end{figure*}

\subsection{Overview of Companion and Companion Candidate Photometry and Astrometry \label{sec:fms}}

In this section, we summarize our recovery of each of the five GAPlanetS companions and companion candidates. In all cases where the candidate recovery is sufficiently robust in H$\alpha$ and/or continuum imagery, we extract astrometry and photometry using BKA as described in Section \ref{sec:results} and report the best-fit values in Table \ref{tab:detections}. 

Figure \ref{fig:allfms} shows Bayesian KLIP astrometry models for all GAPlanetS point-source detections except for HD~142527~B, whose BKA fits we recompute under the unified GAPlanetS framework and report in Table \ref{tab:detections}, but which are similar to those described in detail in \citet{Balmer2022}. Each image panel shows a stamp of the final optimized post-processed image, the corresponding best-fit forward model, and the residuals. 

 Our BKA fits returned ``strong" evidence ratios for all five companions/candidates reported in Section \ref{sec:survey}. For PDS~70~c, LkCa~15~``b", and CS~Cha~``c", no point-source is recovered in the continuum, lending credence to the assertion of a planetary nature. HD~142527~B and HD~100453~B are recovered robustly in both H$\alpha$ and continuum imagery, consistent with their nature as stellar companions. In both cases, planet-to-star contrast is more moderate at H$\alpha$ than in the continuum for all epochs, suggestive of active accretion. 

 Compared to past observations, we find:
 \begin{itemize}
     \item HD~142527~B's astrometry is entirely consistent with the observed position of the companion at similar epochs from other facilities, as discussed in detail in \citet{Balmer2022}. We consider this excellent validation of our instrumental astrometric solution for VisAO. 
     \item Although we have conducted HD~142527 reductions under a more conservative KLIP optimization framework, all astrometry is consistent with \citet{Balmer2022} to within error bars except for two of the three continuum epochs where the companion is recovered at SNR$<$3 (2014 April 8 and 2015 May 15 astrometry disagrees with \citet{Balmer2022} at the 2$\sigma$ level). This suggests that BKA fitting is relatively robust to the choice of KLIP parameters, except in cases where SNR is very low.
     \item  Measurement of the HD~142527~B companion's continuum $\Delta$mag is consistent epoch-to-epoch to within error bars. It is also consistent with the continuum $\Delta$mag measured by SPHERE ZIMPOL \citep{Cugno2019}. This suggests that our photometric extractions for GAPlanetS data are broadly consistent with other HCI instruments. 
     \item Measurement of HD~142527~B's H$\alpha$ $\Delta$mag is lower than the continuum $\Delta$mag at all epochs, with an average brightness increase of 0.9 mag over the continuum, suggestive of ongoing accretion. Our measurements are consistent with the observed H$\alpha$ $\Delta$mag of \citet{Cugno2019} and with one another to within error bars. This is in conflict with the tentative evidence for accretion variability reported in \citet{Balmer2022}, likely as a result of improved photometric extraction procedures described in detail in Section \ref{sec:acclims}. 
     \item  PDS~70~c is detected at the same PA as predicted by \texttt{whereistheplanet} in 2017, but at a significantly ($\sim$5$\sigma$) wider separation. As our astrometric solution has been extensively validated with astrometry of Trapezium cluster members and HD~142527~B, this offset is likely accurate, and the orbital properties of PDS~70~c should be updated. 
    \item PDS~70~c's best-fit H$\alpha$ contrast translates to it being nearly 2 mag brighter relative to the host star in the 2017 epoch than the estimate of \citet{Haffert2019}. Accretion rate variability for young stars is estimated to be on the order of $\sim$0.5 dex \citep{Hartmann2016}, a factor of 2 smaller than the magnitude differential between our observations and \citet{Haffert2019}.
     \item LkCa~15~``b"'s astrometry and photometry is marginally ($\sim$1$\sigma$--2$\sigma$) inconsistent with the values reported for the same dataset in \citet{Sallum2015}, appearing at a tighter separation and lower PA.  Improvements in the centering algorithm, VisAO astrometric solution, and post-processing techniques likely contribute to the observed astrometric offset of LkCa~15~``b" relative to previous estimates, as does the smaller IWA.
 \end{itemize}
 
While our characterization of HD~142527~B is consistent with existing literature, for PDS~70~c and LkCa~15~``b", our measurements of astrometry and photometry present inconsistencies with literature values. There are several possible reasons for this. Perhaps the most likely is that the extracted photometry and astrometry of companions, like many other qualities of post-processed images, are dependent on the choice of pre- (e.g., data quality cut, highpass filter value) and post- (e.g., IWA, \texttt{annuli}, and \texttt{movement}) processing parameters. Based on comparison of our BKA fits of HD~142527~B with those of \citet{Balmer2022} for the same datasets but with different pre- and post-processing choices, we believe that extracted astrometry of high-SNR (SNR$>$3) sources is relatively robust to algorithmic parameters. Photometry, however, appears to be somewhat more sensitive to choices such as the BKA fitting area, especially as they effect the structure of the residuals, to which the photometry is very sensitive\footnote{At the same time, we note that extraction of PDS~70~c astrometry and photometry in both continuum false planet optimized extractions (\texttt{annuli}=1, \texttt{movement}=1, \texttt{numbasis}=50) and direct H$\alpha$ optimized extractions (\texttt{annuli}=17, \texttt{movement}=4, \texttt{numbasis}=20) is consistent within error bars despite extreme variation in KLIP parameters.}. Importantly, uncertainty of this nature is not captured in our error estimates. 

Absolute and relative photometric calibration of post-processed high-contrast images is also notoriously difficult. Though we have validated our photometric extractions relative to published photometry of HD~142527~B, and attempted to rigorously quantify photometric uncertainties, systematic errors remain possible. 

\subsection{Determination of Accretion Rates and Limits \label{sec:acclims}}

BKA-derived photometric contrasts for detected companions and final post-processed contrast limits for nondetected companion candidates from the literature are translated to accretion rates following the procedure described in this section. They appear with uncertainties in Tables \ref{tab:detections} and \ref{tab:nondetections}. 

\subsubsection{Computation of H$\alpha$ Line Luminosities}

In previous work \citep[e.g.,][]{Close2014,Sallum2015,Wagner2015,Follette2017,Rameau2017}, H$\alpha$ line luminosities ($L_{\rm{H}\alpha}$) and limits on this quantity have been computed following the equation:
\begin{equation} \label{eq:1}
L_{\rm{H}\alpha}=4\pi d^2 \; z \; \Delta\lambda \; 10^{(M_{\rm{H}\alpha,*}+\Delta H\alpha)/-2.5}
\end{equation}
where $d$ is the distance in centimeters, $z$ is the instrumental zero-point of the H$\alpha$ filter \citep[$1.733\times10^{-5}$ erg~cm$^2$~sec$^{-1}$~$\mu$m$^{-1}$ as determined by][]{Males2013diss}, $\Delta\lambda$ is the width of the H$\alpha$ narrowband filter ($0.006\mu m$), $M_{\rm{H}\alpha,*}$ is the star's extinction corrected \textit{continuum} magnitude in the H$\alpha$ bandpass, and $\Delta \textrm{H}\alpha$ is the difference between the star and companion brightness at H$\alpha$ in magnitude units (computed as -2.5log$_{10}$(contrast)). 

As the stellar continuum magnitude at H$\alpha$ utilized in Equation \ref{eq:1} is not known, we use the apparent magnitude at the Sloan Digital Sky Survey (SDSS) $r$' band as a proxy. We note that the $r$' band is both centered near H$\alpha$ (making the effect of any continuum slope across the bandpass minimal) and $\sim$25 times wider than the narrowband H$\alpha$  filter width (making the contribution of any stellar H$\alpha$ emission small compared to the overall $r$' band flux). To achieve uniformity in the $r$' band apparent magnitude estimates for our targets, we convert Gaia DR3 $G$, $G_{\rm{BP}}$ and $G_{\rm{RP}}$ photometry \citep{Gaia2022} to SDSS $r$' magnitudes following the best-fit conversions of \citet{Alam2015}. We note that these values are consistent with the values of the Gaia Synthetic Photometry Catalog \footnote{https://doi.org/10.17876/gaia/dr.3/64} of \citep[GPSC;][]{GPSC2022} to within 0.1 mag for all eight GAPlanetS targets that are also in the GPSC. 

$r$'-band photometry is subject to nonnegligible extinction in nearby star forming regions, and this effect should be compensated for in estimating ``true" $r$'-band apparent magnitudes for young stars. We estimate line-of-sight extinction to each GAPlanetS target by computing $E$(BP-RP)=(BP-RP)$_{\mathrm{obs}}$-(BP-RP)$_0$, where (BP-RP)$_{\rm{obs}}$ is the observed Gaia BP-RP color of each system, and (BP-RP)$_0$ is the intrinsic color derived from an update to \citep{Pecaut2013a}\footnote{Available at \url{https://www.pas.rochester.edu/~emamajek/EEM_dwarf_UBVIJHK_colors_Teff.txt}} for a pre-main-sequence star of the same spectral type, assuming the spectral types reported in Table \ref{tab:sample}. $E$(BP-RP) values have been shown to closely approximate literature $E(B-V)$ values \citep[within 0.2mag;][]{Andrae2018}; therefore, we convert  $E$(BP-RP) to an $r$'-band extinction following standard Milky Way extinction laws ($R_V=3.1$, $A_{r'}/A_V=0.758$). We list our $A_{r'}$ estimates, which are subtracted from the derived $r$' band apparent magnitudes to compute a non-extincted $r$` magnitude estimate, in Table \ref{tab:sample}. We note that this estimate is based on stellar photometry alone, and that additional extinction toward companion candidates as a result of intervening circumstellar material is possible. 

In this work, we also implement two minor corrections to the calculation described in Equation \ref{eq:1} for estimating L$_{\rm{H}\alpha}$. First, because the central stars of our targets are themselves still actively accreting, the measured narrowband $\Delta H\alpha$ between the companion and the star is not really a $\Delta$mag relative to the stellar \textit{continuum}, which would be required to compute an H$\alpha$ luminosity. To make it so, we multiply the measured contrast ($L_{\rm{H}\alpha\textrm{,comp}}/L_{\rm{H}\alpha,*}$) by the stellar H$\alpha$-to-continuum scale factor ($L_{\rm{H}\alpha,*}/L_{\rm{Cont,*}}$, determined with aperture photometry as described in Section \ref{sec:preproc} and reported in Table \ref{tab:data}) to compute the companion's H$\alpha$ brightness ratio relative to the stellar \textit{continuum}($L_{\rm{H}\alpha\textrm{,comp}}/L_{\rm{Cont,*}}$). This value is reported in Tables \ref{tab:detections} and \ref{tab:nondetections} as $\Delta$mag. Due to the scale factor correction, it is not a direct magnitude conversion of the observed star-to-companion contrast, but rather a best estimate of the H$\alpha$ excess \textit{unique to the companion}. 

L$_{\rm{H}\alpha}$ is also, properly, a \textit{line} luminosity, meaning the companion's continuum luminosity should be removed from the estimated H$\alpha$ luminosity before using the derived value as a \textit{line} luminosity. In the case where we detect continuum emission from the companion (HD~142527~B and HD~100453~B), this is easily done by substituting $\Delta$Cont for $\Delta H\alpha$ into equation \ref{eq:1} and subtracting the resulting continuum luminosity. In the case where an object is not detected at continuum wavelengths (PDS~70~c, LkCa~15~``b", and CS~Cha~``c"), the contribution of the object photosphere to the H$\alpha$ luminosity is unknown. However, we note that the predicted absolute continuum $r$'-band magnitude of even a very massive, very young planet is extremely faint \citep[$r$'=14 for a 10$M_J$ planet at 1~Myr;][]{Baraffe2015} compared to the H$\alpha$ absolute magnitudes that we estimate for our planetary companions and companion candidates (M$_{\rm{H}\alpha}$ of 7.3, 8.8, and 12.3 for CS~Cha~``c", LkCa~15~``b", and PDS~70~c, respectively). It is thus reasonable to assume that any continuum contribution to the observed H$\alpha$ luminosity for lower-mass companions is negligible. 

In practice, these corrections and approximations make the equation used to compute H$\alpha$ line luminosity:
\begin{equation} \label{eq:2}
L_{\rm{H}\alpha}=4\pi d^2z\Delta\lambda10^{(m_{r',*}A_{r',*}-2.5log(CS))/-2.5}-L_{\rm{cont}}
\end{equation}
where m$_{r',*}$ is the stellar $r$'-band apparent magnitude, A$_{r',*}$ is the estimated $r$'-band extinction, C is the observed H$\alpha$ contrast of the companion, S is the stellar H$\alpha$-to-Continuum scale factor, and L$_{cont}$ is the continuum contribution to the H$\alpha$ luminosity (used only in the case of a continuum detection of the companion, otherwise assumed negligible).

As BKA does not natively propagate absolute photometric uncertainties, we compute our own uncertainty estimates for the final companion $\Delta$mag by propagating the uncertainties on a number of individual quantities into magnitude space. These uncertainties are: the 67\% credibility interval on the best fit scale factor (the ``alpha" value) from the BKA MCMC, uncertainty in the stellar H$\alpha$-to-continuum scale factor (estimated as the standard deviation in the scale factors of individual images in each observing sequence), and uncertainty in the stellar peak used to normalize imagery prior to BKA (estimated as the median photon noise of the best-fit stellar (unsaturated) or ghost (saturated) photometry). We also include uncertainty in the ghost-to-star scale factor in the case of saturated data, a value derived from the standard deviation of the residuals to the linear fit between ghost and stellar Moffat fit peak values in all unsaturated GAPlanetS datasets (230$\pm$50 at H$\alpha$, 245$\pm$50 at the Continuum). 

Uncertainty on the final H$\alpha$ luminosities encompasses uncertainty on $\Delta$mag, as well as uncertainties in the following quantities: the stellar $r$'-band apparent magnitude (assumed to be 0.1 based on the average discrepancy between our estimates and the GPSC), the $r$'-band extinction (estimated at $\pm$0.2 based on the range of estimates for A$_R$ of these objects in the literature), distance (derived from parallax uncertainty in the Gaia catalog), instrumental zero-point (estimated conservatively at 10\%), and the companion's continuum brightness (where detected, estimated following the same procedure as H$\alpha$ photometric uncertainty).

\subsubsection{Estimation of Accretion Luminosities and Rates}

Accretion luminosities are derived from H$\alpha$ line luminosities following the equation: 
\begin{equation}
L_{\rm{acc}}=10^{b+alog_{10}(\frac{L_{\rm{H}\alpha}}{L_{\odot}})}L_{\odot}
\end{equation}
where the coefficients $a$ and $b$ represent an empirically or model-derived scaling law between $L_{\rm{H}\alpha}$ and accretion luminosity. The value of these scaling coefficients is particularly poorly constrained in the substellar regime. We report accretion rates for detected protoplanetary candidates and accretion rate limits for undetected literature companion candidates following both the \citet{Aoyama2021} ($b=1.61\pm0.04$, $a=0.95\pm0.006$, with scatter of 0.3 dex) and \citet{Alcala2017} ($b=1.74\pm0.19$, $a=1.13\pm0.05$, with scatter of 0.5-0.7 dex) scaling relations. The Aoyama relation is theoretically derived from planetary accretion shock models \citep[e.g.,][]{Aoyama2018,Marleau2019}, where the principal difference relative to classical magnetospheric accretion models is the contribution of the (non-fully ionized) postshock region to the line emission. The Alcal\'a relation is an empirically derived $L_{\rm{H}\alpha}-L_{\rm{acc}}$ relation for a large number of T Tauri stars. For the known stellar companions HD~142527~B and HD~100453~B, we report only the Alcal\'a-derived accretion rates, as they fall solidly within the mass regime of that sample.

Mass accretion rates/limits ($\dot{M}$) are derived from total accretion luminosities ($L_{\rm{acc}}$) via the standard relation 
\begin{equation}
\dot{M}=\frac{1.25L_{\rm{acc}}R}{GM}    
\end{equation}
where  and $R$ are the mass and radius of the accreting object \citep{Gullbring1998}. As the masses and radii of protoplanet candidates are poorly constrained or unconstrained, we report most accretion rates and limits as the product of $M$ and $\dot{M}$ and adopt a global value of 2$M_J$ for object radii, which is reasonable for a range of young substellar objects. 

For the stellar companion HD~142527~B, we adopt a mass of 0.26$M_{\odot}$ and a radius of 1.2$R_{\odot}$, following \citet{Balmer2022} in order to estimate an accretion rate. For HD~100453~B, we adopt a mass of 0.2$M_{\odot}$ following \citet{Collins2009} and estimate a radius of 0.5$R_{\odot}$ based on the models of \citet{Baraffe2015} for a 15Myr, 0.2$M_{\odot}$ pre-main-sequence star.

In order to estimate uncertainty in derived accretion rates, we adopt the approach recommended in \citet{Aoyama2021}, which uses the spread in the $L_{\rm{acc}}$--$L_{\rm{H}\alpha}$ relation derived from their theoretical models instead of the formal error to place uncertainties on $L_{\rm{acc}}$ values. We note that all $L_{\rm{H}\alpha}$ values estimated in this work lie in the $L_{\rm{acc}} \lesssim 10^{-4} L_{\odot}$ regime, where the spread in the model relation increases substantially (to 1.5 dex) due to an increased optical depth at $H\alpha$. Combined with a 0.5$R_{\mathrm{Jup}}$ uncertainty on object radii, we estimate accretion rate uncertainties under these models of 2-3 dex.

For the empirical T Tauri stellar relation, \citet{Alcala2017} reported a standard deviation of 0.41 dex around the best fit $L_{\rm{acc}}$--$L_{\rm{H}\alpha}$ scaling. Combined with $\sim$10\% uncertainties on masses and radii, we estimate an uncertainty of $\sim$1 dex on accretion rates estimated under the \citet{Alcala2017} relation. 

In general, given the poorly constrained nature of young substellar objects' masses and radii, a limited understanding of which scaling relations are most appropriate in the substellar regime, and intrinsic photometric uncertainties, we caution that accretion rates should be interpreted as very rough estimates. Their utility lies primarily in comparison with one another; assuming all objects accrete material under the same paradigm, their \textit{relative} accretion rates under a single accretion scaling relation should reflect reality.  


\begin{deluxetable*}{ccccccccccc}
\tablecaption{Results of BKA Forward Model Fitting \label{tab:detections}}
\tablehead{
\colhead{\textbf{Object}} & \colhead{\makecell[t]{\textbf{Date}\\DD-MM-YY}} & \colhead{\makecell[t]{\textbf{Separation}\\(mas)}} & \colhead{\makecell[t]{\textbf{PA}\\(deg)}} & \colhead{$\log$(\textbf{C})\tablenotemark{A}} & \colhead{$\log$(\textbf{${Z_1/Z_0}$})} & \colhead{\textbf{SNR}} &\colhead{\makecell[t]{\textbf{$\Delta\mathrm{mag}$}\tablenotemark{B}\\(mag)}} & \colhead{\makecell[t]{$\log$(\textbf{$L_{\rm{H}\alpha}$})\\($L_{\odot}$)}}& \colhead{\makecell[t]{$\log($\textbf{$M\dot{M}$) A17\tablenotemark{C}}\\($M_{\mathrm{Jup}}^2/yr$)}} & \colhead{\makecell[t]{$\log$(\textbf{$M\dot{M}$) A21}\tablenotemark{D}\\($M_{\mathrm{Jup}}^2/yr$)}}}
\tablecolumns{11}
\startdata
& & & \multicolumn{6}{c}{\textbf{H$\alpha$ Fits to Protoplanet Companions and Candidates}}  & \makecell[c]{ \hspace{5pt} \\ \hspace{5pt} } & \\
\hline
PDS~70~c     & 2017-02-08 & 246.9$\pm$4.4 & 284.2$\pm$0.6 & -2.53$\pm$0.16 & 11  & 4.5  & 6.02$\pm$0.41 & -5.48$\pm$0.19 & -6.5     & -5.6      \\
LkCa~15~``b" & 2014-11-16 & 69.7$\pm$6.1    & 242.6$\pm$2.7 & -1.52$\pm$0.28 & 18   & 3.5  & 3.14$\pm$0.69 & -3.82$\pm$0.29 & -4.6      & -4.1      \\
CS~Cha~``c"  & 2015-05-15 & 68.1$\pm$1.4  & 76.3$\pm$1.1  & -1.29$\pm$0.03 &  11   & 4.3  & 2.34$\pm$0.08 & -3.00$\pm$0.1     & -3.7      & -3.3      \\
\hline
& & &  \multicolumn{6}{c}{\textbf{H$\alpha$ Fits to Stellar Companions}} & \makecell[c]{$\log($\textbf{$\dot{M}$) A17}\\($M_{\mathrm{\odot}}$yr$^{-1}$)} & \\
\hline
HD~142527~B  & 2013-04-11 & 82.7$\pm$1.3  & 128.2$\pm$0.6 & -2.68$\pm$0.01 & 453 & 13.6 & 6.83$\pm$0.06 & -4.07$\pm$0.25 & -10.8      & \nodata \\
HD~142527~B  & 2014-04-08 & 74.5$\pm$1.5  & 120.4$\pm$0.8 & -2.77$\pm$0.02 & 87  & 8.2  & 6.79$\pm$0.05 & -4.07$\pm$0.3  & -10.8      & \nodata \\
HD~142527~B  & 2015-05-15 & 67.4$\pm$1.8  & 110.7$\pm$1.1 & -2.65$\pm$0.09 & 54  & 4.6  & 6.49$\pm$0.4  & -3.86$\pm$0.39 & -10.5      & \nodata \\
HD~142527~B  & 2015-05-16 & 71.6$\pm$1.5  & 107.7$\pm$0.8 & -2.76$\pm$0.13 & 58  & 6.2  & 6.76$\pm$0.46 & -3.92$\pm$0.39 & -10.6      & \nodata \\
HD~142527~B  & 2015-05-18 & 72.6$\pm$1.3  & 109.6$\pm$0.7 & -2.62$\pm$0.01 & 145 & 10.3 & 6.44$\pm$0.3  & -3.93$\pm$0.38 & -10.6    & \nodata \\
HD~142527~B  & 2018-04-27 & 43.7$\pm$1.3  & 54.4$\pm$1.5  & -2.9$\pm$0.07  & 28  & 4.1  & 6.99$\pm$0.36 & -3.99$\pm$0.39 & -10.7      & \nodata \\
HD~100453~B  & 2018-05-02 & 1033.9$\pm$13 & 133.1$\pm$0.2 & -2.96$\pm$0.01 & inf & 30.6 & 7.35$\pm$0.3  & -4.54$\pm$0.19 & -11.8      & \nodata \\
\hline
& & & \multicolumn{6}{c}{\textbf{Continuum Fits to Stellar Companions}} & \makecell[c]{ \hspace{5pt} \\ \hspace{5pt} } & \\
\hline
HD~142527~B  & 2013-04-11 & 82.6$\pm$1.4  & 127.9$\pm$0.6 & -3.01$\pm$0.05 & 162 & 6.4  & 7.53$\pm$0.13 & \nodata      & \nodata & \nodata \\
HD~142527~B  & 2014-04-08 & 71.3$\pm$2.2  & 118.3$\pm$1.5 & -2.99$\pm$0.15 & 14  & 4.2  & 7.47$\pm$0.37 & \nodata      & \nodata & \nodata \\
HD~142527~B  & 2015-05-15 & 63.9$\pm$2    & 112.9$\pm$1.5 & -2.95$\pm$0.2  & 12  & 2.8  & 7.38$\pm$0.6  & \nodata      & \nodata & \nodata \\
HD~142527~B  & 2015-05-16 & 76.6$\pm$2.5  & 107.6$\pm$1.2 & -3.14$\pm$0.31 & 11  & 2.2  & 7.84$\pm$0.83 & \nodata      & \nodata & \nodata \\
HD~142527~B  & 2015-05-18 & 73.7$\pm$1.4  & 110.2$\pm$0.8 & -2.85$\pm$0.02 & 88  & 7.5  & 7.12$\pm$0.29 & \nodata      & \nodata & \nodata \\
HD~142527~B  & 2018-04-27 & 42.1$\pm$2.1  & 54.6$\pm$2.3  & -3.27$\pm$0.56 & 5   & 1.80 & 8.17$\pm$1.42 & \nodata      & \nodata & \nodata \\
HD~100453~B  & 2018-05-02 & 1035.1$\pm$13 & 133$\pm$0.2   & -3.13$\pm$0.01 & inf & 31.1 & 7.81$\pm$0.29 & \nodata      & \nodata & \nodata \\
\enddata
\tablenotetext{A}{Reported uncertainty in contrast reflects only the 67\% credibility interval of the BKA posterior fit to the photometric scale factor. Full photometric errors are reflected in the $\Delta\mathrm{mag}$ column.}
\tablenotetext{B}{For all H$\alpha$ fits, this is the $\Delta\mathrm{mag}$ relative to the stellar continuum, calculated by multiplying the candidate contrast by the H$\alpha$-to-continuum scale factor for the star(given in Table \ref{tab:data}). The uncertainty reported on this quantity reflects a full photometric error accounting, as described in detail in the text.}
\tablenotetext{C}{Accretion rates estimated from the empirical $L_{\rm{acc}}$--$L--{\textrm{H}\alpha}$ scaling law of \citet{Alcala2017}. These are reported as M$\dot{M}$ values in units of $M_{\mathrm{Jup}}^2$/yr for the protoplanets and protoplanet candidates, as their masses are not well constrained. For the stellar companions, the known masses are used to estimate a true accretion rate in $M_{\odot}$/yr. We estimate the uncertainty on accretion rates derived under this model as $\pm$1 dex, as described in detail in the text.}
\tablenotetext{D}{Accretion rates estimated from the theoretical $L_{\rm{acc}}$--$L--{H\alpha}$ scaling law of \citet{Aoyama2021}, reported only for those candidates that are not known stellar companions, as the accretion paradigm applied in these models is planetary in nature. We estimate the uncertainty on accretion rates derived under this model as $\pm$2-3 dex, as described in detail in the text.}
\end{deluxetable*}

\begin{deluxetable*}{ccccccccc}
\tablecaption{Limits on Undetected Protoplanet Candidates}
\tablehead{
\colhead{\textbf{Object}} & \colhead{\makecell[t]{\textbf{Candidate}\\\textbf{Label}}} & \colhead{\makecell[t]{\textbf{Separation}\\(mas)}} & \colhead{\makecell[t]{\textbf{PA}\\(deg)}} & \colhead{\makecell[t]{\textbf{Observation}\\\textbf{Epoch(s)}}} & \colhead{\textbf{Source}} & \colhead{\makecell[t]{\textbf{GAPlanetS}\\\textbf{Best Epoch}}} & \colhead{\makecell[t]{\textbf{Log Contrast}\\\textbf{at planet}}} & \colhead{$\Delta$mag}
} 
\startdata
HD~169142 & O07 & 116$\pm$20 & 250$\pm$5 & Jun-07 & \citet{Okamoto2017} & 8-Apr-14 & -3.85 & $>$9.6\\
HD~169142 & O12-13 & 340 & 175.0 & 2012--2013 & \citet{Osorio2014} & 8-Apr-14 & -4.17 & $>$10.4 \\
HD~169142 & R13 & 156$\pm$32 & 7.4$\pm$11.3 & Jun-13 & \citet{Reggiani2014} & 8-Apr-14 & -3.95 & $>$9.9\\
HD~169142 & B13 & 110$\pm$30 & 0$\pm$14 & Jul-13 & \citet{Biller2014} & 8-Apr-14 & -3.78 & $>$9.5 \\
HD~169142 & B14 & 180 & 33.0 & Apr-14 & \citet{Biller2014} & 8-Apr-14 & -3.99 & $>$10.0 \\
HD~169142 & G15-17A & 115$\pm$15 & 239$\pm$11.5 & 2015--2018 & \citet{Gratton2019} & 8-Apr-14 & -3.84 & $>$9.6 \\
HD~169142 & G15-18B & 189$\pm$8 & 17$\pm$8 & 2015--2018 & \citet{Gratton2019} & 8-Apr-14 & -3.99 & $>$10.0 \\
HD~169142 & G15-18C & 197$\pm$8.5 & 308$\pm$9 & 2015--2018 & \citet{Gratton2019} & 8-Apr-14 & -3.99 & $>$10.0\\
HD~169142 & G15-18D & 317$\pm$7 & 39$\pm$5 & 2015--2018 & \citet{Gratton2019} & 8-Apr-14 & -4.15 & $>$10.4 \\
HD~169142 & B18 & 105.8$\pm$35.3 & 55.5$\pm$ 4.0 & 15-Jul-18 & \citet{Bertrang2020} & 8-Apr-14 & -3.75 & $>$9.4 \\
HD~100546 & Q11 & 480$\pm$40 & 8.9$\pm$0.9 & May-11 & \citet{Quanz2013} & 12-Apr-14 & -4.97 & $>$12.0 \\
HD~100546 & C15 & 131$\pm$9 & 150.9$\pm$2 & Jan-15 & \citet{Currie2015} & 12-Apr-14 & -3.79 &  $>$9.1 \\
HD~100546 & S15-16 & 455$\pm$7 & 11.5$\pm$1 & 2015--2016 & \citet{Sissa2018} & 12-Apr-14 & -4.91 &  $>$11.9 \\
HD~100546 & F15-16 & 964.0 & 10 & 2015--2016 & \citet{Fedele2021} & 12-Apr-14 & -5.30 &  $>$12.9 \\
HD~100453 & G19 & 1074.0$\pm$31.8 & 132.7$\pm$0.8 & Apr-19 & \citet{Gonzalez2020} & 2-May-18-long & -4.76 &  $>$11.6 \\
V1247Ori & W12-13d & 41.5$\pm$6.5 & 305.5$\pm$5.5 & 2012--2013 & \citet{Willson2019} & 11-Dec-15 & -1.94 & $>$4.7 \\
SAO206462 & C16 & 71.1$\pm$5 & 19$\pm$3 & Mar-16 & \citet{Cugno2019} & 12-Apr-14 & -3.06 & $>$7.4 \\
SAO206462 & C19 & 425.9$\pm$1.2 & 212.4$\pm$0.7 & Jul-19 & \citet{Casassus2021} & 12-Apr-14 & -4.54 & $>$11.1\\
CSCha B & G17 & 1316.5$\pm$5 & 261.4$\pm$0.2 & Feb-Jun 2017 & \citet{Ginski2018b} & 15-May-15 & -3.18 & $>$7.1 \\
TW~Hya & I16 & 840.7$\pm$67.2 & 242.5$\pm$2.1 & Dec 2016 & \citet{Ilee2022} & 7-Feb-17 & -4.14 & $>$8.2 \\
TW~Hya & T17 & 865.4$\pm$1 & 237$\pm$1 & May-17 & \citet{Tsukagoshi2019} & 7-Feb-17 & -4.15 & $>$8.2 \\
TW~Hya & H19 & 160$\pm$10 & 190$\pm$1 & 15-Mar-19 & \citet{Huelamo2022} & 7-Feb-17 & -3.11 & $>$5.6 \\
PDS~70 & Z20d & 110 & 310 & Feb-July 2020 & \citet{Zhou2021} & 2-May-18 & -2.64 & $>$6.3\\
\enddata
\tablecomments{Limits for planet candidate nondetections. The ``Candidate Label" column indicates the text marking the candidate in our figures. Separation and PA, as well as the errors on these quantities, are derived from the original detection paper, with separations translated to pixels using the VisAO plate scale of 7.95~mas~pixel$^{-1}$ \citep{Balmer2022}. The ``Observation Epoch" column indicates the date of the original observations used to identify the candidate(s), and the reference that reported it appears in the ``Source" column. The GAPlanetS epoch with the highest achieved contrast at the candidate's separation is indicated in the ``GAPlanetS Best Epoch" column and the logarithm of the achieved contrast at the candidate separation in this epoch is indicated in the ``Log contrast at planet" column. This is translated to a limiting H$\alpha$ magnitude relative to the stellar continuum by multiplying achieved contrast by the stellar H$\alpha$-to-continuum scale factor, as described in detail in the text.}
\label{tab:nondetections}
\end{deluxetable*}

\subsection{Optimal KLIP Parameters}
The products of our optimization processes (outlined in detail in Sections \ref{sec:dqcuts} and \ref{sec:optimize} are, for each dataset: (a) an optimal amount of data to discard to minimize contrast for a fixed choice of KLIP parameters, and (b) a set of values for the \texttt{pyKLIP} parameters \texttt{annuli}, \texttt{movement}, and \texttt{numbasis} designed to maximize the sum of all six normalized image quality metrics across false planets injected into the continuum images between the IWA and control radius. In this section, we discuss trends in optimal parameters. 

\paragraph{Optimal Data Quality Cuts} appear to be more often low than high, as only 19\% ($n$=7) of the optimal cut values were greater than 50\%. Only 18\% ($n$=7) of datasets have an optimal data quality cut of 0\%, indicating that a majority of GAPlanetS reductions are significantly improved by discarding some proportion of images. 

\paragraph{Optimal \texttt{pyKLIP annuli}} values were also more commonly low, with 66\% ($n$=24) of the optimal \texttt{annuli} values at 10 or fewer. 

\paragraph{Optimal \texttt{pyKLIP movement}} optimization showed a strong preference for ``aggressive" values of 1 ($n$=17) or 2 ($n$=8), with only 31\% of the datasets ($n$=11) showing an optimal \texttt{movement} of 3 or greater. 

\paragraph{Optimal \texttt{pyKLIP numbasis}} values showed a peak at 20 KL modes ($n$=13), with values of 50 and 100 somewhat less common ($n$=4 and 2, respectively). Thirty-nine percent ($n$=14) of datasets showed optimal KL mode values of 5 or fewer. 

Somewhat surprisingly, none of these optimized parameters show clear trends with system (e.g. AO wavefront sensor binning), stellar (e.g. $r$'-band magnitude), or atmospheric (e.g. FWHM) variables.

\subsection{Contrasts}

As with any HCI survey, one of the principal products of GAPlanetS is detection limits for accreting protoplanets in all imaged systems. These curves are difficult to interpret in bulk for several reasons. First, achieved contrast is highly sensitive to both the guide star magnitude and atmospheric conditions. This is true to some extent for all AO systems, but is especially true at visible wavelengths. Achieved contrast also varies wildly from dataset to dataset in this regime, even for the same object (see Figure \ref{fig:PSFgallery}). 

\begin{figure*}
    \centering
    \includegraphics[width=\textwidth]{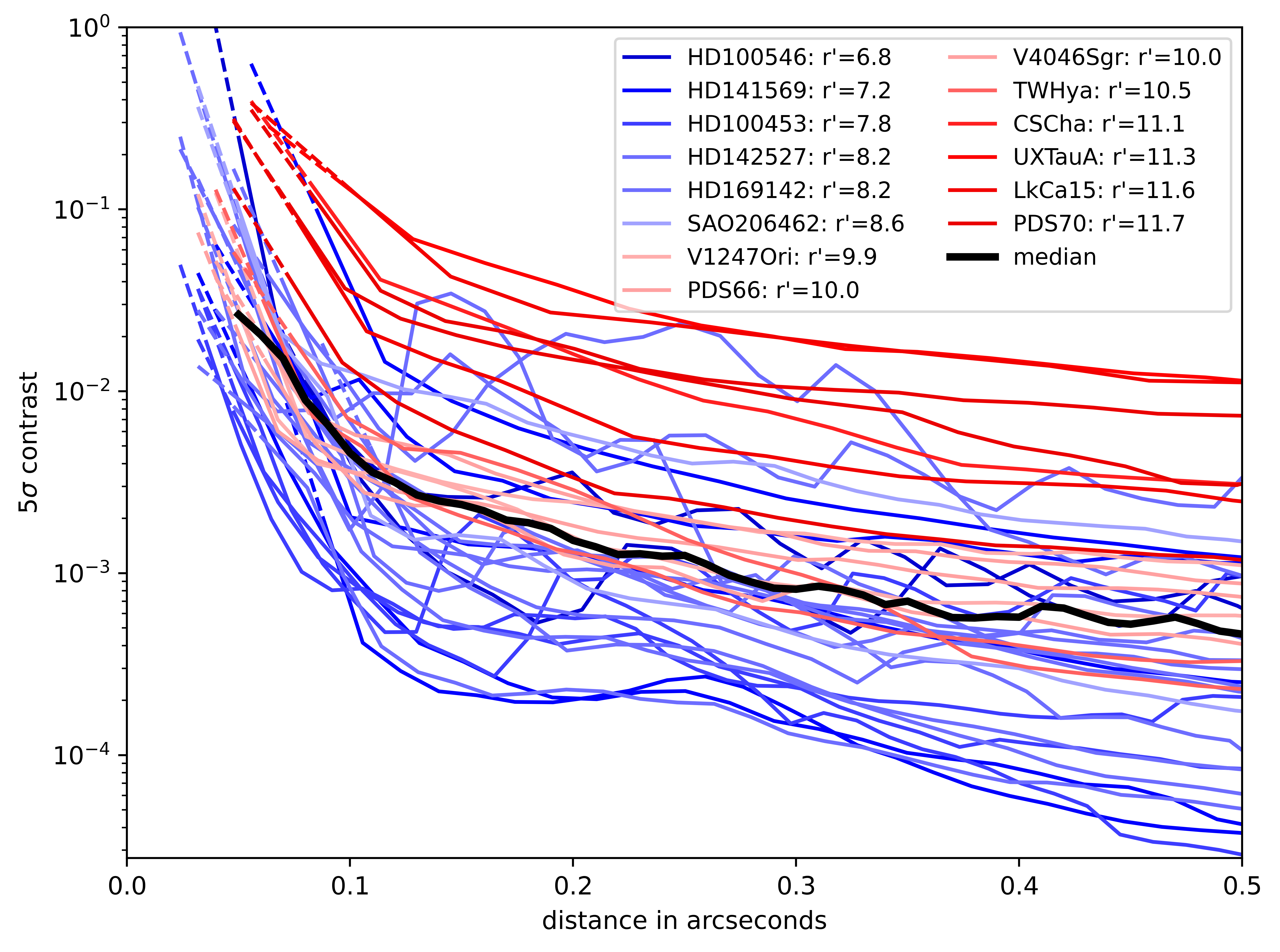}
    \caption{Contrast curves for all GAPlanetS datasets analyzed in this work. These curves were generated following our conservative survey methodology of optimizing on false continuum planets. Curves are colored by the $r$'-band magnitude of the star. While the specific contrast achieved for a given target varies widely with seeing, it is also a strong function of $r$'-band magnitude.}
    \label{fig:ctrst_byrmag}
\end{figure*}

Nevertheless, the bulk contrast curves of the sample can give us a general grasp of performance for this first-generation accreting protoplanet survey and the limits that it places on the prevalence of objects with certain H$\alpha$ contrast ratios embedded in transitional disk cavities, with the caveat that variable accretion will make these boundaries somewhat fuzzy. Figure \ref{fig:ctrst_byrmag} shows optimized contrast curves for the entire GAPlanetS sample colored by stars' $r$'-band magnitude. 

The median achieved contrast for the survey, as well as the best achieved contrast for each target at a range of separations from 0$\farcs$1 to 1$\farcs$5 is also provided in Table \ref{tab:ctrsts}.

\begin{table*}[]
\centering
\setlength{\tabcolsep}{8pt}
\tablecaption{GAPlanetS Best Achieved Contrasts by Target from 0$\farcs$1 to 1$\farcs$5}
\begin{tabular}{l|lllllll}
          & \multicolumn{7}{c}{\textbf{Best Achieved Log Contrast}} \\
\textbf{} &
  \textbf{0$\farcs$1} &
  \textbf{0$\farcs$25} &
  \textbf{0$\farcs$50} &
  \textbf{0$\farcs$75} &
  \textbf{1$\farcs$00} &
  \textbf{1$\farcs$25} &
  \textbf{1$\farcs$50} \\
  \hline\hline
\textit{Survey Median} &
  \textit{-2.34} &
  \textit{-2.90} &
  \textit{-3.34} &
  \textit{-3.45} &
  \textit{-3.62} &
  \textit{-3.69} &
  \textit{-3.76} \\
  \hline
HD100546  & -2.34  & -2.89  & -3.02 & -3.25 & -3.44 & -3.40 & -3.48 \\
HD100453  & -3.04  & -3.41  & -4.08 & -4.27 & -3.97 & -4.32 & -4.48 \\
HD142527  & -2.44  & -3.15  & -3.36 & -3.43 & -3.55 & -3.65 & -3.82 \\
HD169142  & -2.68  & -3.03  & -3.48 & -3.70 & -3.84 & -3.94 & -3.98 \\
SAO206462 & -2.26  & -3.18  & -3.76 & -4.06 & -4.25 & -4.36 & -4.41 \\
V1247Ori  & -2.44  & -2.95  & -3.24 & -3.33 & -3.42 & -3.44 & -3.46 \\
HD141569  & -1.95  & -2.73  & -2.94 & -3.12 & -3.42 & -3.72 & -3.94 \\
V4046Sgr  & -2.46  & -3.01  & -3.39 & -3.64 & -3.72 & -3.73 & -3.76 \\
PDS66     & -2.22  & -2.71  & -3.06 & -3.24 & -3.33 & -3.36 & -3.40 \\
TWHya     & -2.26  & -3.06  & -3.48 & -3.58 & -3.64 & -3.69 & -3.69 \\
UXTauA    & -0.88  & -1.62  & -1.94 & -2.09 & -2.17 & -2.23 & -2.27 \\
CSCha     & -1.09  & -2.01  & -2.51 & -2.70 & -2.75 & -2.76 & -2.80 \\
LkCa15    & -1.44  & -2.30  & -2.61 & -2.69 & -2.76 & -2.76 & -2.78 \\
PDS70     & -1.14  & -1.92  & -2.13 & -2.30 & -2.47 & -2.62 & -2.72
\end{tabular}%
\label{tab:ctrsts}
\end{table*}

\subsection{Detection Rates}
Computation of robust survey statistics for GAPlanetS is difficult, as this is not a large or unbiased sample, and the nature of some of the candidates is unclear. Nevertheless, the detection rate relative to other HCI surveys is striking and is a result of the highly targeted nature of the GAPlanetS sample (transitional disks; see Section \ref{sec:obs}). We detect two accreting stellar companions across our sample of 14 objects, one at very tight separation (HD~142527~B, $\sim$0$\farcs$1) and one more distant companion (HD~100453~B, $\sim$1"). We also detect four protoplanets or robust protoplanet candidates  (PDS~70~b, PDS~70~c, LkCa~15~``b", and CS~Cha ``c") inside the cavities of three additional systems. This makes a total of five systems with accreting companion candidates out of 14 systems targeted, a detection rate of $\sim$36$^{+26}_{-22}$\% assuming binomial statistics \footnote{More specifically, the Wilson score interval with continuity correction for a detection rate of 0.36 gives a 95\% confidence interval of [0.14,0.64]}.

This is substantially higher than has been found by previous exoplanet direct imaging surveys in the NIR. In a meta-analysis of first-generation direct imaging survey results, \citet{Bowler2018} found that the occurrence rate for planets with masses $\sim$ 5--13$M_J$ and separations $\sim$ 5--500 AU was around $\%1$. More recently, the Gemini Planet Imager Exoplanet Survey (GPIES) detected nine companions (six planetary and three brown dwarf) in six systems from among a sample of 300, yielding a planet occurrence rate estimate of 9$^{+5}_{-4}$\% for planets between 5 and 13$M_J$ and 10--100au separation around stars greater than 1.5 solar masses. For brown dwarfs companions, GPIES yielded an even lower occurrence rate: only 0.8 $^{+5}_{-4}$ for brown dwarfs between 13 and 80$M_J$ and 10--100au \citep{Nielsen2019}.

GAPlanetS is not the only survey for protoplanets in the literature. Previously published H$\alpha$ direct imaging surveys have yielded no new confirmed planets, though several have recovered HD~142527~B \citep{Cugno2019,Zurlo2020} at high SNR, and/or present new yet-to-be-confirmed candidates \citep[e.g.,][]{Cugno2019,Huelamo2022}. The largest previous survey for accreting companions, which had similar selection criteria to GAPlanetS, was the 11-object VLT-SPHERE survey of \citet{Zurlo2020}. They recovered HD~142527~B, but detected no new accreting candidates aside from HD98800 Ba, one of the stellar members of the HD~98800~BaBb circumbinary transitional disk and therefore akin to our (also accreting) central transitional disk host stars. They hypothesize that the prevalence of strong residual speckles in the inner 0$\farcs$2 of their post-processed images contributes to their low detection rate, and we note that three of the systems in which we have detected candidates are within this region. When expanding their sample to include three additional archival H$\alpha$ observations, as well as PDS~70, their detection rate rises to 2/15 objects, or $\sim$13$^{+29}_{-11}$\% assuming binomial statistics, consistent with our detection rate within uncertainties \footnote{More specifically, the Wilson score interval with continuity correction for a detection rate of 0.20 gives a 95\% confidence interval of [0.02,0.42]}.

\section{Conclusions} \label{sec:conclusion}

\subsection{Summary of Results}
In this work, we present observational results from the GAPlanetS HCI campaign, a targeted, multiepoch H$\alpha$ adaptive optics study of 14 transitional disk systems with MagAO. Of these targets, we robustly recover previously reported accreting stellar and planetary companions/candidates in four systems: HD~100543, HD~142527, PDS~70, and LkCa~15. We do not robustly recover H$\alpha$ emission from previously reported planet candidates in five additional systems, namely: CS~Cha (the ``B" companion), HD~100546, HD~169142, SAO~206462, and TW~Hya. In the remaining five systems (HD~141569, PDS~66, UX~Tau~A, V1247~Ori, and V4046~Sgr), we do not find evidence of accreting planetary-mass companions. We also report the detection of a single new accreting candidate companion, CS~Cha~``c", bringing the detection rate to 5/14, or $\sim$36\% for the GAPlanetS sample. 

While we do not detect many of the previously reported protoplanet candidates from the literature, we note that nondetections here do not speak to the robustness of those previous detections so much as to the limitations of our data. Even relatively high-mass protoplanets are likely to have accretion luminosities below our detection threshold (see Table \ref{tab:nondetections}) in many cases.  Furthermore, very little small grain dust is required to extinct at H$\alpha$, so in the case of planet candidates that do not lie in highly cleared NIR cavities, H$\alpha$ emission from protoplanets may be obscured by intervening small grain material. In other words, the absence of an H$\alpha$ signal at the location of a candidate does not imply its nonexistence, or even that it is not accreting.

GAPlanetS results underscore the unique scientific capabilities and challenges of visible-light adaptive optics protoplanet imaging. Differential imaging that exploits the enhanced luminosity of emission lines from accreting companions is a powerful tool to isolate planetary signals and distinguish them from stellar and disk contributions. However, it is not without scientific and technical complexities, including challenges in distinguishing companion emission from reflected circumstellar disk light and mitigating the effects of extreme PSF variation across observations. 

Similar to the 11-target H$\alpha$ transitional disk survey conducted by \citet{zurlo19}, we are able to recover known accreting companions, but the discovery of new accreting systems is seemingly rare, even in these highly targeted disk systems that feature multiple signposts of planet formation. At the same time, this first-generation protoplanet survey achieved only moderate contrasts ($10^{-2}--10^{-3}$) in cleared disk regions for most of the targets in the sample, where only the most massive and actively accreting protoplanets are likely to be detectable \citep{Mordasini2017}. We also note the variable SNR of companion recoveries epoch to epoch, a likely result of both variations in observation quality and intrinsic accretion variability, which may explain the many nondetections of planet candidates associated with H$\alpha$ searches like GAPlanetS.

On a more technical note, we use new systematic and robust methodologies of image post-processing optimization to improve PSF subtraction and contrast in visible light high-contrast imagery and apply these methods to the GAPlanetS sample. We utilize a contrast-curve minimization strategy to select an optimal amount of raw data to discard prior to full post-processing analysis. We also implement a data-driven strategy to minimize false positives and recover only the most robust candidates, optimizing KLIP parameters for recovery of false planets injected into continuum images. We uniformly apply this technique to the full survey sample. This strategy removes the need for subjective parameter choices often made in direct imaging, without requiring a single uniform set of PSF subtraction parameters for the entire survey sample that may be poorly matched to some datasets. Our data demonstrate the need for a PSF subtraction methodology that is tuned to the conditions (seeing, PSF variability, total rotation) of the dataset, and our approach offers an unbiased method for conducting uninformed companion searches throughout a region of interest in HCI data. We find that this methodology can lead to both detections of new candidates and recoveries of previously known companions.

\subsection{Future Work and Opportunities}
\label{sec:future}

Surveying transitional disks for H$\alpha$ protoplanet emission remains a powerful and viable method to search for new exoplanets and conduct reconnaissance of the earliest stages of planet formation. To this end, future hardware improvements -- i.e., the next-generation of SDI instrumentation -- will greatly enhance our ability to discover and characterize such systems.  New coronagraph technologies, faster wavefront control, and the use of customized beamsplitters to maximize H$\alpha$ throughput are just some of the near-future improvements. The newly commissioned 2040 actuator MagAO-X high-contrast visible light SDI imager \citep{Males2020} should improve H$\alpha$ contrasts by a factor of 10-100x over MagAO, particularly for brighter targets \citep{Close2020}. As a result, the MagAO-X system will be able to place more stringent constraints on the population of protoplanets inside transitional disk gaps. Other visible light instruments are also performing quite well in this regime, including SCExAO's Visible Aperture Masking Polarimeter Imager for Resolved Exoplanetary Structures \citep[VAMPIRES;][]{Uyama2020} and SPHERE's Zurich Imaging Polarimeter \citep[ZIMPOL;][]{Schmid2018}.

The improved stability of space-based facilities (e.g., HST, \textit{Roman}) will also contribute significantly to our understanding of more distant accreting companions, and indeed has already begun to \citep[e.g.,][]{Zhou2022, Sanghi2022}.

On the ground and in space, hardware upgrades and improvements in algorithms for PSF subtraction and optimization will reveal many higher-contrast planetary accretion signals than reported here. The ability to detect lower-mass and/or more weakly accreting companions will in turn provide more, and more robust, tests of planet formation and accretion theories.  The future of protoplanet imaging is bright.


\acknowledgments
\par We would like to thank the anonymous referee for providing supportive and constructive suggestions for improving this manuscript. 
\par K.B.F., W.O.B., J.A.R., J.M., and C.S. acknowledge funding from NSF-AST-2009816. KBF's work on this project was also supported by a NASA Sagan fellowship. L.M.C. and K.B.F.'s work was supported in part by NASA Exoplanets Research Program (XRP) grants 80NSSC18K0441 and 80NSSC21K0397 and NSF-AAG-1615408. The MagAO system would not have been possible without construction support from the NSF MRI and ATI programs (MRI-0321312, ATI-206422, ATI-1506818). W.O.B. thanks the LSSTC Data Science Fellowship Program, which is funded by LSSTC, NSF Cybertraining Grant \#1829740, the Brinson Foundation, and the Moore Foundation; their participation in the program has benefited this work. K.M.M.'s work has been supported by the NASA XRP by cooperative agreement NNX16AD44G. K.W. acknowledges support from NASA through the NASA Hubble Fellowship grant HST-HF2-51472.001-A awarded by the Space Telescope Science Institute, which is operated by the Association of Universities for Research in Astronomy, Incorporated, under NASA contract NAS5-26555.
\par This paper includes data gathered with the 6.5 meter Magellan Telescopes located at Las Campanas Observatory, Chile. We thank TJ Rodigas for his help in collecting some of the data reported in this work. 
\par This work has made use of data from the European Space Agency (ESA) mission {\it Gaia} (\url{https://www.cosmos.esa.int/gaia}), processed by the {\it Gaia} Data Processing and Analysis Consortium (DPAC, \url{https://www.cosmos.esa.int/web/gaia/dpac/consortium}). Funding f'or the DPAC has been provided by national institutions, in particular the institutions participating in the {\it Gaia} Multilateral Agreement.
\par We would like to acknowledge the land that the observations used in this paper were taken from. Las Campanas Observatory, and the Magellan Clay Telescope, are built on Diaguita land. More on the Diaguita is available from the Museo Chileno de Arte Precolombino: \url{http://precolombino.cl/en/culturas-americanas/pueblos-originarios-de-chile/diaguita/}.

%

\vspace{5mm}
\facilities{Magellan (MagAO)}


\software{\texttt{astropy} \citep{astropy:2013, astropy:2018}, \texttt{emcee} \citep{emcee}, \texttt{photutils} \citep{photutils}, \texttt{ptemcee} \citep{ptemcee}, \texttt{pyklip} \citep{Wang2015}}

\clearpage


\appendix

\section{Individual Target Summaries \label{sec:targets}}

This appendix contains brief literature reviews for each of the 14 targets observed as part of the GAPlanetS Campaign. Targets are discussed in the order in which they appear in Section \ref{sec:results}. The reviews focus on: (a) stellar age, mass, and moving group membership estimates, (b) disk morphological characteristics, and (c) previously reported direct and indirect evidence for planet candidates in these systems.

\subsection{HD~142527}
HD~142527~A is a young \citep[$5.0\pm1.5$Myr;][]{Mendigutia2014} transition disk host. Its Gaia DR3 position and motion (see Table\ref{tab:sample}) are consistent with membership in the Upper Centaurus Lupus star forming region \citep[membership probability of 92.10\% per Banyan $\Sigma$;][]{Gagne2018}. The central star has a mass of $\mathrm{M_A}=2.0\pm0.3M_{\mathrm{\odot}}$ \citep{Mendigutia2014} and is a F6III-V type Herbig Ae/Be star $R$=6~mag\citep{Ofek2008}. HD~142527~A is actively accreting \citep{Mendigutia2014} from an unresolved inner disk that is likely replenished by gas flowing through the massive cavity ($\sim$ 30au to $\sim$ 140au\citep{Avenhaus2017,Avenhaus2014}). At sub-mm wavelengths, the cavity shows complex spiral arm and horseshoe structures observed \citep[e.g.,][]{Ohashi2008, Boehler2017, Garg2021}. Similarly complex and asymmetric scattered light structures \citep[e.g.,][]{Fukagawa2006, Avenhaus2014, Hunziker2021} have been observed in the NIR. 

The low-mass stellar companion HD~142527~B was first detected via Sparse Aperture Masking at the Very Large Telescope at $H$, $K$ and $L$', and its mass estimated at $\sim$0.1-0.4$M_{\odot}$ via pre-main-sequence model fitting of its infrared photometry \citep{Biller:2012}. The first noninterferometric direct detection of the low-mass stellar companion was made with MagAO in 2013 \citep{Close2014}. At a separation of just 86~mas, HD~142527~B was detected in both H$\alpha$ and continuum emission using simple classical angular differential imaging. It was found to be 1.2~mag brighter at H$\alpha$ than in the continuum.  These data served as the first proof of concept that direct H$\alpha$ emission from a companion could be isolated at $<$0$\farcs$1 separations with visible light HCI, and led to the development of the GAPlanetS campaign. For a full review of the literature surrounding the relationship between the HD~142527~B companion and the wide central cavity, see \citet{Balmer2022}. 

\subsection{PDS~70}
PDS~70 is a 0.82$M_{\odot}$ K7 star \citep{Riaud2006}. Its Gaia DR3 position and motion (see Table\ref{tab:sample}) are consistent with membership in the Upper Centaurus Lupus star forming region \citep[membership probability of 98.7\% per Banyan $\Sigma$;][]{Gagne2018}. The companion PDS~70~b was discovered by \citet{Keppler2018} using SPHERE SHINE data in $L$', $K$, and $H$ bands, and subsequently recovered in archival NICI imagery. Comparison with various hot- and warm-start evolutionary models suggest a mass of between 5$M_J$ and 14$M_{J}$ for the companion. Comparison of PDS~70~b's location on the H-R diagram with pre-main-sequence evolutionary models suggest that the system has a significantly younger age \citep[5.4$\pm$1Myr;][]{Keppler2018} than is typical of Upper Centaurus Lupus \citep[16$\pm$2Myr;][]{Pecaut2016}. The companion was detected at a separation of 0$\farcs$195 (22AU), well inside the cleared central cavity of PDS~70. 

A second companion, PDS~70~c, was detected first in H$\alpha$ emission by \citet{Haffert2019}, and then recovered in the NIR in reanalyzed VLT SPHERE observations by \citet{Mesa2019}. Both planets were detected in VLTI/GRAVITY observations of the system by \citet{Wang2021}, who found that their orbital properties were consistent with being in 2:1 mean motion resonance. A compact sub-mm continuum signal suggestive of a circumplateary disk has also been recovered with ALMA for PDS~70~c \citep{Benisty2021, Isella2019}, providing further evidence of its protoplanetary nature.

The transitional disk of PDS~70 has been resolved in both NIR scattered light \citep{Hashimoto2012,Keppler2018} and in large grain thermal emission in the sub-mm \citep{Hashimoto2015,Long2018}. The NIR cavity extends to 0$\farcs$39, well beyond the observed location of PDS~70~b, and the sub-mm cavity extends even farther to 0$\farcs$7 \citep{Long2018}. \citet{Long2018} also found evidence for a inner disk extending to $\sim$ 0$\farcs$11. The breadth of the PDS~70 cavity, as well as the variation in cavity radius with wavelength/grain size, is consistent with its nature as a multiplanetary system.

\subsection{LkCa~15}

LkCa~15 is a 1.25$\pm$0.10$M_{\odot}$ star \citep{Donati2019} at a distance of 157.2 $\pm$0.7pc \citep[GAIA DR3;][]{Gaia2022}. Its Gaia DR3 position and motion (see Table\ref{tab:sample}) are consistent with membership in the Taurus-Auriga star forming region \citep[membership probability of 88.2\% per Banyan $\Sigma$;][]{Gagne2018}, though evolutionary model fits suggest an age of $\sim$5~Myr, somewhat older than the canonical age of  Taurus-Auriga \citep[1--2Myr;][]{Kenyon1995}.

Despite its very faint primary star, which makes natural guide star adaptive optics imaging difficult, The LkCa~15 system is among the most well-studied transitional disks because it was the first with a reported protoplanet candidate inside of its disk gap. This object was first identified in NIR Non-Redundant Masking (NRM) data by \citet{Kraus:2012}, and subsequently argued by \citet{Sallum2015} to have been at least two separate protoplanets (LkCa~15~``b" and ``c") that were coincidentally aligned during the first detection epoch. Only one of those protoplanetary candidates (LkCa~15~b) was detected at H$\alpha$ in the initial epoch, but both were detected in multiple LBT NRM epochs \citep{Sallum2016}. Another planet candidate, LkCa~15~d, was detected in one of the LBT epochs. 
    
The system has a well-established inner cavity interior to $\sim$0$\farcs$3 \citep[$\sim$40AU;][]{Thalmann2010, Thalmann2015, Thalmann2016, Currie2019}. This cavity is not entirely cleared, however, as an inner disk component has been directly imaged in polarized scattered light in the optical with SPHERE ZIMPOL by \citet{Thalmann2015} and in the NIR with SPHERE IRDIS by \citet{Thalmann2016}. Photometric and spectroscopic monitoring of LkCa~15 suggests the presence of an inner accretion disk component near the corotation radius that is more highly inclined than the outer disk, as well as magnetospheric accretion funnels that impact the star at high latitudes and an accretion rate of 7.4$\pm$2.8 $\times10^{-10}M_{\odot}/yr$ \citep{Alencar2018}.

\subsection{HD~169142}
HD~169142 is a 1.85$\pm0.25M_{\odot}$ F0V SpT star \citep{Gratton2019} at a distance of 111.6 $\pm$0.4 pc \citep[GAIA DR3;][]{Gaia2022}. This is significantly closer than its previously assumed 145~pc distance, and properties from the literature have been updated in this work to reflect this change where necessary. The Gaia DR3 position and motion of HD~169142 (see Table\ref{tab:sample}) do not suggest membership in a young moving group \citep[99.9\% probability of being a field star per Banyan $\Sigma$;][]{Gagne2018}, and its age is estimated at 6$^{+6}_{-3}$~Myr \citep{Grady2007}. The mass accretion rate onto the primary star is estimated at 1.5--2.7$\times10^{-9}M_{\odot}yr^{-1}$ based on fits to the Pa$\beta$ and Br$\gamma$ lines obtained with SpeX on the IRTF \citep{Wagner2015}. 

 The complex and asymmetric morphology of HD~169142 is highly consistent with the presence of multiple planets. The disk has a cleared central cavity at r$<$15AU with a millimeter and NIR bright cavity rim at $\sim$20AU showing E/W asymmetry, suggesting a possible dust trap to the west of the star \citep{Quanz2013, Osorio2014, Bertrang2018, Momose2015}. The disk also hosts an annular gap from $\sim$30-55AU that is heavily depleted at millimeter wavelengths and less depleted in NIR scattered light and millimeter gas tracers \citep{Fedele2017, Momose2015}. The outer disk contains a second ring of large grain material extending from $\sim$55-85AU and a more extended small grain dust and gas disk that reaches $\sim$1$\farcs$2-1$\farcs$7 ($\sim$200AU) \citep{Quanz2013,Fedele2017}. The outer dust ring was recently resolved with ALMA into three separate narrow rings at 57, 64, and 76 AU \citep{Perez2019}. The inner regions of the disk host several spiral arms resolved in scattered light \citep{Gratton2019}. 

Given its morphological complexity, it is unsurprising that a number of point-sources have been reported inside the cleared regions of the HD~169142 disk. In the inner dust cavity, \citet{Reggiani2014} reported a 12.2$\pm$0.5mag $L$' source at 0$\farcs$156$\pm$0$\farcs$032 and a PA of 7.4$^o\pm$11.3$^o$, and \citet{Biller2014} independently found an L' point-source at a consistent location within error bars (0$\farcs$11$\pm$0$\farcs$03, PA=0$^o\pm$14$^o$) as well as another candidate at 0$\farcs$18 and PA of 33$^o$. \citet{Ligi2018} imaged the disk in the NIR with VLT/SPHERE in both total and polarized intensity. In total intensity with ADI processing, they found several clumps at similar separation ($\sim$0$\farcs$18), but the structures were fairly continuous in PDI and RDI imagery, suggesting that they may be part of an inhomogeneous dust ring at 0$\farcs$18. They also found a structure at 0$\farcs$10 consistent with the overlapping \citet{Reggiani2014} and \citet{Biller2014} $L$' candidate; however, it appeared extended at longer wavelengths, and they interpreted it as a potential second inner dust ring. This hypothesis is further supported by recent Keck/NIRC2 $L$' observations, which are consistent with an inner $\sim$7AU small grain dust ring \citep{Birchall2019}.  The 0$\farcs$18 structures observed by \citet{Ligi2018} were followed up with SPHERE in \citet{Gratton2019} and shown to have astrometry consistent with Keplerian orbital motion. A new clump was also identified inside the outer annular gap at a separation of 0$\farcs$335 and a position angle of 35 exhibiting photometry consistent with a $\sim$2.2$M_J$ planet.

Potential circumplanetary disk detections have also been reported around HD~169142. \citet{Momose2017} observed an $N$-band mid-IR clump to the west of the star at 0$\farcs$116$\pm$0$\farcs$020 and a PA of 250$^o\pm$5$^o$. In the outer annular gap, \citet{Osorio2014} reported a compact 5$\sigma$ excess in 7mm emission at a separation of 0$\farcs$34 (PA$\sim$175) that they interpreted as a possible circumplanetary disk with an estimated mass of $\sim$0.6$M_J$. 

Modeling of disk structures have also led to predictions for masses and locations of planets embedded in HD~169142. \citet{Kanagawa2015} estimated the mass of the planet clearing the 40-70AU gap to be $\ge$0.4$M_J$ based on an analytical relationship between planet mass and gap depletion, while \citet{Dong2017} estimate the mass to be 0.2-2.1$M_J$ from hydrodynamical + radiative transfer simulations of gap opening. Hydrodynamical modeling by \citet{Perez2019} suggests that a single mini-Neptune ($M~<~$10$_{\earth}$) migrating inward from $\sim$69~au to $\sim$64~au is consistent with the triple-ringed structure of the outer disk. 

\subsection{HD~100546}
 HD~100546 is an A0 star \citep{Gray2017}. Its Gaia DR3 position and motion (see Table\ref{tab:sample}) are consistent with membership in the Lower Centaurus Crux star forming region \citep[membership probability of 98.90\% per Banyan $\Sigma$;][]{Gagne2018}
 The disk around HD~100546 is morphologically complex, with evidence for an inner clearing, multiple spiral arms, and two planet candidates. The HD~100546 ``b" protoplanet candidate was first reported by \citet{Quanz2013} and confirmed by \citet{Quanz2015} and \citet{Currie2015}, but more recent attempts to recover the planet have failed \citep{Rameau2017}. 
 
 For a full review of the complex multiwavelength morphology of this disk and limits on candidate protoplanets in this system, see \citet{Follette2017}. The GAPlanetS data for HD~100546 were analyzed in detail in \citet{Follette2017} and \citet{Rameau2017}, but are revisited with improved processing in this study.

\subsection{SAO~206462}
SAO~206462, also referred to in the literature as HD~135344B, is a $1.7^{+0.2}_{-0.1}M_{\odot}$ F4Ve star \citep{Muller2011} at a distance of 135 $\pm$0.4pc \citep[GAIA DR3;][]{Gaia2022}. Its Gaia DR3 position and motion (see Table\ref{tab:sample}) are consistent with membership in the Upper Centaurus Lupus star forming region \citep[membership probability of 99.5\% per Banyan $\Sigma$;][]{Gagne2018}, which has an age of 16$\pm$2Myr \citep{Pecaut2016}. The disk hosts two potentially planet-induced spiral arms, first seen in scattered light by \citet{Muto:2012}. 

Computational models with a massive outer companion have been shown to create qualitatively similar spiral features ($\sim$6$M_J$ at $r=0\farcs6$ and $PA=10^{\circ}$ per \citet{Dong2015}, $\sim$10$M_J$ at $\sim$100au if S1 is the primary arm or $\sim$15$M_J$ at $\sim$150AU if S2 is primary, per \citet{Bae2016}). However, direct imaging searches for the presence of massive perturbers in the outer disk of SAO~206462 have ruled out the presence of companions more massive than $\sim6M_J$ beyond the spirals, suggesting that the perturber is either located interior to the arms or is less massive than predicted \citep{Maire2017}. 

At millimeter wavelengths, high-resolution ALMA imagery of the disk reveals a contiguous ring centered at $\sim0\farcs4$ as well as a more distant azimuthally asymmetric dust crescent centered around $\sim0\farcs6$ with multiwavelength properties consistent with predictions for dust vortices \citep{Cazzoletti2018}. The location of the millimeter overdensity is coincident with some predictions for the location of the perturber responsible for the S1 spiral arm \citep{Muto2012, Stolker2016a}, suggesting that the arm may be incited by dust overdensity at this location \citep{Perez2014, Cazzoletti2018} rather than a point-source perturber. However, computational models have shown that planetary perturbers in the outer disk can also incite similar overdensities \citep{Bae2016}. \citet{vanderMarel2015} constrained the large grain dust cavity wall to $\sim$40AU with a heavily depleted interior, but also show that the disk hosts a smaller gas cavity in $^{13}$CO and C$^{18}$O that is heavily but not fully depleted (r$\sim$30AU as in the NIR, $\delta_{\rm{gas}}=2\times10^{-4}$). The 10au inconsistency in gap radii between large thermally emitting grains, NIR-scattering grains, and gas is suggestive of the presence of planets in the gap. 

\citet{Casassus2021} imaged the disk in high-resolution  J=(2--1) CO isotopologues and adjacent continuum, and in doing so, detected a fine continuum filament (sep $\sim0\farcs468$, PA $\sim216.1\circ$) connecting the inner disk and outer crescent. They also reanalyzed the NIR data from \citet{Stolker2017} to derive a best-fit location for the perturber, which they found to be radially shifted $\> 0\farcs00421$ from the center of the filament. 

\citet{Cugno2019} used SPHERE to search for H$\alpha$ signals inside the disk gap of SAO~206462 and in the outer disk regions where, they point out, the presence of small dust grains is likely to heavily extinct H$\alpha$ emission from any embedded planets. They place a limit of $<$2.4$\times$10$^{-12}M_J/yr$ for a $\sim$10.2$M_J$ planet (derived from the detection limit of \citet{Maire2017}) at the outer radius of the scattered light cavity (0$\farcs$18). They also reported a tentative detection of a low SNR point-source at a separation of 71mas and a PA of 19$^{\circ}$.  

\subsection{TW~Hya}
TW~Hya is the nearest \citep[60.1~$\pm$~0.1 pc, GAIA DR3;][]{Gaia2022} disk-bearing T Tauri star to Earth, and its Gaia DR3 position and motion (see Table\ref{tab:sample}) are consistent with membership in the eponymous TW~Hya association \citep[membership probability of 99.9\% per Banyan $\Sigma$;][]{Gagne2018}. Pre-main-sequence model fits to the high-resolution NIR spectrum of TW~Hya suggest that it is an M0.5 (0.6$\pm$0.1M$_{/odot}$) star with an age of 8$\pm$3Myr \citep{Sokal2017}. 

High-resolution ALMA observations by \citet{Andrews2016} revealed concentric dark rings/gaps centered at 1, 24, 41, and 48au, and fainter gaps at 13, 31 and 34au. The sensitivity of their observations was such that disk emission was not detectable beyond $\sim$60--70~au.

The highest-resolution scattered-light images to date were reported in \citet{vanBoekel2017}. They revealed three concentric, moderately depleted (50-80\%) scattered light gaps centered at $\sim$7~au, 22~au, and 90~au, as well as a spiral feature in the outer disk beyond the $\sim$90~au gap. \citet{Teague2019} searched for comparable structure in the gas disk and found spiral substructure in both the gas velocity and temperature maps. 

Estimates for the masses of planets responsible for the $\sim$20 and $\sim$90au scattered light gaps range from 0.05-0.5$M_J$ and 0.01-3$M_J$, respectively \citep{Rapson2015, Dong2017b, vanBoekel2017, Mentiplay2019}. Results of the numerical simulations of \citet{Dong2017b} suggested that the shallow millimeter gaps at 41 and 48au could be generated by a single $\sim$30M$_{\oplus}$ planet located between them at $\sim$ 44au, and \citet{Bae2017} suggested that the shock generated by a secondary planet-induced spiral arm might also carve the gap at $\sim20$au. 

Recently, \citet{Tsukagoshi2019} reported the presence of a 12$\sigma$ millimeter continuum excess emission located at 52~au and a PA of -133$^{\circ}$ that they suggest is consistent with either a dust clump/vortex or a circumplanetary disk around a Neptune-mass planet, though there is no known gap at this location. \citet{Ilee2022} confirmed this detection of with an 8$\sigma$ millimeter excess at nearly the same location. 

\subsection{HD~100453}
HD~100453A is a Herbig A9.5Ve (1.7$M_{\odot}$) star with a M4V stellar companion (HD~100453~B, $\sim0.3M_{\odot}$) at 1$\farcs$05 separation \citep{Collins2009, Wagner2018b}. The system, located at a distance of 103.8 $\pm$0.2pc \citep[GAIA DR3;][]{Gaia2022}, has a Gaia DR3 distance and motion (see Table\ref{tab:sample}) consistent with membership in the 15$\pm$3 Myr \citep{Pecaut2016} Lower Centaurus Crux association \citep[membership probability of 99.3\% per Banyan $\Sigma$;][]{Gagne2018}. 

Scattered light imagery has revealed a depleted small grain dust cavity extending from the inner working angle of 0$\farcs$09 to 0$\farcs$14, as well as striking spiral arm features and a number of time-variable ``shadows'' suggestive of an unresolved inner disk component that is misaligned with the outer disk \citep{Benisty2017,Long2018}. 
The disk is resolved in CO~(2--1) line emission from 0$\farcs$23 to 1$\farcs$10, while in scattered light, the spiral arms in the disk extend to 0$\farcs$37 and the inner ring of emission extends from 0$\farcs$18 to 0$\farcs$25 \citep{VanderPlas2019, Wagner2015b}.
Millimeter continuum emission reveals an inner cavity extending from the resolution limit of 0$\farcs$09 out to 0$\farcs$22, and an annular gap from 0$\farcs$40 to 0$\farcs$48. The observed spiral features are consistent with being driven by the known outer stellar companion HD~100453~B\citep{Dong2016}. 

\subsection{CS~Cha}
CS~Chamaeleonis is a young $2\pm2$ Myr  \citep{Luhman2008} spectroscopic binary system \citep{Guenther2007} comprised of two T-Tauri stars with spectral type K2Ve \citep{Manara2014}. Its distance of 168.8 $\pm$1.9 pc \citep[GAIA DR3;][]{Gaia2022} and sky position suggest membership in the Chamaeleon I association \citep{Ginski2018b}. CS~Cha hosts a smooth, low-inclination disk with an outer radius of $0\farcs312$ in polarized light \citep[$i=24.2^o\pm 3.1^o$;][]{Ginski2018b} and an inner cavity that is estimated to extend to $18^{+6}_{-5}$~au based on SED modeling \citep{Ribas2016}. 

The CS~Cha system contains a comoving polarized ($13.7\pm0.4\%$ in J-band) companion, CS~Cha B, which lies beyond the outer radius of the circumbinary disk at a projected separation of 1\farcs19 \citep{Ginski2018b}. In order to explain the photometry of the system, \citet{Ginski2018b} originally proposed that the companion was either a heavily-extincted brown dwarf ($\sim20M_{jup}$) or a planetary-mass companion with an unresolved disk or dust envelope; however, follow-up observations with with VLT MUSE by \citet{Haffert2020} suggested that CS~Cha B is most likely a heavily disk-obscured mid-M type stellar companion ($M$=0.07-0.71$M_{odot}$). \citet{Haffert2020} resolved the H$\alpha$ emission line of the companion and found that CS~Cha B is still actively accreting, with an estimated accretion rate of 4$\times$10$^{-11\pm0.4}M{\odot}$yr$^{-1}$. They estimate the continuum brightness of the companion at $\sim$10 magnitudes fainter than the primary (a contrast of 10$^{-4}$). 

\subsection{HD~141569}
HD~141569A is a young \citep[$\sim$5Myr;][]{Merin2004}, $2.39^{+0.04}_{-0.05}$ $M_{\odot}$ \citep{White2016}, Herbig A2Ve star \citep{Gray2017} at a distance of 111.6 $\pm$0.4~pc \citep{Gaia2022}. Its Gaia DR3 position and motion (see Table\ref{tab:sample}) do not suggest membership in a young moving group \citep[99.9\% probability of being a field star per Banyan $\Sigma$;][]{Gagne2018}. \citet{Aarnio2008} conducted a search for a comoving group for HD~141569 and found that the system likely formed in isolation. HD~141569A is the primary star of a hierarchical triple with two M-dwarf companions that lie $\sim9"$ beyond the circumstellar disk. 

The disk of HD~141569~A is generally classified as a debris disk \citep{Hughes2018} and it is significantly depleted in millimeter grains \citep{Wyatt2015a}. It has four concentric scattered-light gaps between $0\farcs25-0\farcs4, 0\farcs43-0\farcs52, 0\farcs60-0\farcs69, 1\farcs-2\farcs$ \citep{Perrot2016, Konishi2016}, as well as a narrow ring of millimeter emission centered at 2". The distribution of $^{13}$CO~(2--1) gas emission in the system is asymmetrical, with a peak 1$\farcs$1 from the star at a PA of $\sim$-36$^{\circ}$ \citep{Miley2018}.

\subsection{PDS~66}
Also known as MP Mus, PDS~66 is a K1Ve \citep{DaSilva2009a} $1.4\pm0.1 M_\odot$ \citep{Avenhaus2018} star with Gaia DR3 position and motion (see Table\ref{tab:sample}) consistent with membership in the 15$\pm$3 Myr \citep{Pecaut2016} Lower Centaurus Crux association \citep[membership probability of 97.5\% per Banyan $\Sigma$;][]{Gagne2018}. It should be noted that previously reported distance measurements are significantly different ($\sim10$ pc) than the most recent \textit{Gaia} distance of 97.9$\pm$0.1 pc \citep[GAIA DR3;][]{Gaia2022}. The moderately inclined ($i=31^{\circ}\pm2^{\circ}$) disk surrounding PDS~66 has a resolved small grain dust gap from $0\farcs46$ to $0\farcs81$, as reveled in NIR polarized intensity imagery \citep{Wolff2016}. \citet{Kastner2010} used the CO line emission profile to constrain the gas disk's outer radius to $\approx$ 120 AU. \citet{Cortes2009} detected 3~mm and 12~mm continuum emission towards PDS~66; however, it has not yet been resolved.

\subsection{UX~Tau~A}
UX~Tau~A is a K2Ve star with two companions- UX Tau B at $\sim5\farcs8$ (itself a tight binary with $\sim0\farcs1$ separation) and UX Tau C at $\sim2\farcs7$ \citep{Kraus2009, Schaefer2014}. The system, at a distance of 142.2$\pm$0.7 pc \citep[GAIA DR3;][]{Gaia2022}, has Gaia DR3 position and motion (see Table\ref{tab:sample}) consistent with membership in the 1-2 Myr \citep{Kenyon1995} Taurus-Auriga Association \citep[membership probability of 98.1\% per Banyan $\Sigma$;][]{Gagne2018} Submillimeter Array (SMA) images have revealed a disk of large grains with a cavity interior to $\sim{0\farcs18}$, a peak near $\sim{0\farcs23}$ and an outer extent of $\sim{0\farcs34}$ \citep{Andrews2011a}. Sub-mm gas emission \citep[1.3mm, CO (2-1) ALMA images;][]{Akeson2019} traces these structures closely. HiCIAO polarimetric imaging of the disk suggests that it is moderately inclined ($i=46^{\circ}\pm2^{\circ}$) and extends from an IWA of 0\farcs16 to 0\farcs86, but the inner gap detected in thermal emission is not resolved \citep{Tanii2012}. 

\subsection{V1247~Ori}
V1247 Orionis is a single F0V star with a mass of $1.86\pm0.02 M_{\odot}$ and an age of $7.4\pm0.4~Myr$ \citep[estimated from PMS evolutionary tracks;][] {Kraus2013}. It resides at a distance of 401.3$\pm$3.2 pc \citep[GAIA DR3;][]{Gaia2022} and resides within the $\epsilon$ Ori association \citep{Caballero2008}. In 2016, \citet{Ohta2016} observed the star in scattered NIR light, detecting an arc-like structure at $0\farcs28\pm0\farcs09$ spanning position angles from $60^{\circ}$ to $210^{\circ}$. \citet{Kraus2017} used ALMA to resolve the disk in $870 \mu$m continuum, CO 3-2, and H$_{12}$CO 4-3 emission. These images revealed an asymmetrical crescent at a separation of $0\farcs38$. The arc-like structure revealed in \citet{Ohta2016} lies interior to the millimeter arc and may represent an accretion stream onto a planet. \citet{Kraus2017} hypothesized that this emission represents a spiral arm inclined relative to the inner disk by approximately $17^{\circ}$, and simulations assuming a planet mass of 3 $M_{\rm{Jup}}$ were able to reproduce these features to high accuracy.

\subsection{V4046~Sgr}
V4046 Sagittarii is a young binary system composed of two K-type T~Tauri stars \citep[K5Ve/K7Ve, 0.9/0.85 $M_\odot$;][]{Nefs2012,Rosenfeld2012} at a distance of 71.5$\pm$ 0.1pc \citep[GAIA DR3;][]{Gaia2022} with Gaia DR3 position and motion (see Table\ref{tab:sample}) consistent with membership in the $24\pm3$ Myr \citep{Bell2015} $\beta$ Pictoris association \citep[membership probability of 98.4\% per Banyan $\Sigma$;][]{Gagne2018}. The disk was imaged in 15 molecular gas tracers by \citet{Kastner2018}, who found that the morphology varied by spectral line. Some tracers exhibited sharp (e.g HC$_3$N and C$_2$H) or diffuse (eg. DCN, H$^{13}$CO$^+$) ring-like features, while others showed smooth disks (e.g. CO, HCN), with detectable emission extending as far as $4\farcs0$ ($^{12}$CO). Scattered light images reveal a cavity interior to $0\farcs19$ and a brighter northern edge \citep{Rapson2015}. In the 1.3 millimeter continuum, a narrow ring-like feature has been identified at $0\farcs18$ and a thicker outer ring from $0\farcs34$ to $0\farcs84$ \citep{Martinez–Brunner2022}.

\clearpage

\section{Image Galleries for all Epochs\label{sec:allepochs}}

\begin{figure*}[h]
    \centering
\includegraphics[width=\textwidth]{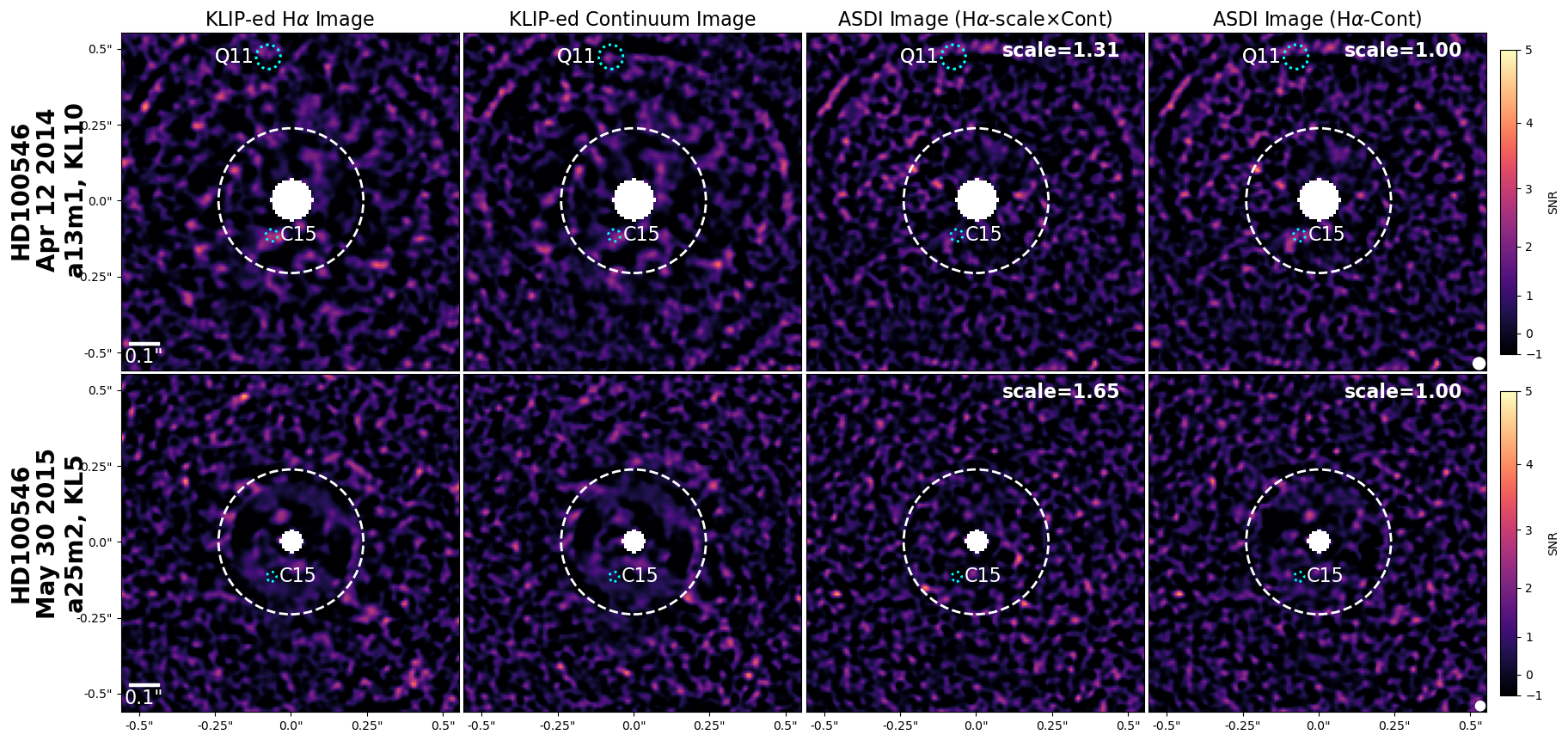}
    \caption{Final KLIPed H$\alpha$ (left), continuum (middle left), stellar H$\alpha$/continuum-scaled ASDI (middle right), and 1:1 scaled ASDI (right) imagery for both HD~100546 epochs. \texttt{pyKLIP} reduction parameters (indicated in the text labels to the left of each image panel) have been optimized for recovery of false continuum planets injected between the IWA and control radius, as described in detail in the text. The AO control radius of the images is indicated with a white dashed circle. The \citet{Currie2015} and \citet{Quanz2013} planet candidate locations at the original detection epoch are marked with dashed cyan circles labeled ``C15" and ``Q11", respectively}
    \label{fig:hd100546}
\end{figure*}

\begin{figure*}
    \centering
    \includegraphics[width=\textwidth]{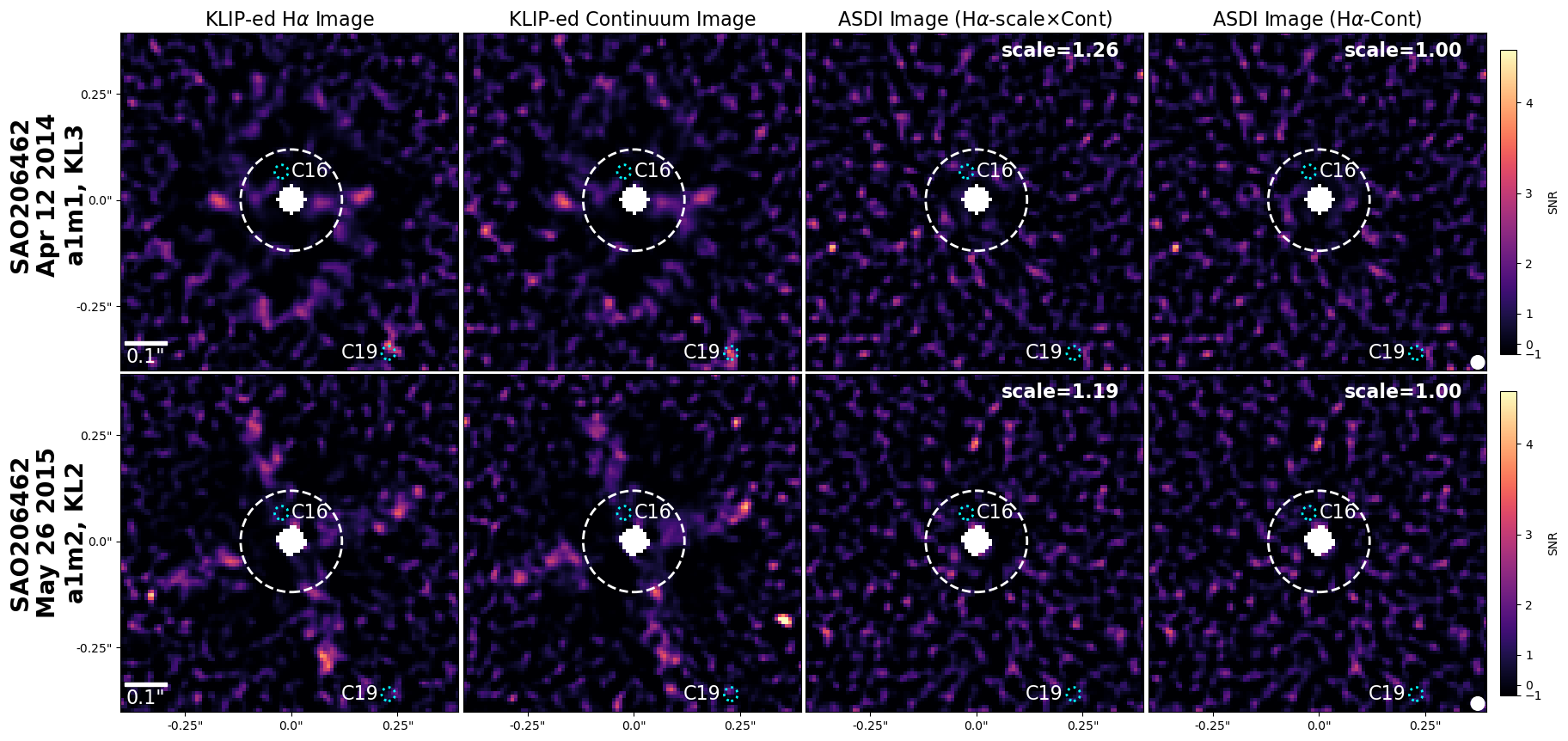}
    \caption{Final KLIPed H$\alpha$ (left), continuum (middle left), stellar H$\alpha$/continuum-scaled ASDI (middle right), and 1:1 scaled ASDI (right) for both SAO~206462 epochs. \texttt{pyKLIP} reduction parameters (indicated in the text labels to the left of each image panel) have been optimized for recovery of false continuum planets injected between the IWA and control radius, as described in detail in the text. The AO control radius of the images is indicated with a white dashed circle. The \citet{Cugno2019} and \citet{Casassus2021} planet candidate locations at the original detection epoch are marked with dashed dashed cyan circles labeled ``C16" and ``C19", respectively}
    \label{fig:sao206462}
\end{figure*}

\begin{figure*}
    \centering
    \includegraphics[width=\textwidth]{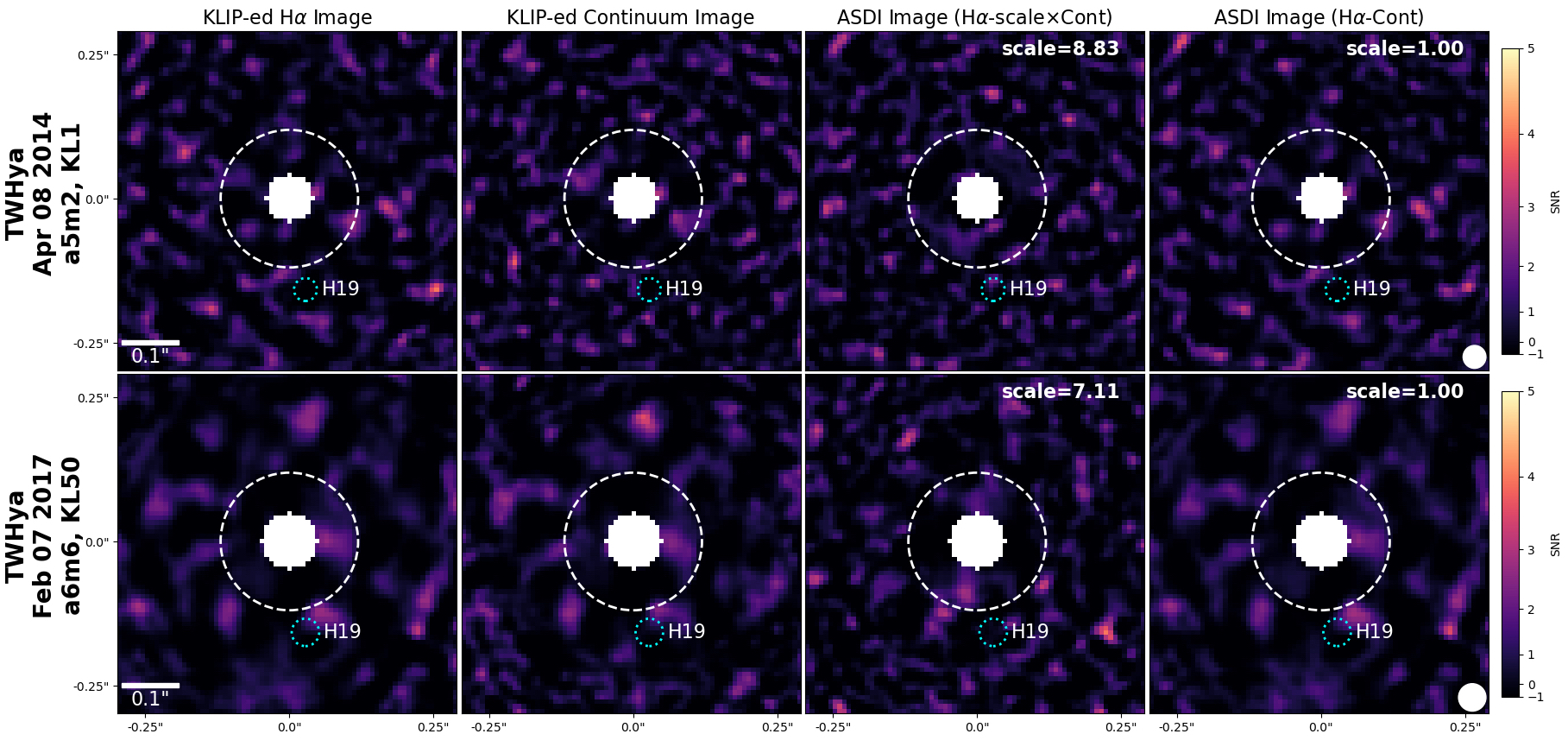}
    \caption{Final KLIPed H$\alpha$ (left), continuum (middle left), stellar H$\alpha$/continuum-scaled ASDI (middle right), and 1:1 scaled ASDI (right) imagery for both TW~Hya epochs. \texttt{pyKLIP} reduction parameters (indicated in the text labels to the left of each image panel) have been optimized for recovery of false continuum planets injected between the IWA and control radius, as described in detail in the text. The AO control radius of the images is indicated with a white dashed circle. The \citet{Huelamo2022} point-source candidate (a suspected artifact according to the authors) is marked with a dashed cyan circle labeled ``H19".}
    \label{fig:twhya}
\end{figure*}

\begin{figure*}
    \centering
    \includegraphics[width=\textwidth]{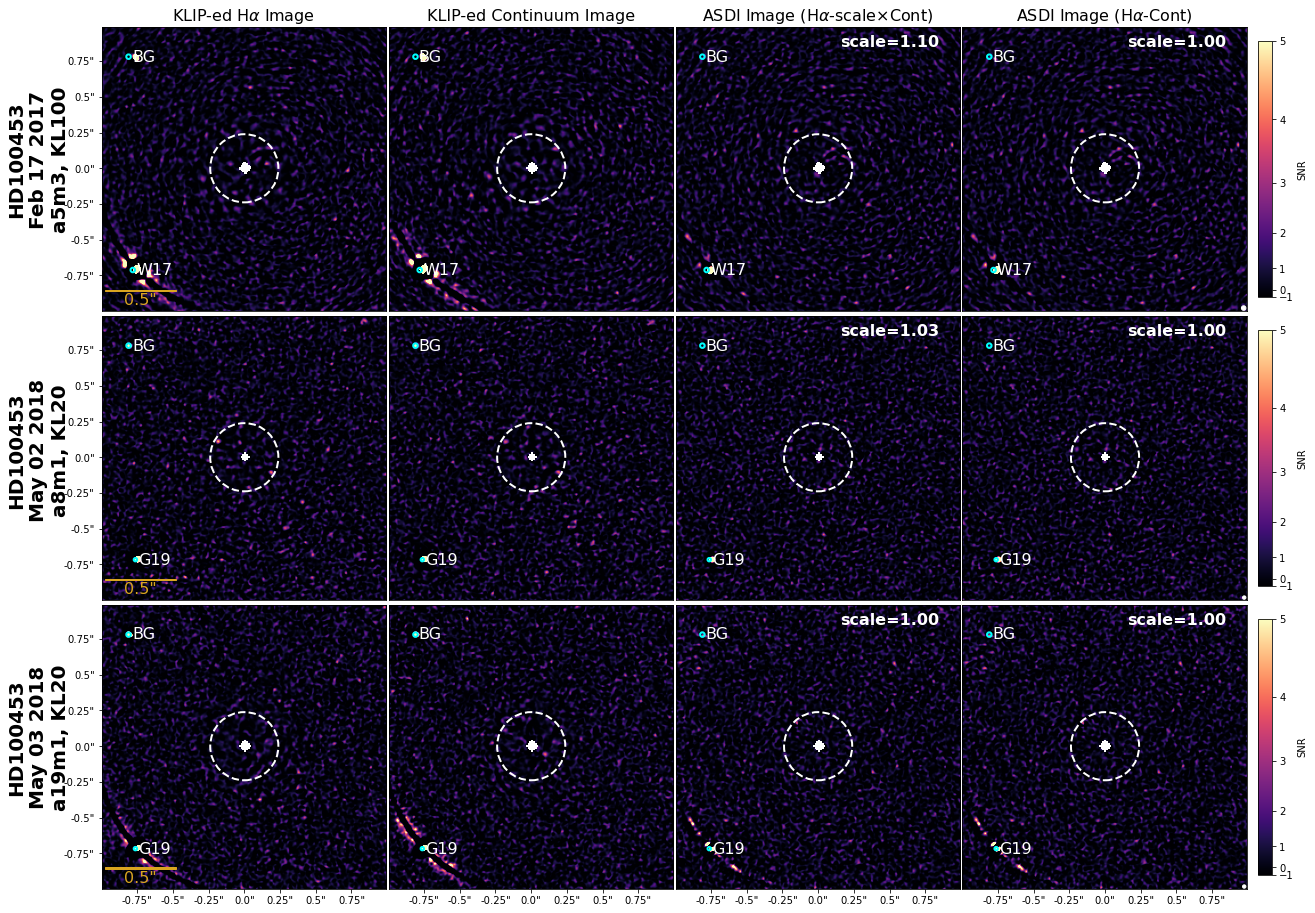}
    \caption{Final KLIPed H$\alpha$ (left), Continuum (middle left), stellar H$\alpha$/continuum-scaled ASDI (middle right), and 1:1 scaled ASDI (right) imagery for all three HD~100453 epochs. \texttt{pyKLIP} reduction parameters (indicated in the text labels to the left of each image panel) have been optimized for recovery of false continuum planets injected between the IWA and control radius, as described in detail in the text. The AO control radius of the images is indicated with a white dashed circle. The companion HD~100453~B, is clearly visible in the lower left of each image panel, and its locations from \citet{Wagner2018b} and \citet{Gonzalez2020} are marked with cyan circles labeled ``W17" and ``G19", respectively. A known background star lies in the upper left-hand corner of each H$\alpha$ and continuum image panel (except in the 2018 May 2 epoch, when it is saturated and therefore not completely removed through SDI subtraction), but is absent in the SDI imagery because it is not actively accreting.}
    \label{fig:hd100453}
\end{figure*}

\begin{figure*}
    \centering
    \includegraphics[width=0.7\textwidth]{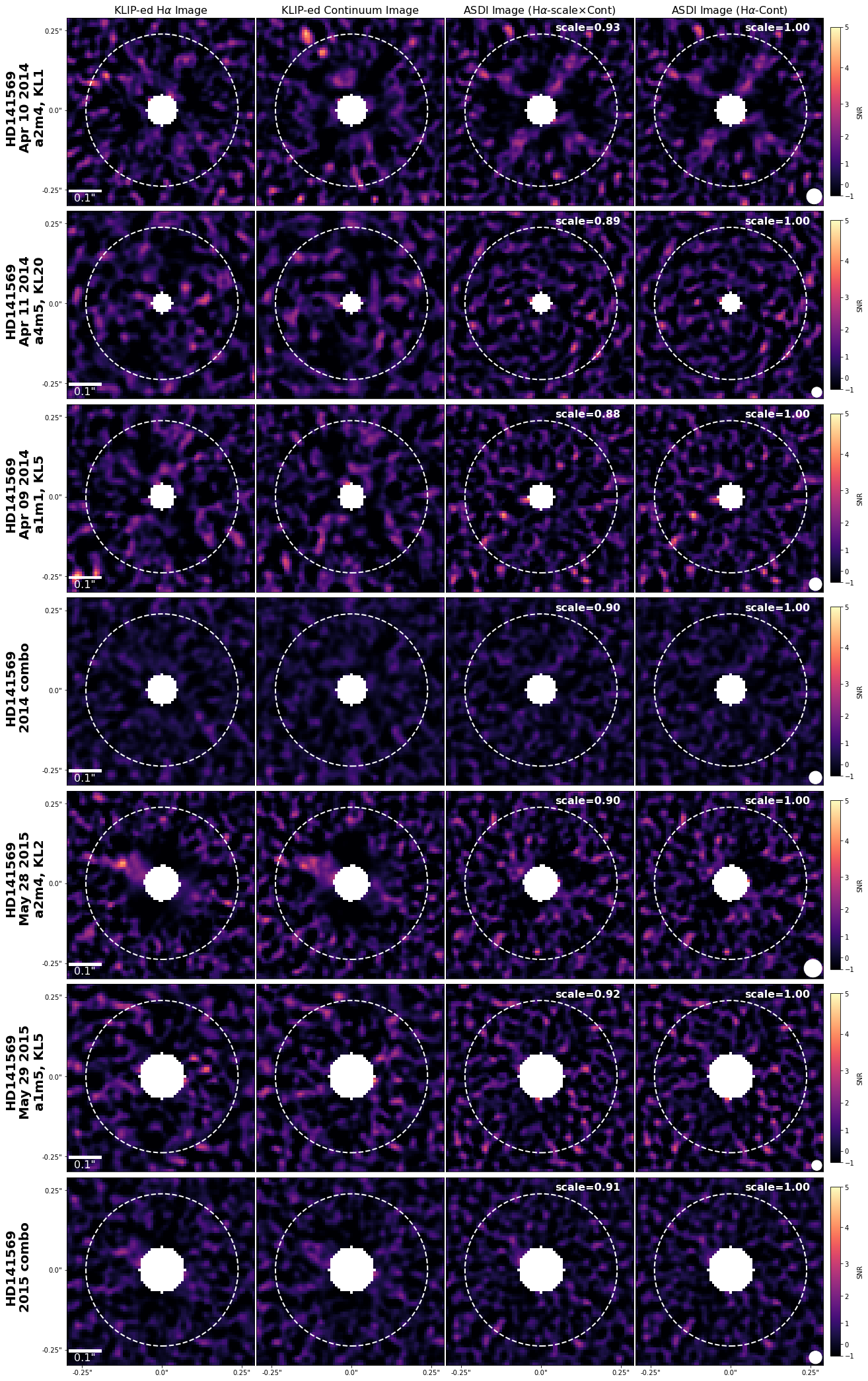}
    \caption{Final KLIPed H$\alpha$ (left), continuum (middle left), stellar H$\alpha$/continuum-scaled ASDI (middle right), and 1:1 scaled ASDI (right) imagery for all five HD~141569 epochs, as well as the combination of the three 2014 epochs (fourth panel) and the two 2015 epochs (seventh panel). \texttt{pyKLIP} reduction parameters (indicated in the text labels to the left of each image panel) have been optimized for recovery of false continuum planets injected between the IWA and control radius, as described in detail in the text. The AO control radius of the images is indicated with a white dashed circle.}
    \label{fig:hd141569}
\end{figure*}

\begin{figure*}
    \centering
    \includegraphics[width=\textwidth]{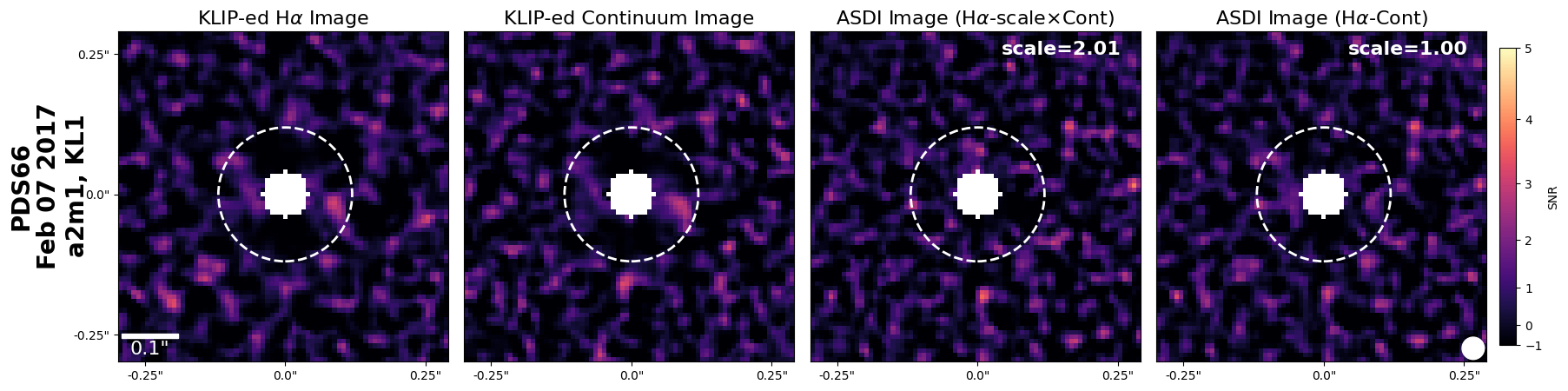}
    \caption{Final KLIPed H$\alpha$ (left), continuum (middle left), stellar H$\alpha$/continuum-scaled ASDI (middle right), and 1:1 scaled ASDI (right) imagery for the single PDS~66 epoch. \texttt{pyKLIP} reduction parameters (indicated in the text labels to the left of the image panel) have been optimized for recovery of false continuum planets injected between the IWA and control radius, as described in detail in the text. The AO control radius of the images is indicated with a white dashed circle.}
    \label{fig:pds66}
\end{figure*}

\begin{figure*}
    \centering
    \includegraphics[width=\textwidth]{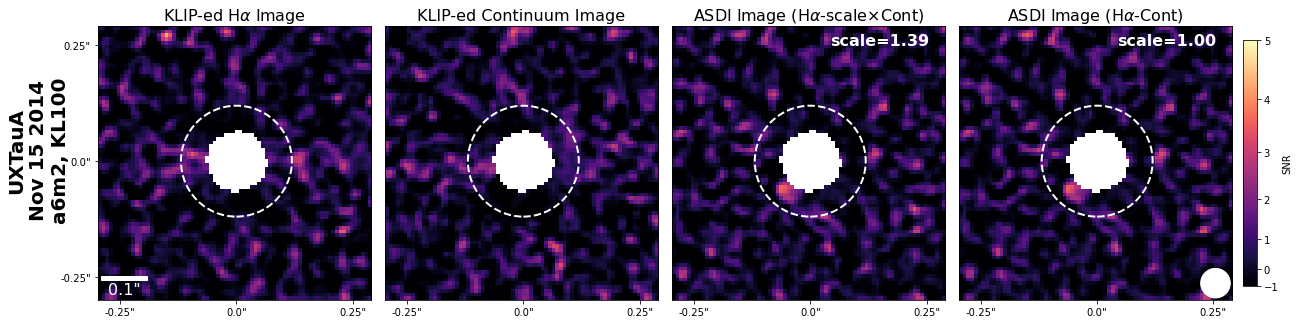}
    \caption{Final KLIPed H$\alpha$ (left), continuum (middle left), stellar H$\alpha$/continuum-scaled ASDI (middle right), and 1:1 scaled ASDI (right) imagery for the single UX~Tau~A epoch. \texttt{pyKLIP} reduction parameters (indicated in the text labels to the left of the image panel) have been optimized for recovery of false continuum planets injected between the IWA and control radius, as described in detail in the text. The AO control radius of the images is indicated with a white dashed circle.}
    \label{fig:uxtaua}
\end{figure*}

\begin{figure*}
    \centering
    \includegraphics[width=\textwidth]{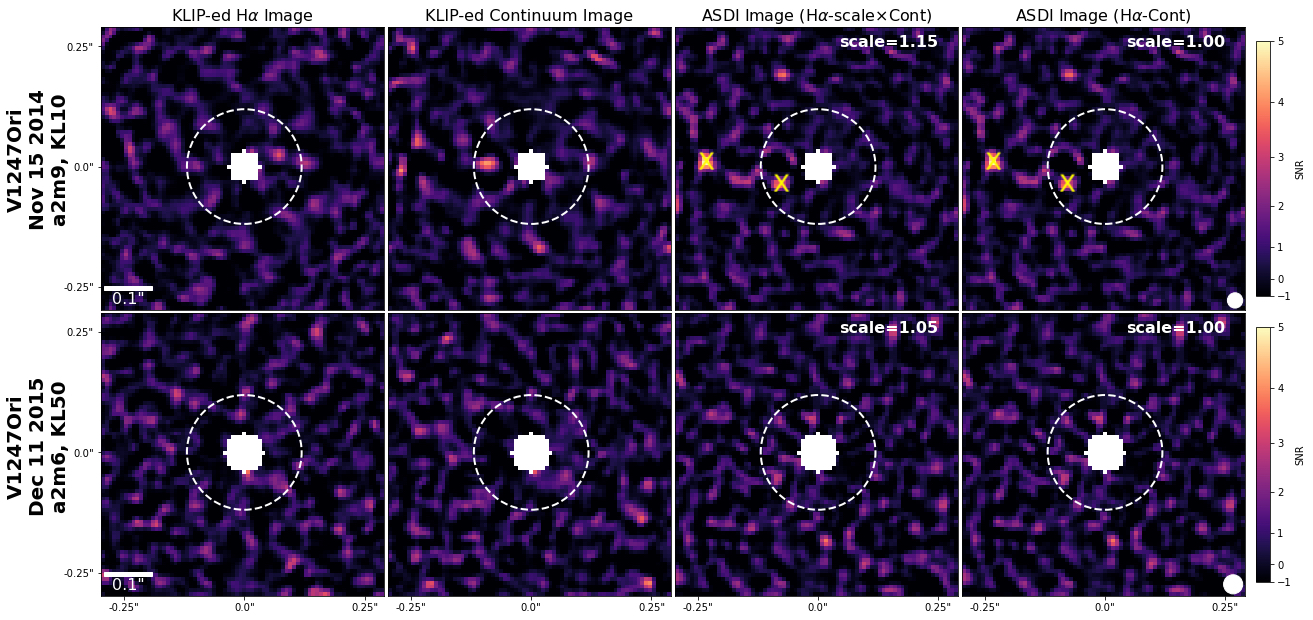}
    \caption{Final KLIPed H$\alpha$ (left), continuum (middle left), stellar H$\alpha$/continuum-scaled ASDI (middle right), and 1:1 scaled ASDI (right) for both V1247~Ori epochs. \texttt{pyKLIP} reduction parameters (indicated in the text labels to the left of each image panel) have been optimized for recovery of false continuum planets injected between the IWA and control radius, as described in detail in the text. The AO control radius of the images is indicated with a white dashed circle.}
    \label{fig:v1247ori}
\end{figure*}

\begin{figure*}
    \centering
    \includegraphics[width=\textwidth]{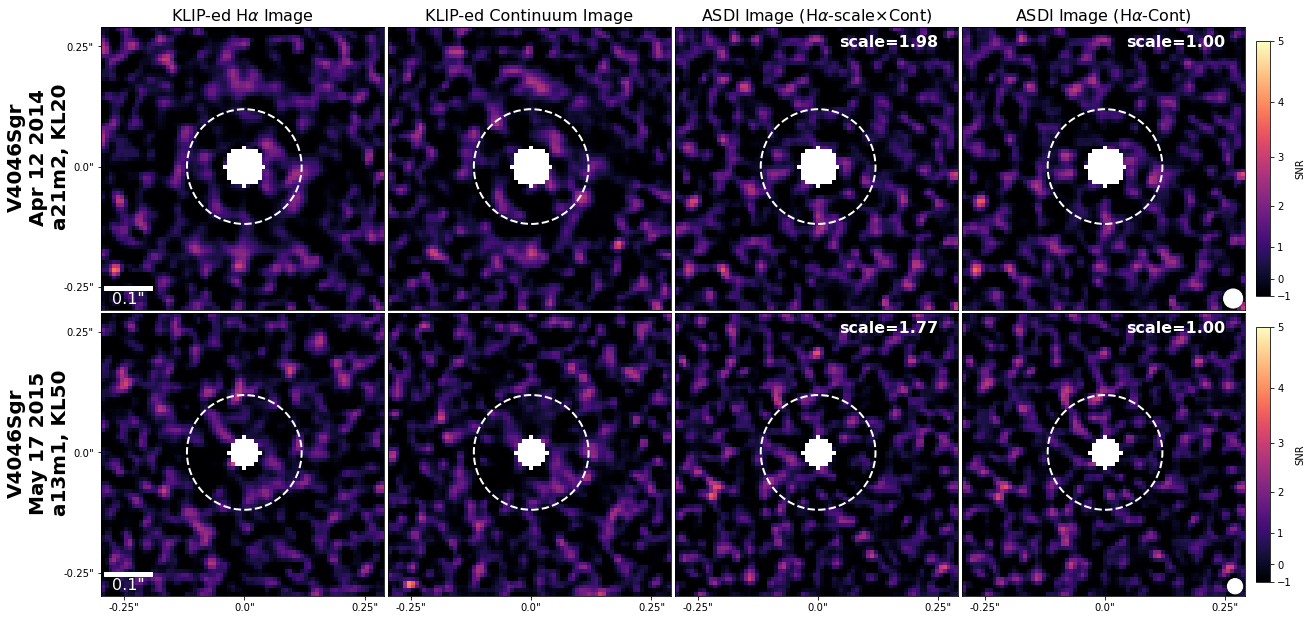}
    \caption{Final KLIPed H$\alpha$ (left), continuum (middle left), stellar H$\alpha$/continuum-scaled ASDI (middle right), and 1:1 scaled ASDI (right) imagery for both V4046~Sgr epochs. \texttt{pyKLIP} reduction parameters (indicated in the text labels to the left of each image panel) have been optimized for recovery of false continuum planets injected between the IWA and control radius, as described in detail in the text. The AO control radius of the images is indicated with a white dashed circle.}
    \label{fig:v4046sgr}
\end{figure*}

This appendix includes Figures \ref{fig:hd100546}--\ref{fig:v4046sgr}, in which we provide H$\alpha$, continuum, and SDI reductions (conservative and 1:1 scaled) for all GAPlanetS datasets not shown in the main text, sorted by object. Although we consider all of these datasets nondetections, we note that there are a number of $>$3$\sigma$ point-sources in many/most datasets. These sources were deemed less compelling in our survey analysis because they either (a) appear in SDI imagery without a clear H$\alpha$ counterpart, or (b) do not appear in a consistent location in epochs that are closely spaced in time and similar in quality. At the same time, given the variable image quality and stochastic nature of accretion processes, we note that nondetection in subsequent epochs is not equivocal proof of a false positive. We provide full reductions of all datasets here for future reference, as some of these candidates may prove to be \textit{bona fide} companions.

\clearpage




\bibliography{ref.bib}



\end{document}